\documentclass[namedreferences]{solarphysics}

\usepackage[hyperref,optionalrh]{spr-sola-addons} 
\usepackage{graphicx}        
\usepackage{xcolor}
\usepackage{subcaption}
\usepackage{amssymb}        
\usepackage{color}           
\usepackage{longtable}
\usepackage{lscape}
\usepackage{geometry}
 \usepackage[
singlelinecheck=false 
]{caption}
\usepackage[normalem]{ulem}

\chardef\us=`\_

\begin{document}

\begin{article}
\begin{opening}

\title{Coronal Elemental Abundances during A-class Solar Flares Observed by Chandrayaan-2 XSM}

\author[addressref={aff1},corref,email={lakshithanama@gmail.com, lakshitha.nm@res.christuniversity.in}]{\inits{L.}\fnm{Lakshitha}~\lnm{Nama}}
\author[addressref={aff2},corref,email={biswajit70mondal94@gmail.com}]{\inits{B.}\fnm{Biswajit}~\lnm{Mondal}}
\author[addressref={aff3}]{\inits{S.}\fnm{S.}~\lnm{Narendranath}}
\author[addressref={aff1}]{\inits{KT.}\fnm{KT.}~\lnm{Paul}}

\address[id=aff1]{Department of Physics and Electronics, CHRIST (Deemed to be University), Hosur Main Road, Bangalore, India}
\address[id=aff2]{Physical Research Laboratory, Navrangpura, Ahmedabad, Gujarat, India}
\address[id=aff3]{Space Astronomy Group, U R Rao Satellite Centre, ISRO, Bengaluru, India}
\address[id=aff1]{Department of Physics and Electronics, CHRIST (Deemed to be University), Hosur Main Road, Bangalore, India}

\runningauthor{L. Nama et al.}
\runningtitle{Coronal Elemental Abundances during A-class Solar Flares}

\begin{abstract} 
The abundances of low First Ionisation Potential (FIP) elements are three to four times higher (FIP bias) in the closed loop active corona than in the photosphere, known as the FIP effect. Observations suggest that the abundances vary in different coronal structures. Here, we use the soft X-ray spectroscopic measurements from the Solar X-ray Monitor (XSM) on board the Chandrayaan-2 orbiter to study the FIP effect in multiple A-class flares observed during the minimum of solar cycle 24. Using time-integrated spectral analysis, we derive the average temperature, emission measure, and the abundances of four elements - Mg, Al, Si, and S. We find that the temperature and emission measure scales with the flares sub-class while the measured abundances show an intermediate FIP bias for the lower A-flares (e.g., A1), while for the higher A-flares, the FIP bias is near unity. To investigate it further, we perform a time-resolved spectral analysis for a sample of the A-class flares and examine the evolution of temperature, emission measure, and abundances. We find that the abundances drop from the coronal values towards their photospheric values in the impulsive phase of the flares, and after the impulsive phase, they quickly return to the usual coronal values. The transition of the abundances from the coronal to photospheric values in the impulsive phase of the flares indicates the injection of fresh unfractionated material from the lower solar atmosphere to the corona due to chromospheric evaporation. However, explaining the quick recovery of the abundances from the photospheric to coronal values in the decay phase of the flare is challenging.

\end{abstract}
\end{opening} 

\section{Introduction}
     \label{intro} 
     
Understanding the elemental abundances at different layers of the Sun can provide crucial inputs to model the solar atmosphere. This may aid our understanding of the mass and energy flow from the photosphere to the corona via the chromosphere, which is most likely transported through coronal loops. \cite{1963Pottasch} discovered that elements with low First Ionization Potential (FIP $\leq$ 10eV) are more abundant in the closed loop active corona than in the photosphere (termed as the FIP bias). This phenomenon has come to be known as the FIP effect, and research in the years since its discovery has validated it (e.g., \citealp{Meyer1985, Feldman1992, Fludra1999, Schmelz2012, Phillips2012, Narendranath2014, Dennis_2015}). It is also established that the FIP bias differs depending on the features being observed in the corona. The quiet coronal regions have a FIP bias of 1-2 (e.g., \citealp{Feldman1992, Feldman2000, Feldman2003, Baker2013}), while new loops in active regions have a photospheric composition and slowly develop the coronal FIP bias of 3 to 4 \citep{Widing_2001, Baker2015}. 
Using the sun-as-a-star observations, 
\cite{Brooks2017} and \cite{Pipin2018} have also shown that the FIP bias is highly correlated with the solar cycle phase.
The FIP fractionation in the solar wind is observed to be dependent on its speed \citep{Zurbuchen1999, VonSteiger2000}. The slow solar wind has a FIP bias in the range 2-5, while the fast solar wind has a significantly smaller degree of FIP bias \citep{Feldman1998b, Bochsler2007a}.

Abundance studies of larger flares and surges have also been carried out starting from \cite{Feldman1990, Feldman1992}, who derived the relative abundances of elements such as Mg and O.
\cite{Fludra1999} derived the absolute abundances (with respect to H) of S, Ca, and Fe for 57 large flares.
Similarly, \cite{DelZanna2013, Warren2014, Sylwester2014, Sylwester2015, Dennis_2015} used line to continuum measurements with improved spectroscopic instruments
to report that the abundances of low FIP elements during 
the flares are near to their photospheric values. \cite{Phillips2003} derived the absolute abundance of K and relative abundances of Ar and S 
for four long-duration flares of GOES class M or higher. \cite{Narendranath2014} 
derived the absolute abundances of Fe, Ca, Si and S for 20 flares. Both \cite{Phillips2003} and \cite{Narendranath2014} found that the abundance of low FIP elements is enhanced by a factor of two in the corona while the intermediate FIP and high FIP elements retain their photospheric abundances during the flare peak. It should, however, be noted that there is a large scatter in the reported values of FIP bias from these studies due to the uncertainties arising from the different analysis techniques used~\citep{zanna_2018}. Recently, inverse FIP (IFIP effect: depletion of low FIP elements in the corona with respect to the photospheric composition) was reported for Si, S and Ar during the peak of 4 X-class flares by \cite{Katsuda2020}.
\cite{Baker2019} had also observed the IFIP effect at the footpoints of loops during two confined M-class flares in an active region, with the flaring loop tops showing a FIP effect. 

A few attempts are made to study the evolution of the abundances during large solar flares; e.g., \cite{Sylwester1984} reported spectroscopic evidence for the variation of Ca abundance during high-temperature solar flare plasmas.
\cite{Narendranath_2020} found a variation of the FIP bias during larger M class flares for the elements Fe, Ca, Si, and S. They noted that the variation is lowest for the mid-FIP element Sulfur. Further, they showed that stronger solar flares tend to have lower FIP bias.

The Solar X-ray Spectrometer (XSM) on board the Chandrayaan-2 orbiter is one of the next generation spectrometers having a higher resolution and cadence compared to earlier instruments. The XSM started solar observations in September 2019, which was during the minimum of solar cycle 24. Observations from XSM have previously been used to study the quiet Sun \citep{Vadawale_2021b} and microflares \citep{Vadawale_2021} outside active regions. \cite{Mondal_2021} derived the abundance evolution of low FIP elements - Mg, Al, Si and S during nine B-class flares. They have shown a transition of the abundances from preflare coronal to the near photospheric during the impulsive phase of the flare and again from near photospheric to the coronal values in the decay phase. To complement the XSM observations, \cite{Zanna_2022ApJ} carried out a companion study of abundances for an active region and a B-class flare using simultaneous observations of XSM along with AIA \citep{Lemen2012} onboard Solar Dynamics Observatory \citep{Pesnell2012}, as well as XRT \citep{Golub2007} and EIS \citep{culhane2007} onboard Hinode \citep{Kosugi2007}.

Though the earlier spectroscopic observations reported the variation of the abundances (or FIP bias) for the larger flares as well as smaller flares up to B-class, a detailed study for the much smaller A-class (or below) flares were difficult because of their small signals. The study of these smaller flares is of great interest in the entire solar physics community, as they are very common in active regions and may contribute to the properties of quiescent active region emission. In this work, we have used the XSM observations of much smaller flares of GOES A-class to study the elemental abundance and its evolution by performing spectroscopic analysis. 

The rest of the paper is organized as follows: in Section~\ref{obs} we provide details of the XSM observation, data analysis, and identification of events. In Section~\ref{spec}, we present the details of the spectroscopic analysis. After discussing the results in Section~\ref{res}, we summarize the article in Section~\ref{sum}.

\section{Observations and data analysis}
\label{obs}

The Solar X-ray Monitor (XSM:~\citealp{Vadawale2014, Munuswamy2020}) on board the Chandrayaan-2 orbiter~\citep{Vanitha2020} measures the Sun's disk integrated spectra from a lunar orbit, with a cadence of 1 second in the energy range of 1-15 keV \citep{Mithun_2020}. The unique design of XSM makes it possible to measure the X-ray intensity of the Sun over a dynamic range from below A-class to X-class activity~\citep{mithun2021}. It uses a Silicon Drift Detector to achieve an energy resolution better than 180 eV at 5.9 keV~\citep{Mithun_2020}, which was the highest energy resolution broadband X-ray spectrometer observing the Sun during the minimum of solar cycle 24 covering the year 2019-2020.

During the minimum of solar cycle 24, from September 12 to November 20, 2019, and February 14 to May 19, 2020, when the Sun was in the continuous field-of-view (FOV) of XSM, it observed repeatedly flaring (GOES range of A to B class) activity within active regions. 
The objective of the present work is to identify all of the small A-class flares from the XSM observed light curve and then perform spectroscopic analysis to estimate the plasma parameters associated with them. A flare list of all the flares whose peak flux lies between $10^{-8}$ to $10^{-7}$ $Wm^{-2}$ (in the 1-8 $\textup{\AA}$ range) was compiled by visual inspection using the daily light curves available at the official website of XSM\footnote{\hyperlink{https://www.prl.res.in/ch2xsm/}{https://www.prl.res.in/ch2xsm/}}. A total of 72 events satisfying this criterion were found. 

\begin{figure}    
\centerline{\includegraphics[width=\textwidth,clip=]{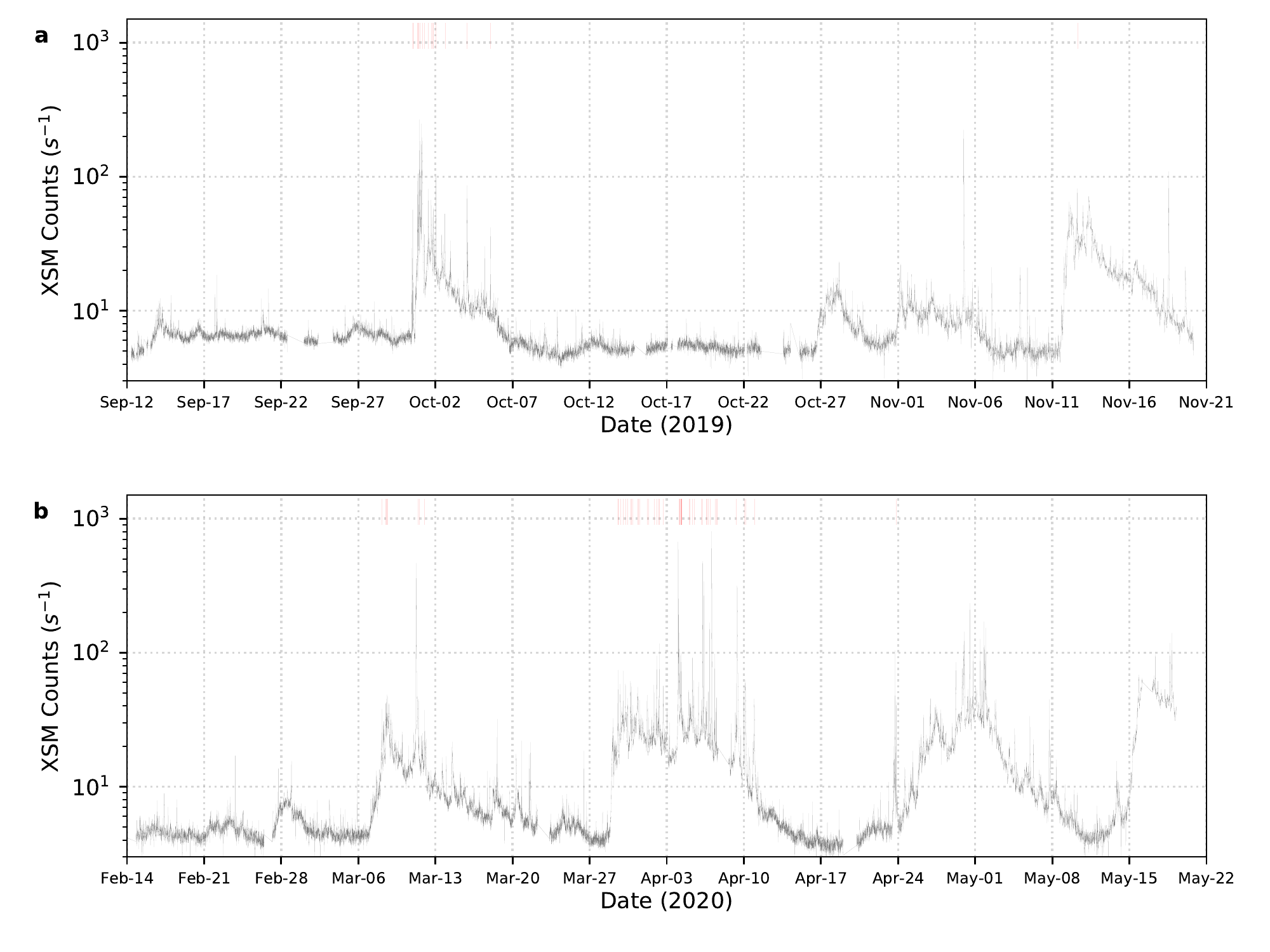}}
\caption{Panels {\bf a} and {\bf b} show the X-ray light curve in the 1-15 keV energy range with a time cadence of 120 s, as measured by XSM from September 12 to November 20, 2019, and February 14 to May 19, 2020, respectively. Vertical red lines represent the peak time of all the observed A-class flares.}
   \label{fig1}
   \end{figure}

XSM data used in this work is publicly available from the ISRO Space Science Data Archive (ISDA)\footnote{\hyperlink{https://pradan.issdc.gov.in/pradan/}{https://pradan.issdc.gov.in/pradan/}}. XSM data consists of day-wise raw (level-1) and calibrated (level-2) files. As XSM has a wide FOV of $\pm$~ 40$^\circ$ and it observes the Sun from a lunar orbit, there are times when the Sun is out of FOV of XSM or is occulted by the moon. Such time periods are used for generating non-solar background observations. Using the XSM level-1 data along with the XSM Data Analysis Software (XSMDAS:~\citealp{Mithun_2021}), we have generated the effective area corrected XSM daily light curve in the energy range of 1-15 keV with a cadence of 120 seconds, as shown in Figure~\ref{fig1}. All the 72 A-class flares during this period are marked by red vertical lines. From the daily light curves, the duration of A-class flares is selected for the generation of spectra as discussed in Section~\ref{spec}.

\begin{figure} 
    
    \subcaptionbox{}{{\includegraphics[width=0.48\textwidth]{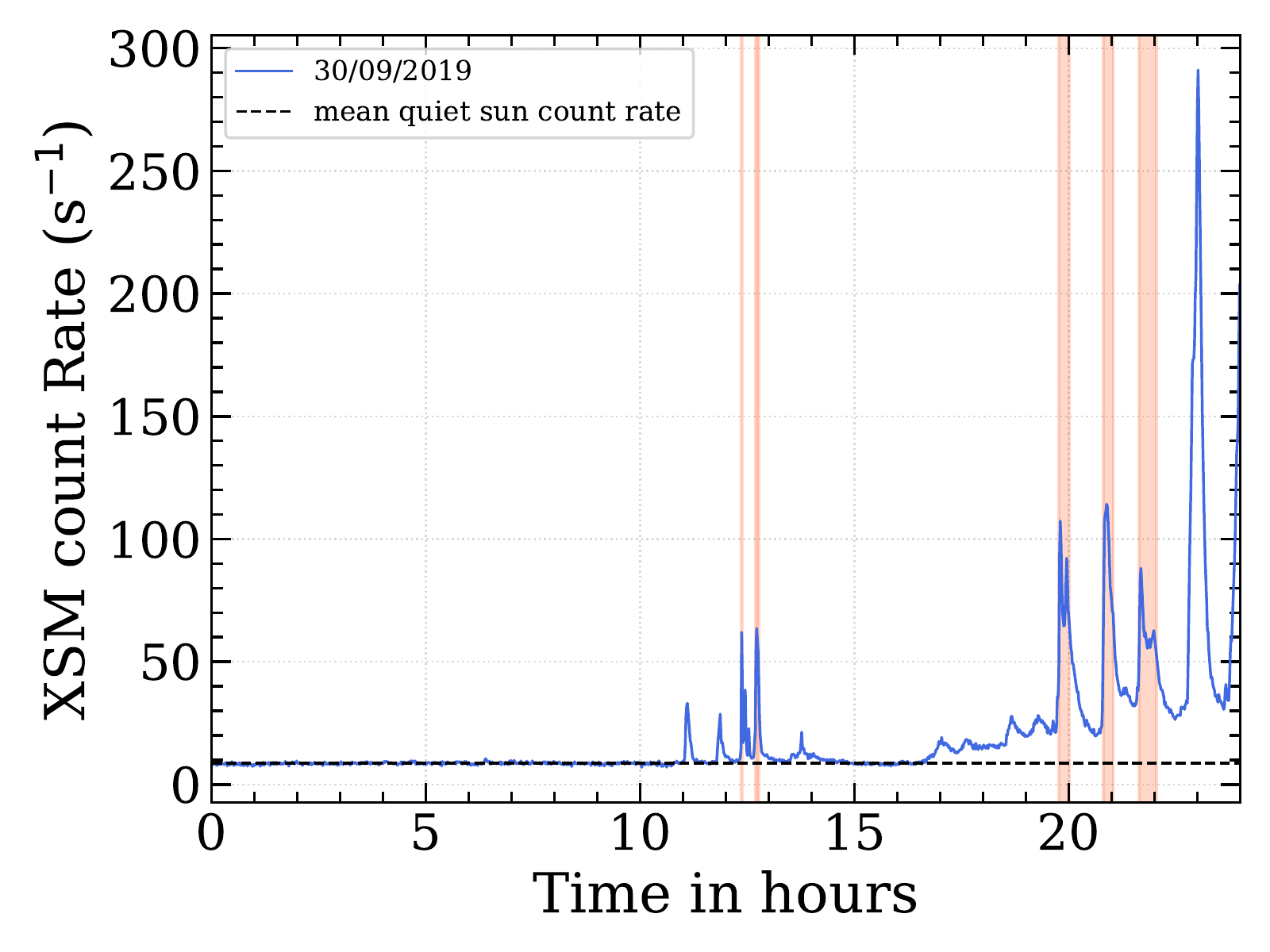} }}
    \hfill
    \subcaptionbox{}{{\includegraphics[width=0.48\textwidth]{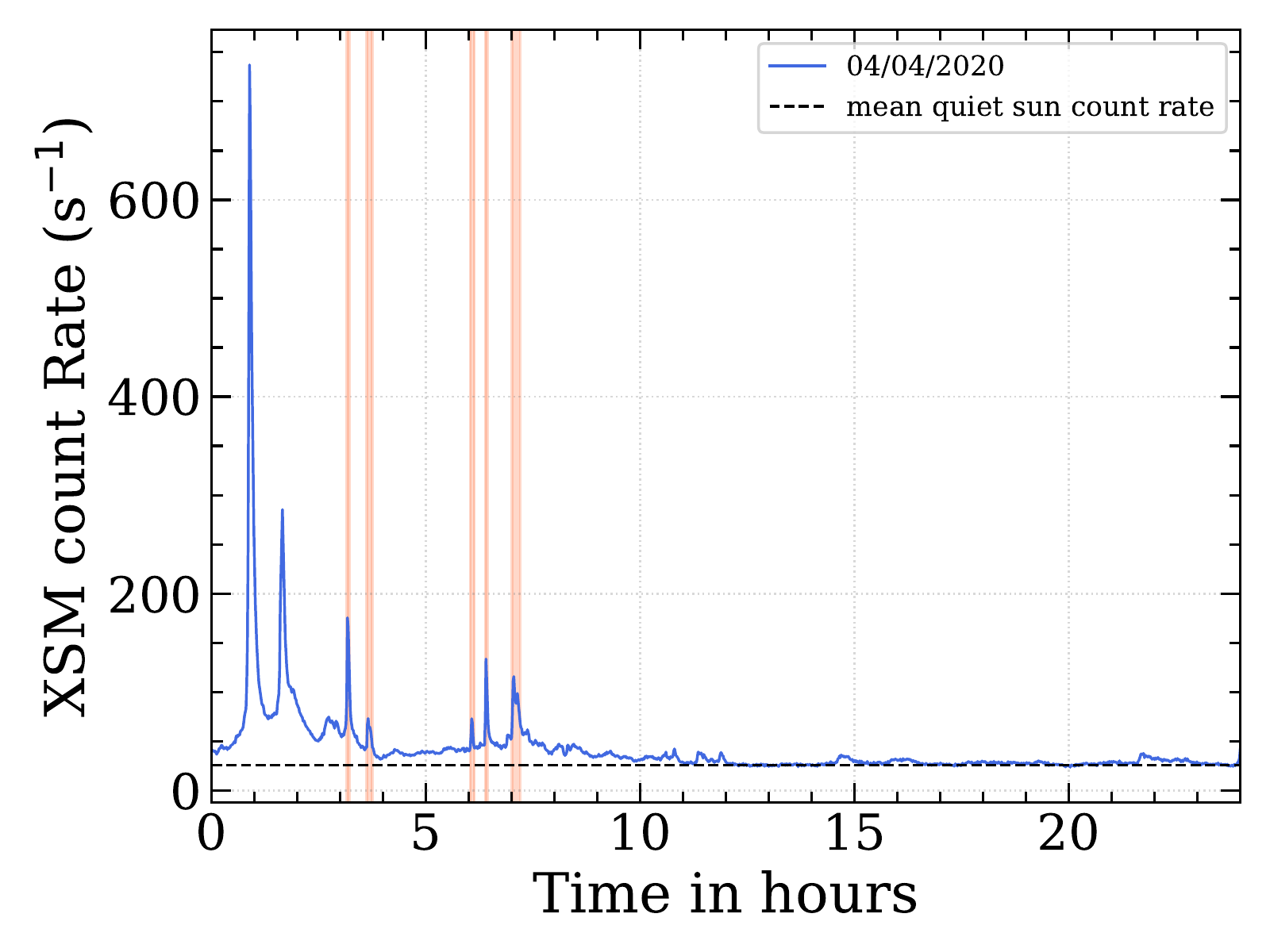} }}
    \caption{Full day XSM 1-15 keV solar X-ray light curve on 2020-09-30 (a) and 2020-04-04 (b). The red shaded regions represent the time durations used to generate the individual flaring emission spectra used for the one-temperature spectral fitting.}
    \label{fig2}
\end{figure}

\section{Spectral analysis}
\label{spec}

Soft X-ray spectrum from the Sun in the energy range of 1-15 keV contains both the continuum and line emissions. The continuum is mainly produced by free-free and free-bound emissions, with contributions also from the two-photon radiative process, whereas the line emissions primarily originate due to the atomic transitions between the different ionization states of the atoms~\citep{zanna_2018}. Modelling the XSM spectrum, containing both continuum and emission lines, allows us to estimate the plasma parameters such as temperature, emission measure, and the abundances of the various elements. 
In this work, we have used the spectral fitting model, `chisoth' \citep{Mondal_2021} for the spectral analysis. `chisoth' is a local model in the X-ray spectral fitting package (XSPEC: \citealp{Arnaud1999}), which uses the CHIANTI database version 10 \citep{DelZanna2021} to generate the synthetic spectra. The input parameters of this model are the logarithm of temperature, volume emission measure and the abundances of the elements with Z=2 to Z=30.

We perform the time-integrated spectral analysis of all the chosen flares as described in Section \ref{stat}. We have also performed the time-resolved spectral analysis (Section~\ref{time}) of some of the A-class flares to understand the temporal evolution of temperature, emission measure and abundances.

\subsection{Time integrated spectral analysis} 
      \label{stat}      

In this section, we discuss the time-integrated spectral analysis of A-class flares in the soft X-ray energy band observed by XSM. 
We first defined the flare start and end time when the count rate came down to 60\% of the XSM peak count rate to ensure that the flaring plasma is significantly separated from the background emissions. For eight of the 72 flares observed, the background active region emission is found to be more than 60\% of the peak rate. Thus, excluding these, the remaining 65 flares are selected for further analysis. Time-integrated spectrum, along with the corresponding Ancillary Response File (ARF), for the selected duration was generated using the `xsmgenspec' module of the XSMDAS by selecting the Good Time Intervals (GTI). Figure~\ref{fig2} shows the light curves for two representative days of 2019-09-30 (panel {\bf a}) and 2020-04-04 (panel {\bf b}) where the red-shaded regions represent durations of the A-class flares for which the time-integrated spectra is generated.

\begin{figure} 
    
    \subcaptionbox{}{{\includegraphics[width=0.47\textwidth]{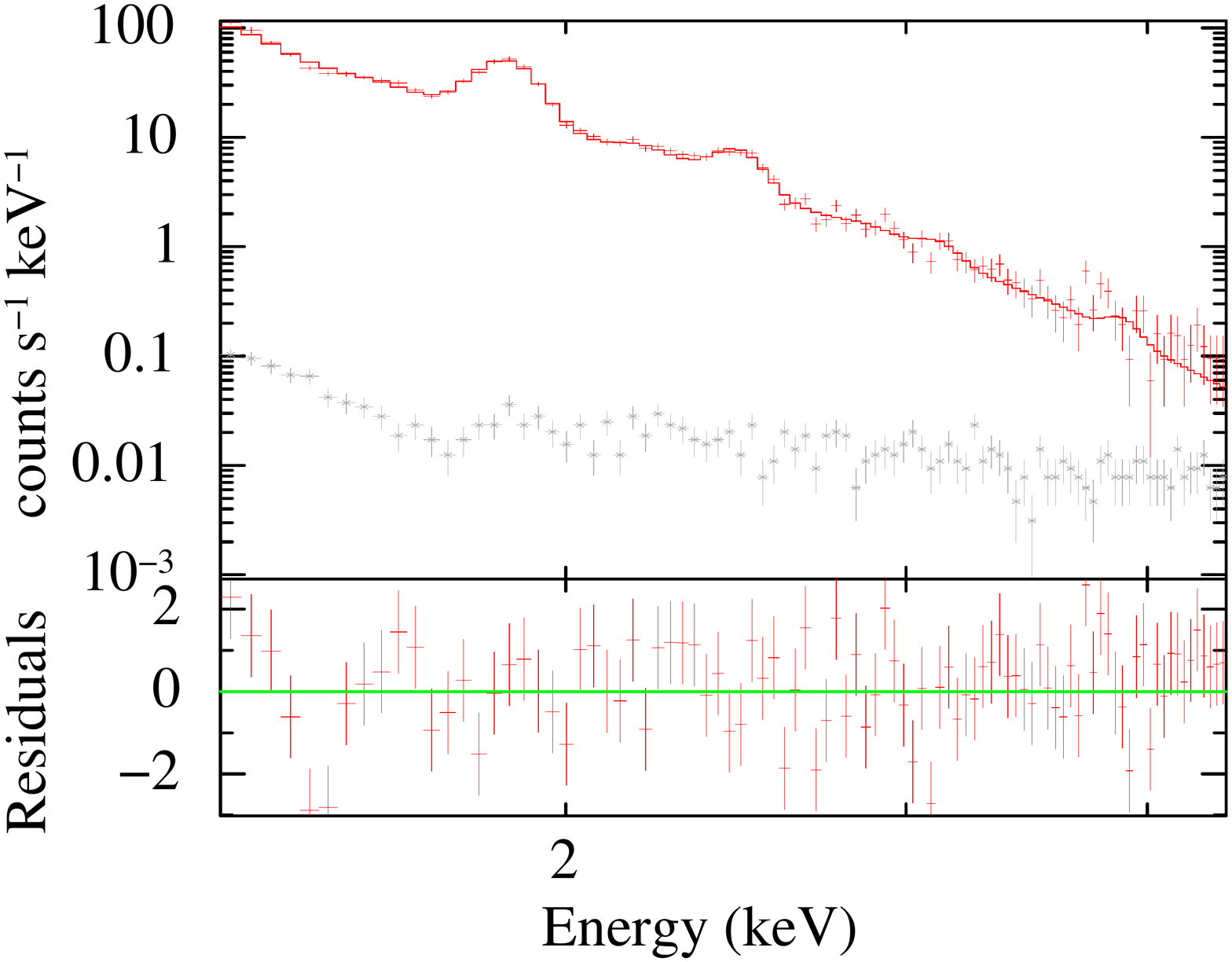} }}
    \hfill
    \subcaptionbox{}{{\includegraphics[width=0.47\textwidth]{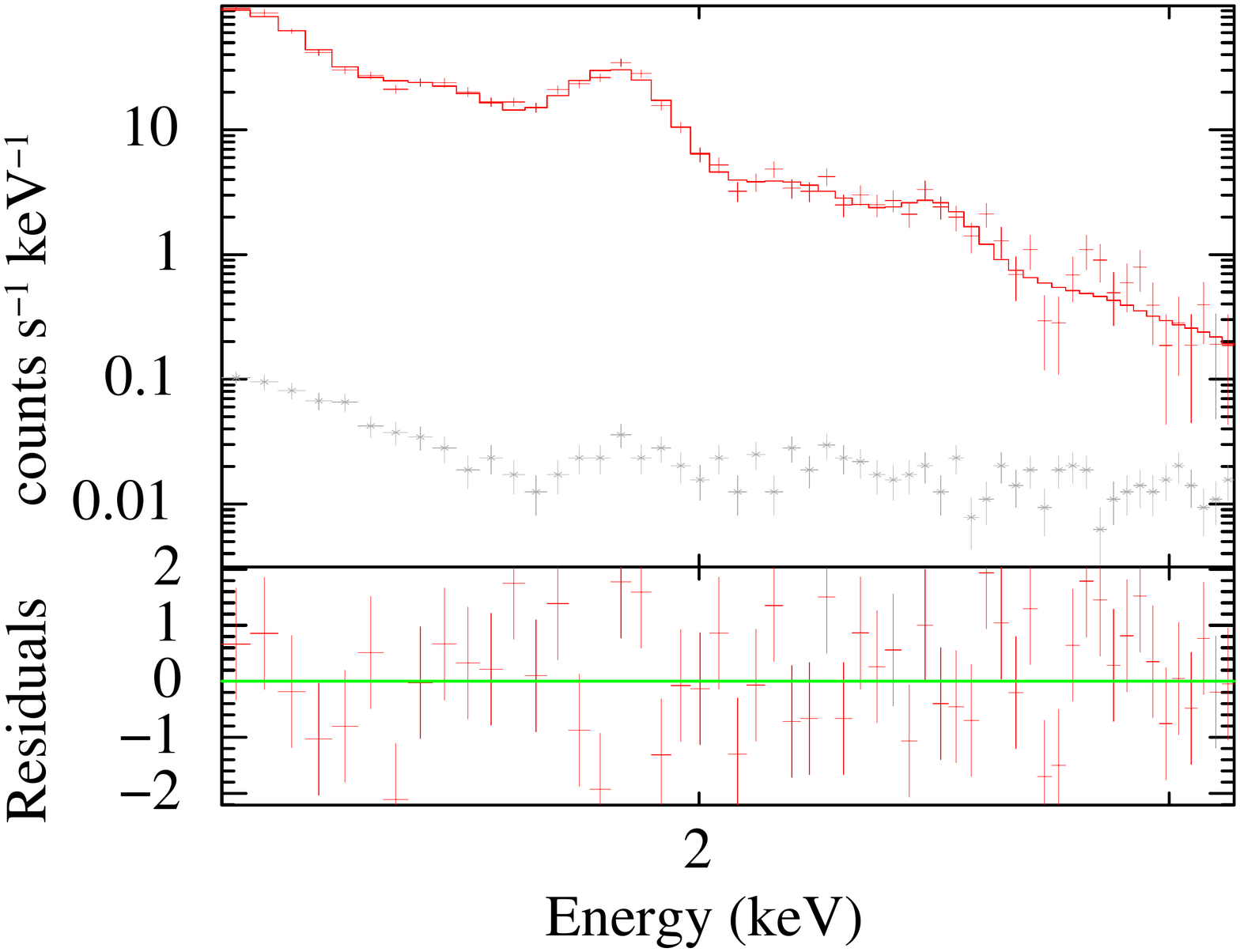} }}
    \caption{Fitted 1T-spectrum (red) for the Flare SOL20190930T19:47 (a) and SOL20200404T06:03 (b). The grey points represent the non-solar background spectrum generated for the times when the Sun was out of the FOV of XSM. The spectral fits' energy ranges were limited to those where the solar X-ray spectrum dominates over the non-solar background.}
    \label{fig3}
\end{figure}

\begin{figure} 
    
    \subcaptionbox{}{{\includegraphics[width=0.3\textwidth]{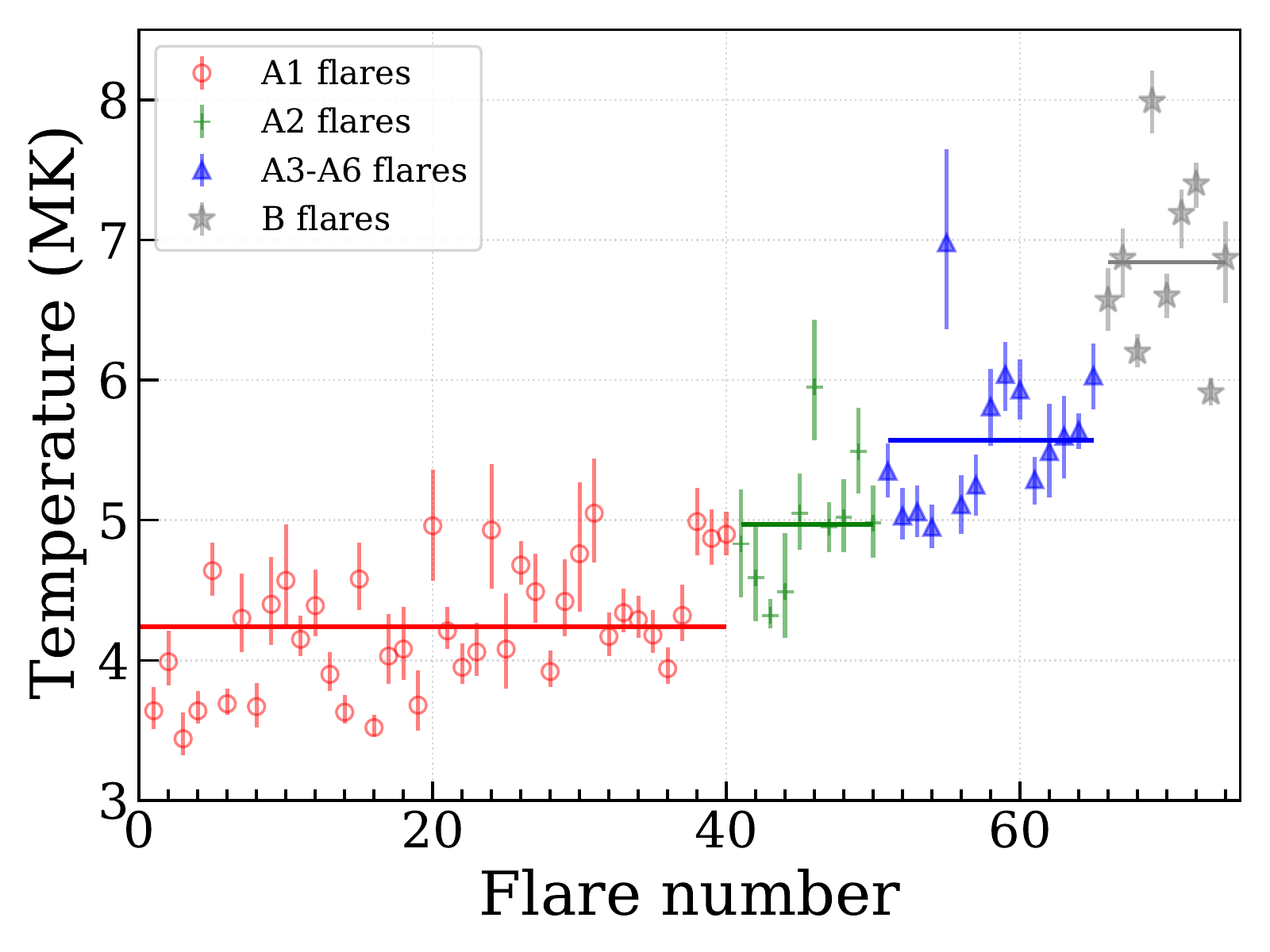} }}
    \hfill
    \subcaptionbox{}{{\includegraphics[width=0.3\textwidth]{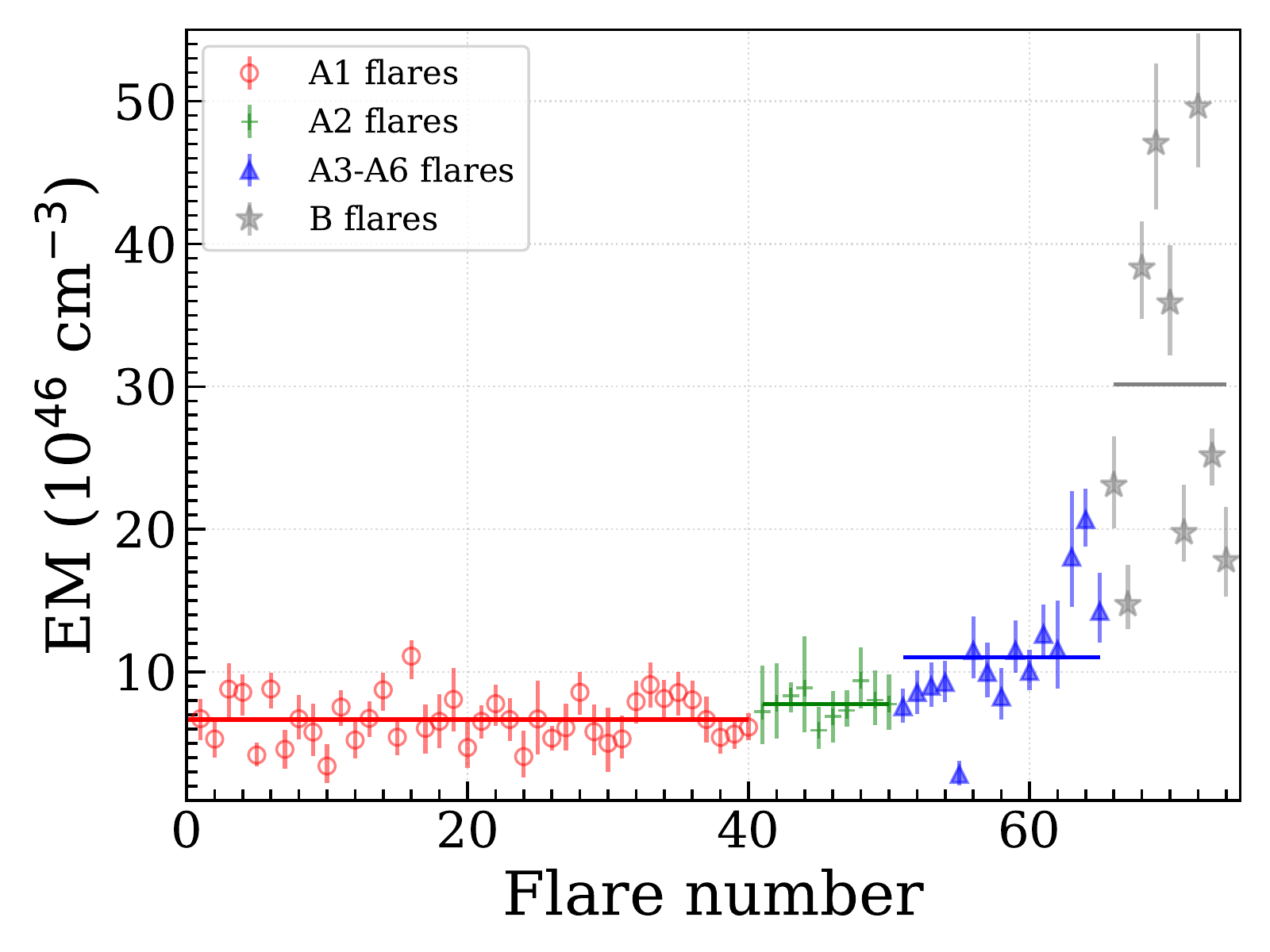} }}
    \hfill
    \subcaptionbox{}{{\includegraphics[width=0.3\textwidth]{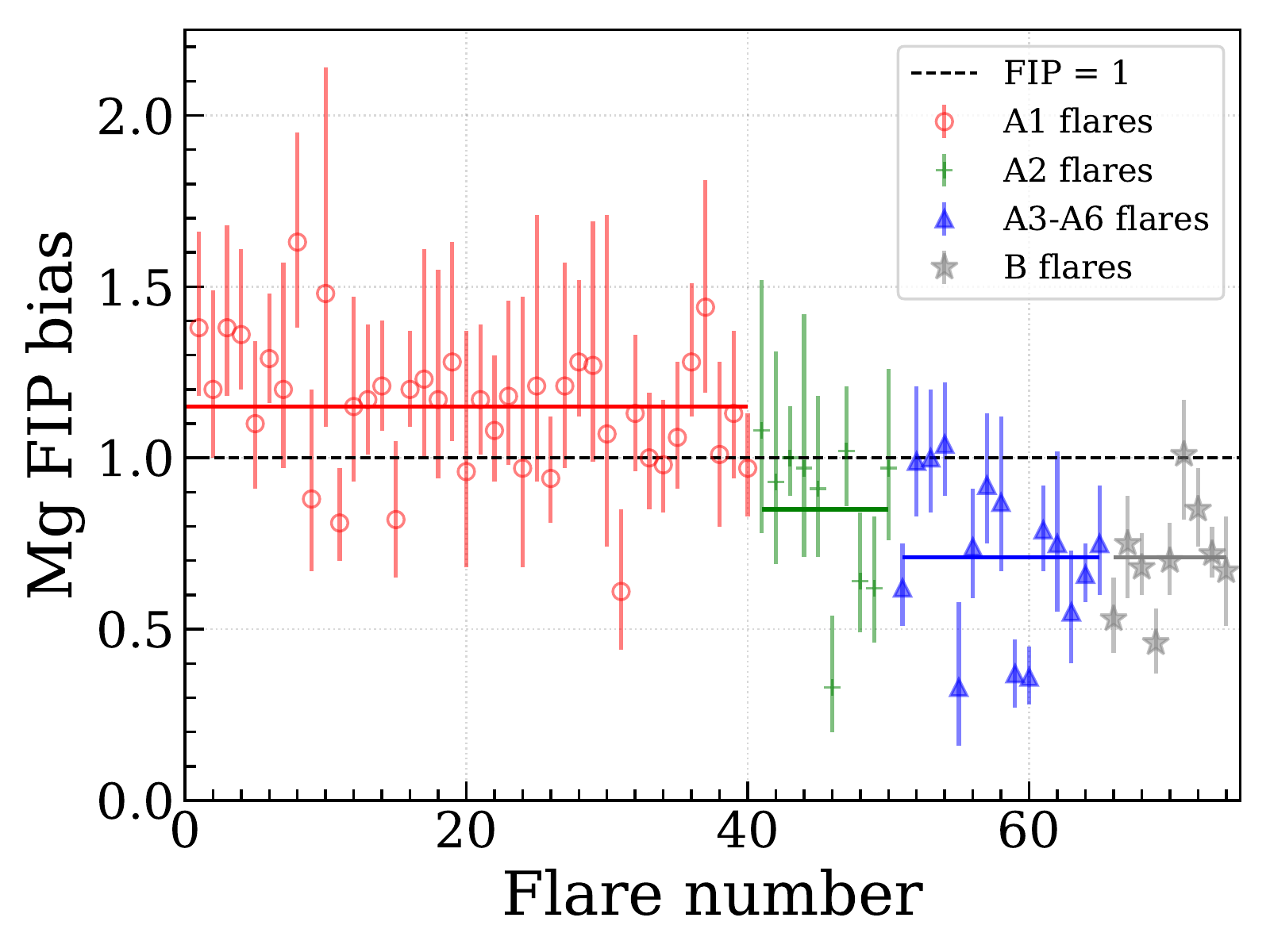} }} 
    \subcaptionbox{}{{\includegraphics[width=0.3\textwidth]{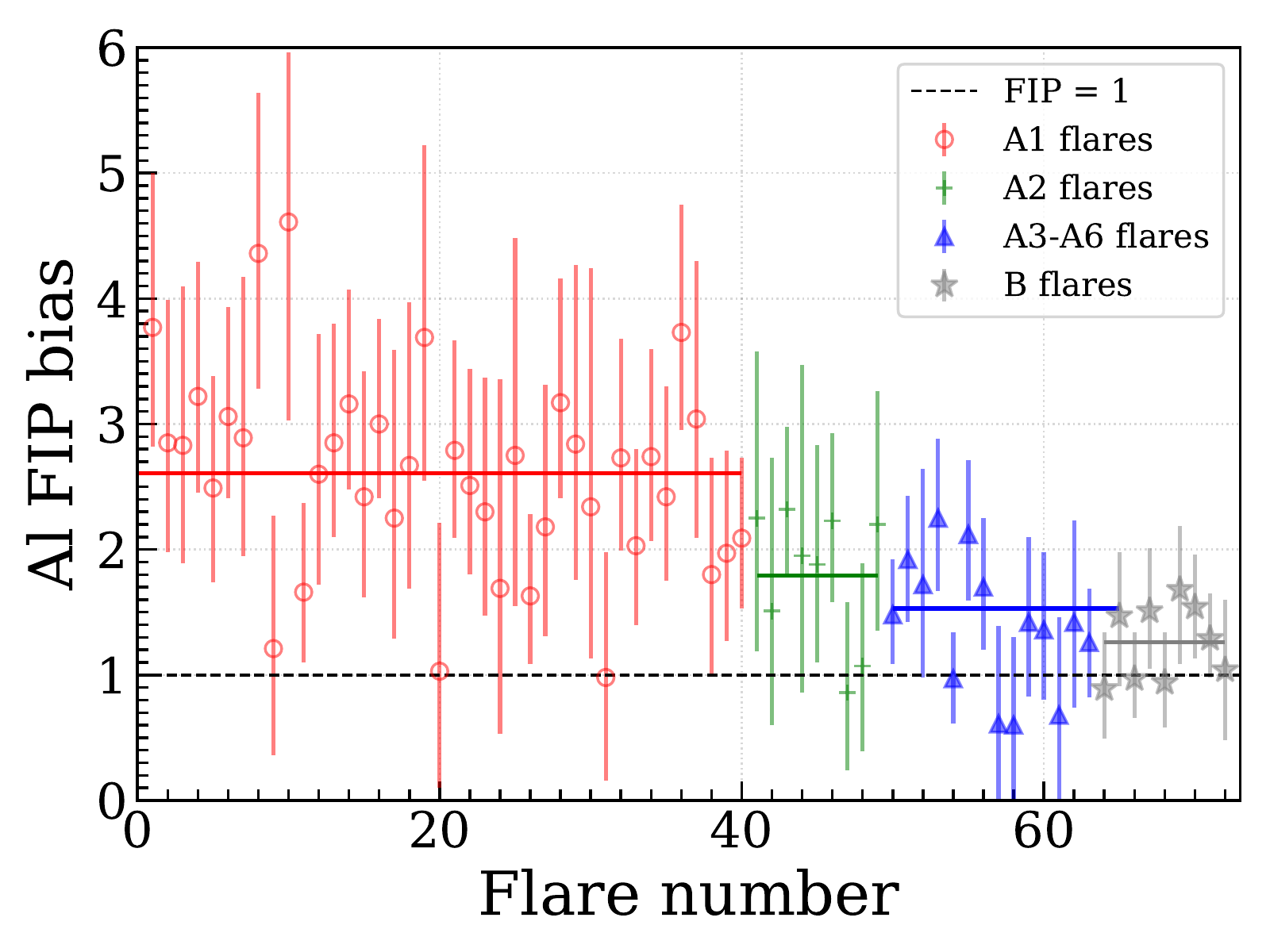} }}
    \hfill
    \subcaptionbox{}{{\includegraphics[width=0.3\textwidth]{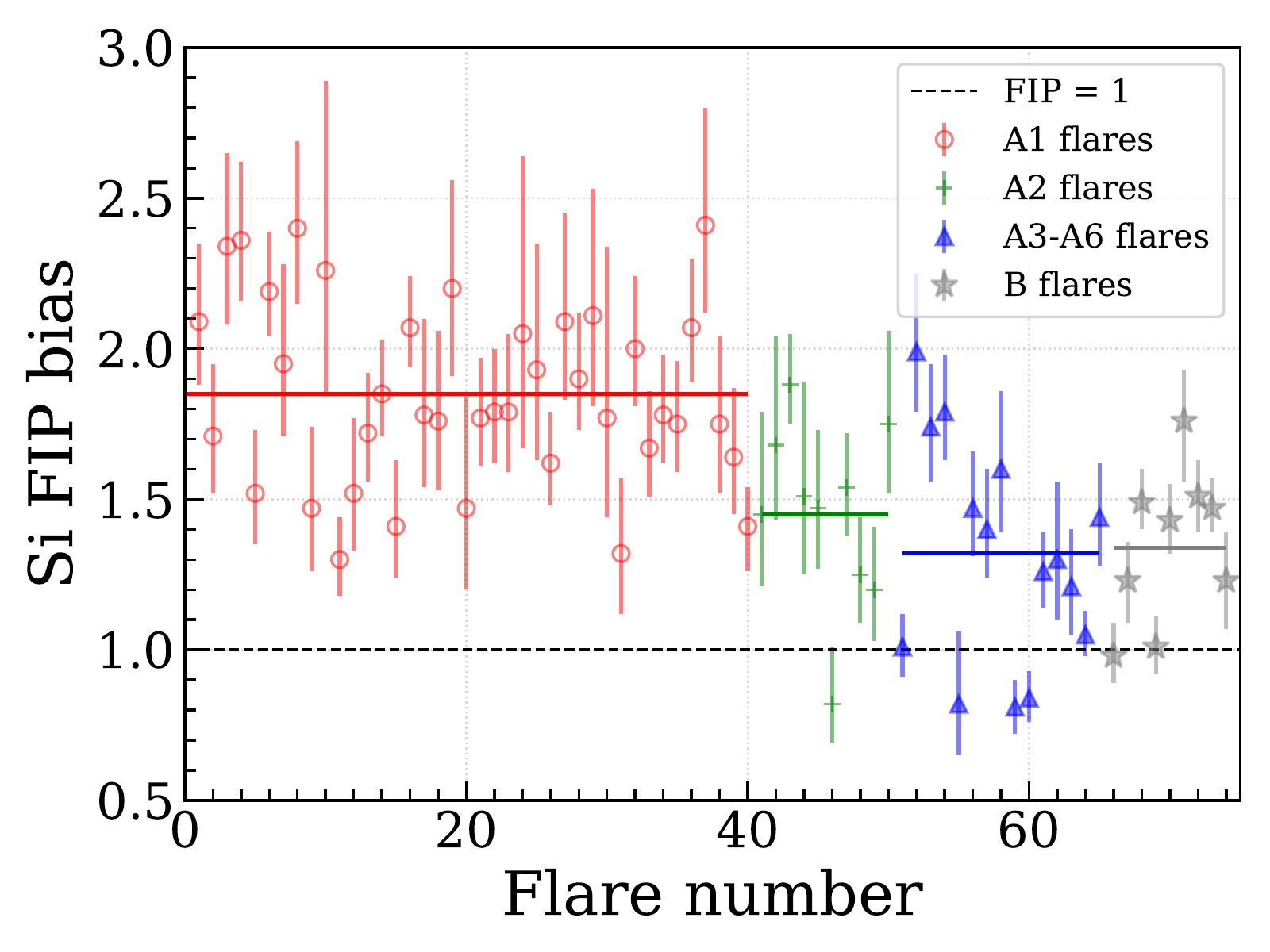} }}
    \hfill
    \subcaptionbox{}{{\includegraphics[width=0.3\textwidth]{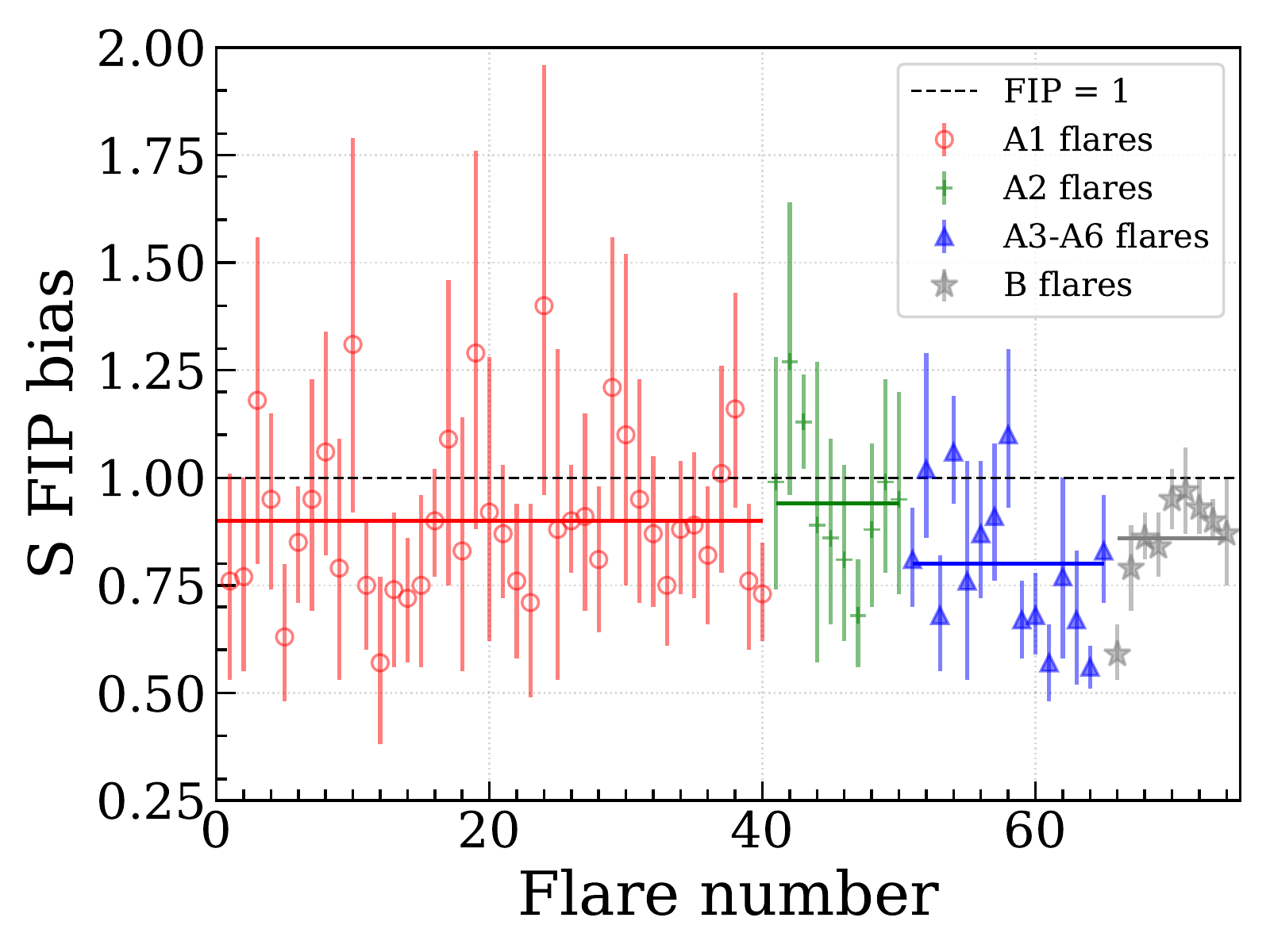} }}
    \caption{Results from the 1T time-integrated spectral analysis for all the 65 A-class flares considered in this study and 9 B-class flares studied in \cite{Mondal_2021}. Panels {\bf a} and {\bf b} show the temperature and emission measure values. Panels {\bf c-f} show the values of FIP bias of Mg, Al, Si and S, respectively. Red circles represent A1 flares, green cross represents A2 flares, blue triangles represent A3-A6 flares, and grey stars represent B-class flares. The horizontal lines represent the average parameter values of each sub-class. The horizontal black dashed line marks a FIP bias of 1 in panels {\bf c-f}.}
    \label{fig4}
\end{figure}

For the analysis of the flare spectra, the energy range of below 1.3 keV is ignored due to the uncertainty of XSM response to these low energies~\citep{Mithun_2021}.
The higher energy is also limited to where the solar X-ray spectrum dominates over the non-solar background.  
The modelling of the spectra is carried out by assuming that the emission is coming from plasma at a single-temperature, as shown in Figure~\ref{fig3} for two representative flares, where the grey points show the background spectra. The abundances of Mg, Al, Si and S (in most cases), whose emission complexes are visible clearly, along with temperature and emission measure are kept as free parameters. All other elemental abundances are frozen to their coronal values taken from \cite{Schmelz2012}. The best-fitted parameters for all the flares, along with the one sigma uncertainties estimated using the standard procedure in XSPEC, are given in Table 1. The same results are plotted against the flare number (flare numbers are IDs given to the flares according to their flare sub-class in ascending order) in Figure~\ref{fig4}. For comparison, we have carried out the time-integrated analysis of the 9 B-class flares studied in \cite{Mondal_2021} and plotted the plasma parameters in the same figure (Figure~\ref{fig4}). 

\begin{figure}    
\centerline{\includegraphics[width=0.44\textwidth,clip=]{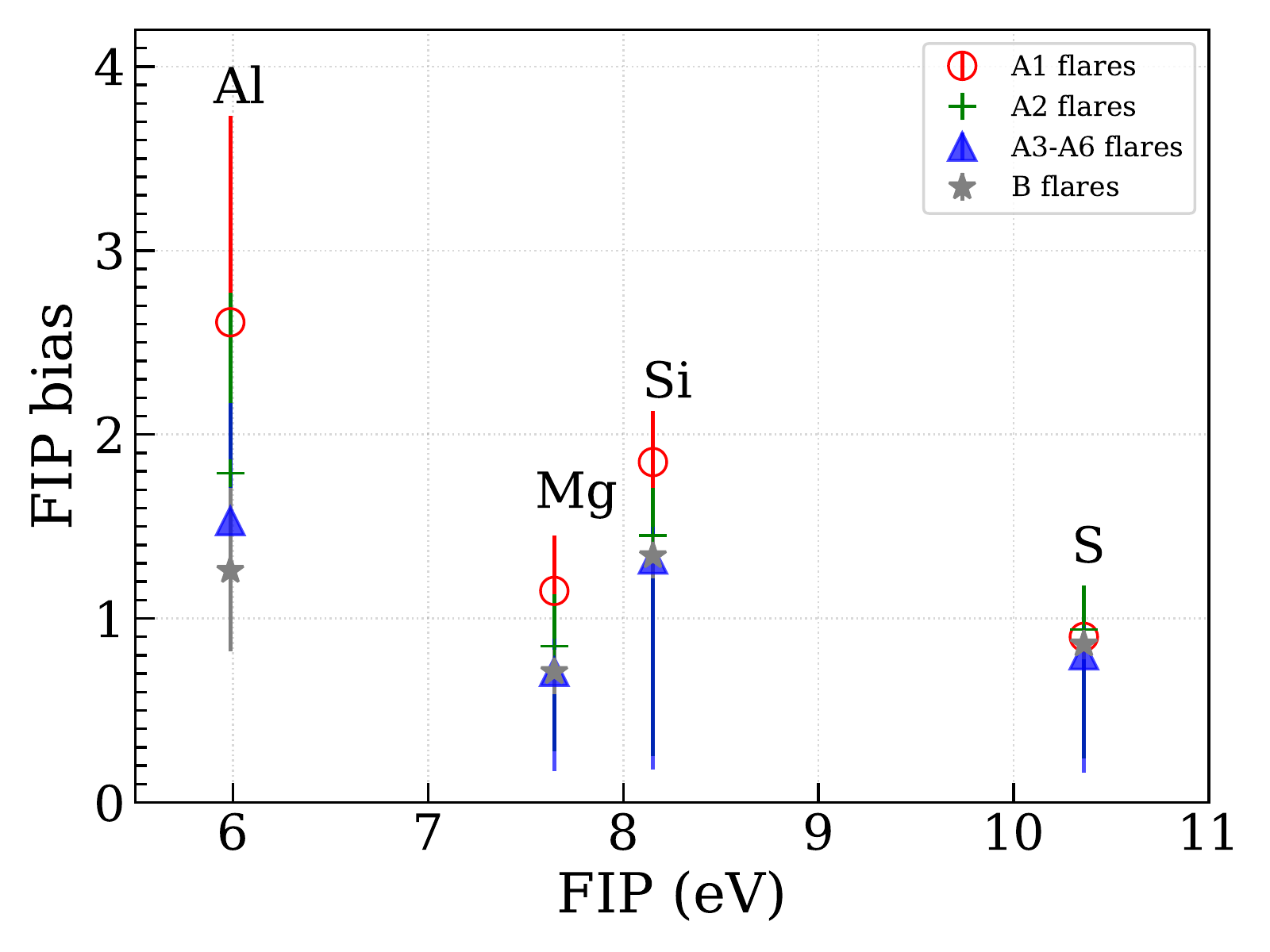}}
\caption{Average FIP bias values of A1, A2, A3-A6 flares (red circles, green cross, blue triangles respectively: this study), B-class flares (grey stars: \citealp{Mondal_2021}) plotted against the first ionization potential. }
   \label{fig5}
\end{figure}

\subsection{Time resolved spectral analysis} 
      \label{time}      

\begin{figure} 
    
    \subcaptionbox{}{{\includegraphics[width=0.3\textwidth]{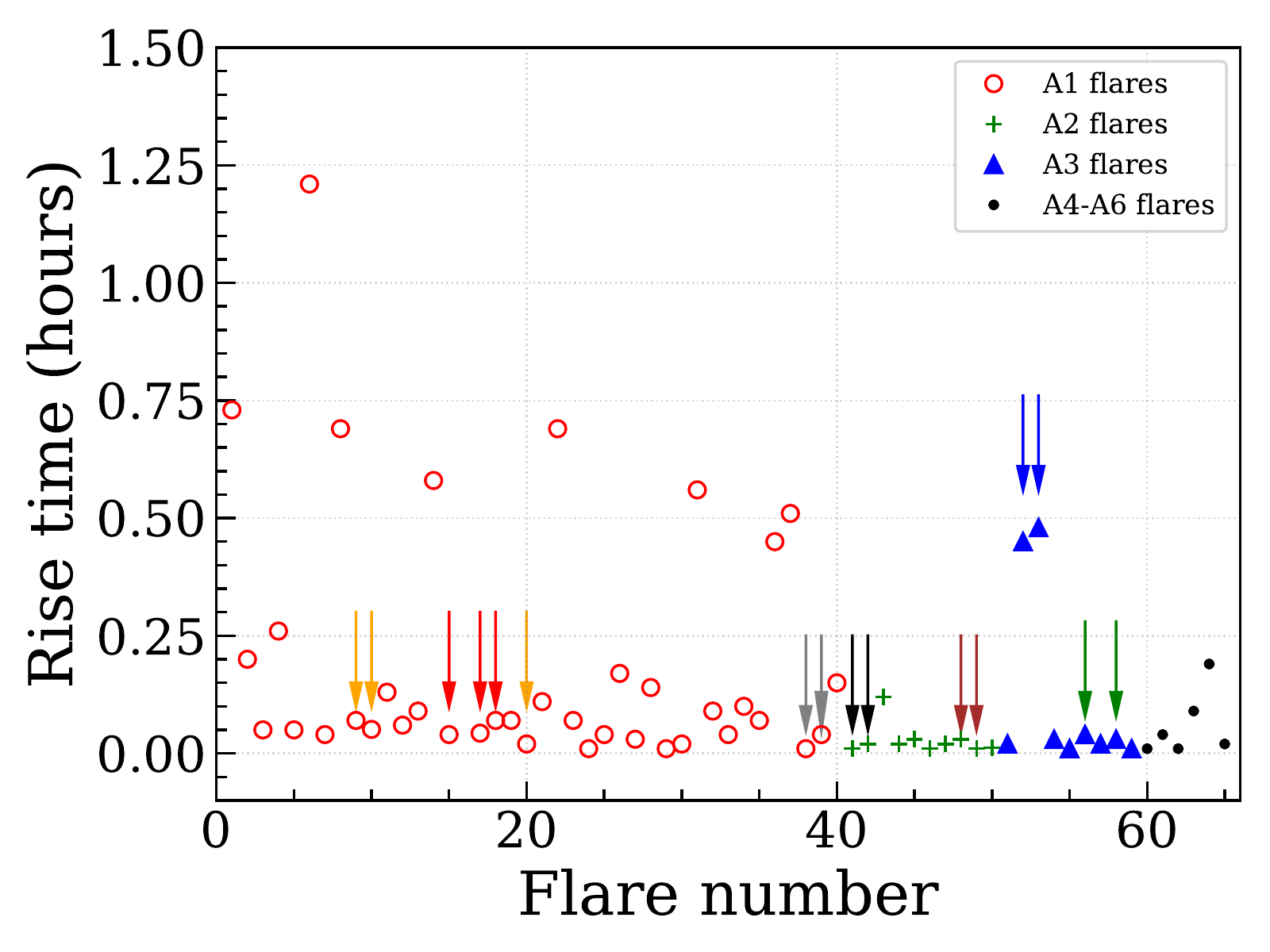} }}
    \hfill
    \subcaptionbox{}{{\includegraphics[width=0.3\textwidth]{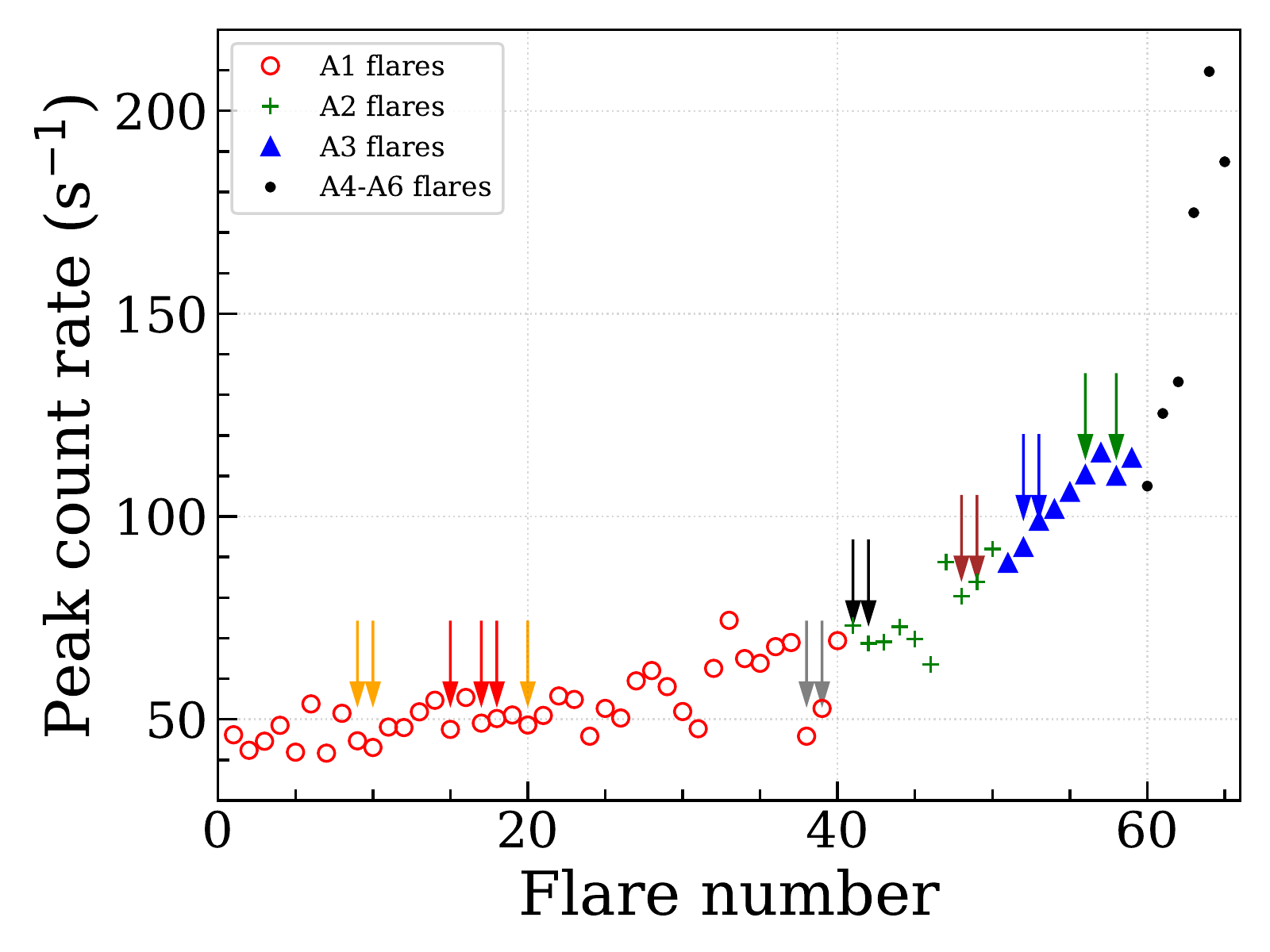} }} 
    \hfill
    \subcaptionbox{}{{\includegraphics[width=0.3\textwidth]{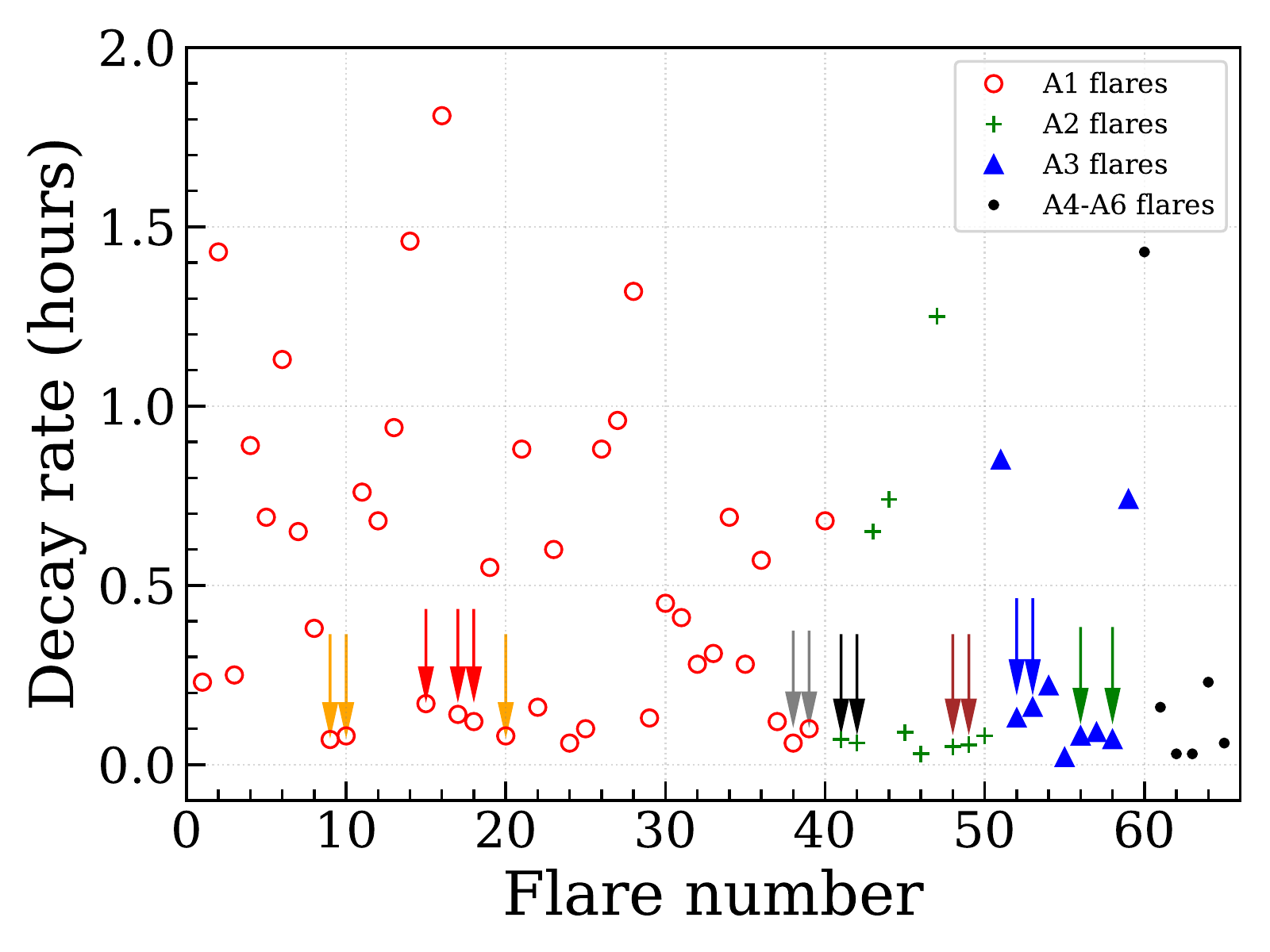} }}
    \caption{Panels {\bf a, b} and {\bf c} show the peak count rate, rise time and decay time (in hours) versus flare number for all the 65 flares considered in this study. Seven sets of flares having similar values in all three plots, as well as similar flaring light curves, are selected for performing time-resolved spectral analysis as described in Section~\ref{time}. All the flares of a set are marked with arrows of the same colour (set 1 - blue, set 2 - brown, set 3 - red, set 4 - grey, set 5 - orange, set 6 - black, set 7 - green).}
    \label{fig6}
\end{figure}

\begin{figure} 
    
    \subcaptionbox{}{{\includegraphics[width=0.3\textwidth]{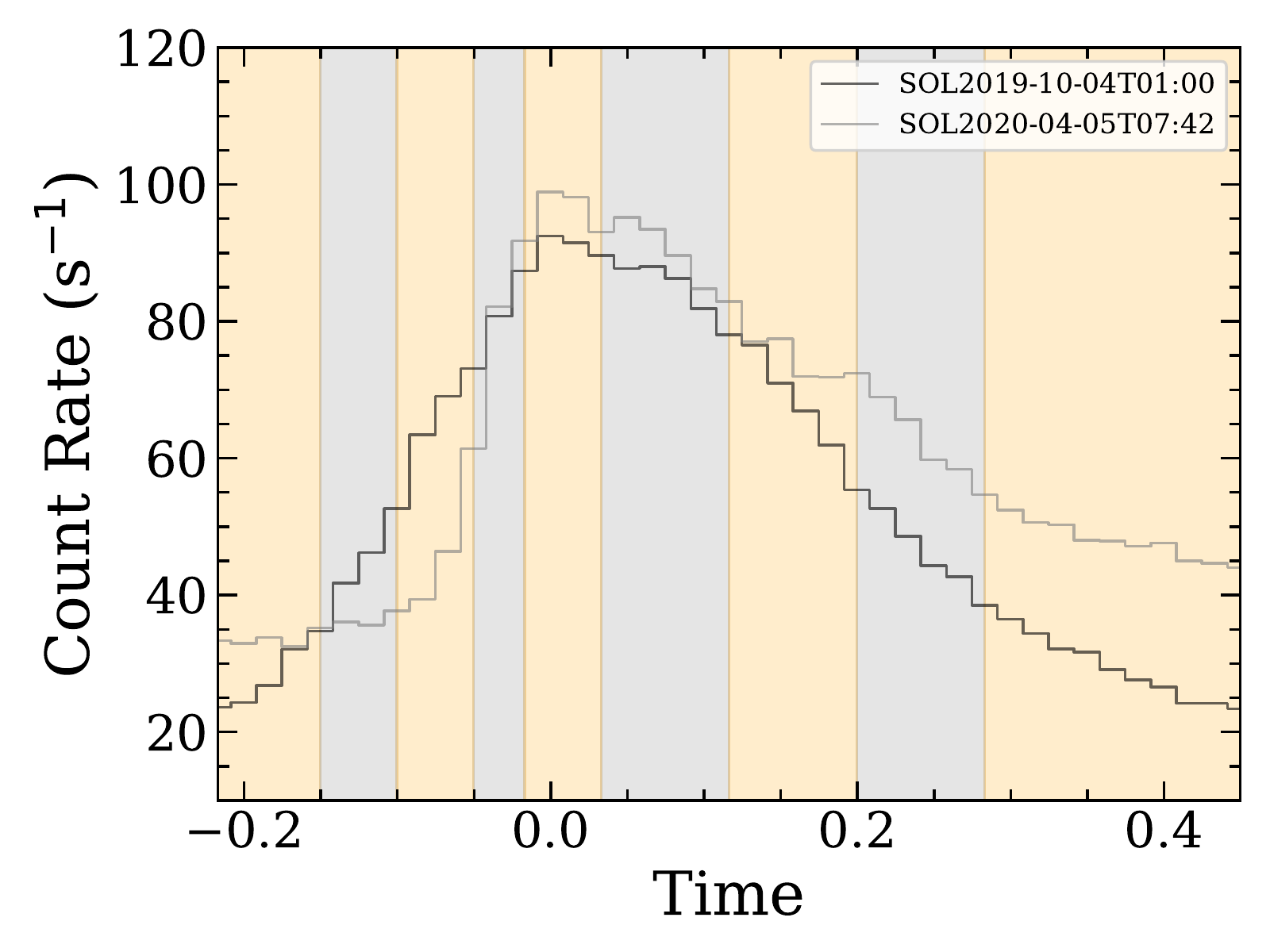} }}
    \hfill
    \subcaptionbox{}{{\includegraphics[width=0.3\textwidth]{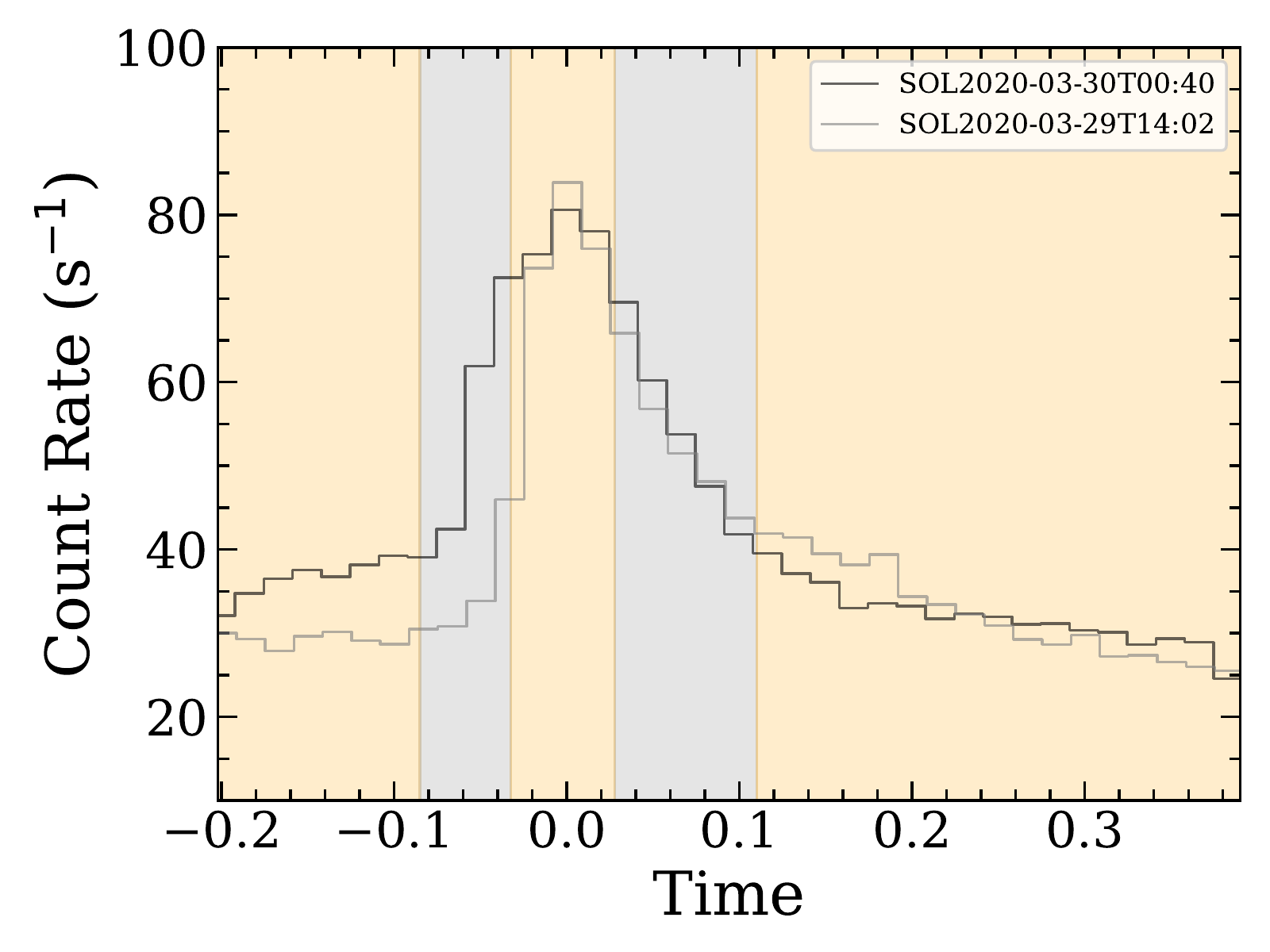} }}
    \hfill
    \subcaptionbox{}{{\includegraphics[width=0.3\textwidth]{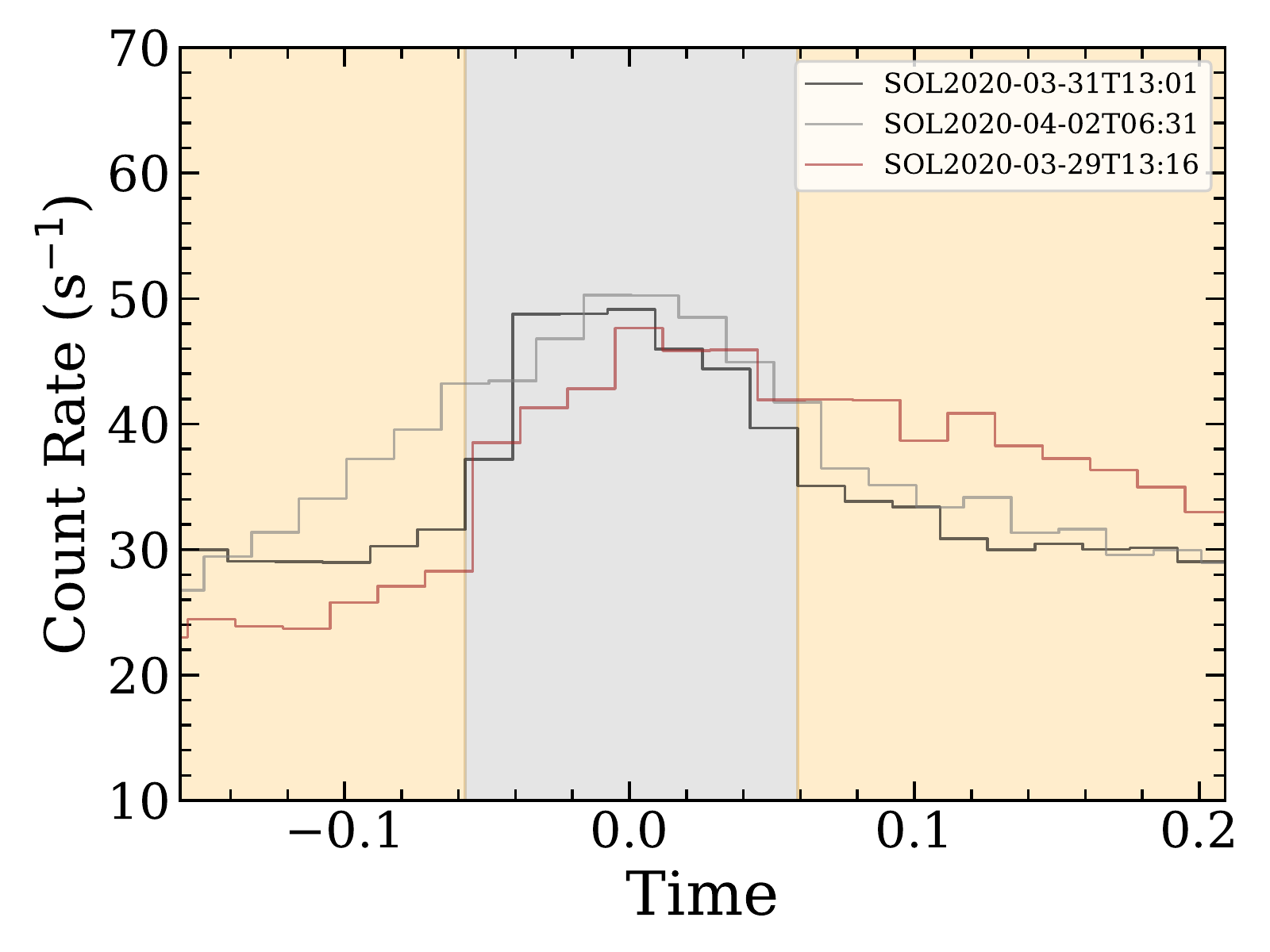} }} 
    \caption{The 1-15 keV XSM X-ray light curves for three representative sets (set 1, set 2 and set 3) of flares. Background orange and grey shaded regions demarcate durations for which the integrated XSM spectra have been generated for each of the flares in a set and then added to carry out the time-resolved spectroscopy as described in Section~\ref{res}. The flare IDs for all the flares of a set are given in the legend in the format SOLyyyy-mm-ddThh:mm.}
    \label{fig7}
\end{figure}

We have carried out the time-integrated spectral analysis of all the chosen flares as discussed in Section~\ref{stat}. However, carrying out the time-resolved spectroscopy for these individual flares is challenging due to their weak signals. 
In order to increase the signal-to-noise ratio of these small flares, we have added the spectra of multiple flares having similar rise time, peak counts, and decay time. 
Figure~\ref{fig6} shows the rise time (RT: Panel {\bf a}), the peak count rate (PCR: Panel {\bf b}) and the decay time (DT: Panel {\bf c}) of all the flares.
We have identified flares having approximately equal values of RT, PCR, and DT. 
The selected sets of flares were over-plotted by co-aligning the flare peak at zero seconds. Further, by visual inspection, we kept only those flares in a set which followed a similar evolutionary profile.
This selection resulted in seven sets of near-identical flares. All the flares of a particular set are marked with arrows of the same colour (set 1 - blue, set 2 - brown, set 3 - red, set 4 - grey, set 5 - orange, set 6 - black, set 7 - green in Figure~\ref{fig6}. The properties of all the flares considered in each set are given in Table~\ref{tab1}. 
The over-plotted light curves of three representative sets (set 1, set 2 and set 3) of flares out of the seven are shown in panels {\bf a-c} of Figure~\ref{fig7}. Panel {\bf a} contains a set of two A3-class flares, Panel {\bf b} contains a set of two flares of class A2.6 and A2.7 and Panel {\bf c} contains a set of three A1.2-class flares.
As A1-class flares are very weak, three flares are required in this set in order to constrain the spectral parameters. 
We have divided these sets of flares into multiple time bins, as shown by the alternative orange and grey shaded regions.

\begin{table}

\caption{Flare properties of all the flares considered in each set of similar flares. Flare IDs corresponding to peak times of the flares is given in the format SOLyyyy-mm-ddThh:mm. Approximate GOES classes are mentioned according to the peak flux of the flare in the 1-8  \textup{\AA} XSM flux.
Peak count rate, rise time and decay time are calculated from the XSM light curve.}
\begin{tabular} {lccccc} \\ \hline

{\bf Set number} & {\bf Flare IDs} & {\bf Flare class} & \multicolumn{1}{p{3cm}}{\centering \bf Peak count rate \\(s$^{-1}$) } & \multicolumn{1}{p{3cm}}{\centering \bf Rise time \\ (Hours)} & \multicolumn{1}{p{2cm}}{\centering \bf Decay time \\ (Hours)} \\ [0.5ex] \hline \hline \\

 Set 1 & SOL2019-10-04T01:00 & A3.0 & 92.4 & 0.45 & 0.13  \\
       & SOL2020-04-05T07:42 & A3.0 & 98.83 & 0.48 & 0.16  \\ [0.2cm]
 \hline \\
 Set 2 & SOL2020-03-29T14:02 & A2.7 & 83.83 & 0.01 & 0.05 \\ 
       & SOL2020-03-30T00:40 & A2.6 & 80.31 & 0.03 & 0.05 \\ [0.2cm]
 \hline \\
 Set 3 & SOL2020-03-31T13:01 & A1.2 & 49.05 & 0.04 & 0.14  \\
       & SOL2020-04-02T06:31 & A1.2 & 50.19 & 0.07 & 0.12  \\
       & SOL2020-03-29T13:16 & A1.2 & 47.52 & 0.04 & 0.17 \\ [0.2cm]
 \hline \\
 Set 4 & SOL2020-04-10T00:22 & A1.8 & 61.23 & 0.03 & 0.12 \\
       & SOL2019-10-01T20:33 & A1.9 & 65.1 & 0.02 & 0.18  \\ [0.2cm]
 \hline \\
 Set 5 & SOL2020-04-07T11:36 & A1.1 & 44.7 & 0.07 & 0.07 \\
       & SOL2020-03-08T03:15 & A1.2 & 43.06 & 0.05 & 0.08 \\
       & SOL2020-03-11T11:50 & A1.3 & 48.6 & 0.02 & 0.08 \\ [0.2cm]
 \hline \\
 Set 6 & SOL2020-04-04T03:38 & A2.0 & 73.15 & 0.01 & 0.07 \\
       & SOL2020-04-09T07:14 & A2.0 & 68.72 & 0.02 & 0.06  \\ [0.2cm] 
 \hline \\
 Set 7 & SOL2020-04-02T02:04 & A3.1 & 109.9 & 0.03 & 0.07 \\
       & SOL2019-10-02T00:33 & A3.4 & 110.3 & 0.04 & 0.08 \\ [0.2cm]
 \hline
 \end{tabular}
\label{tab1}
\end{table}

\begin{figure} 
    \subcaptionbox{}{{\includegraphics[width=0.25\textwidth]{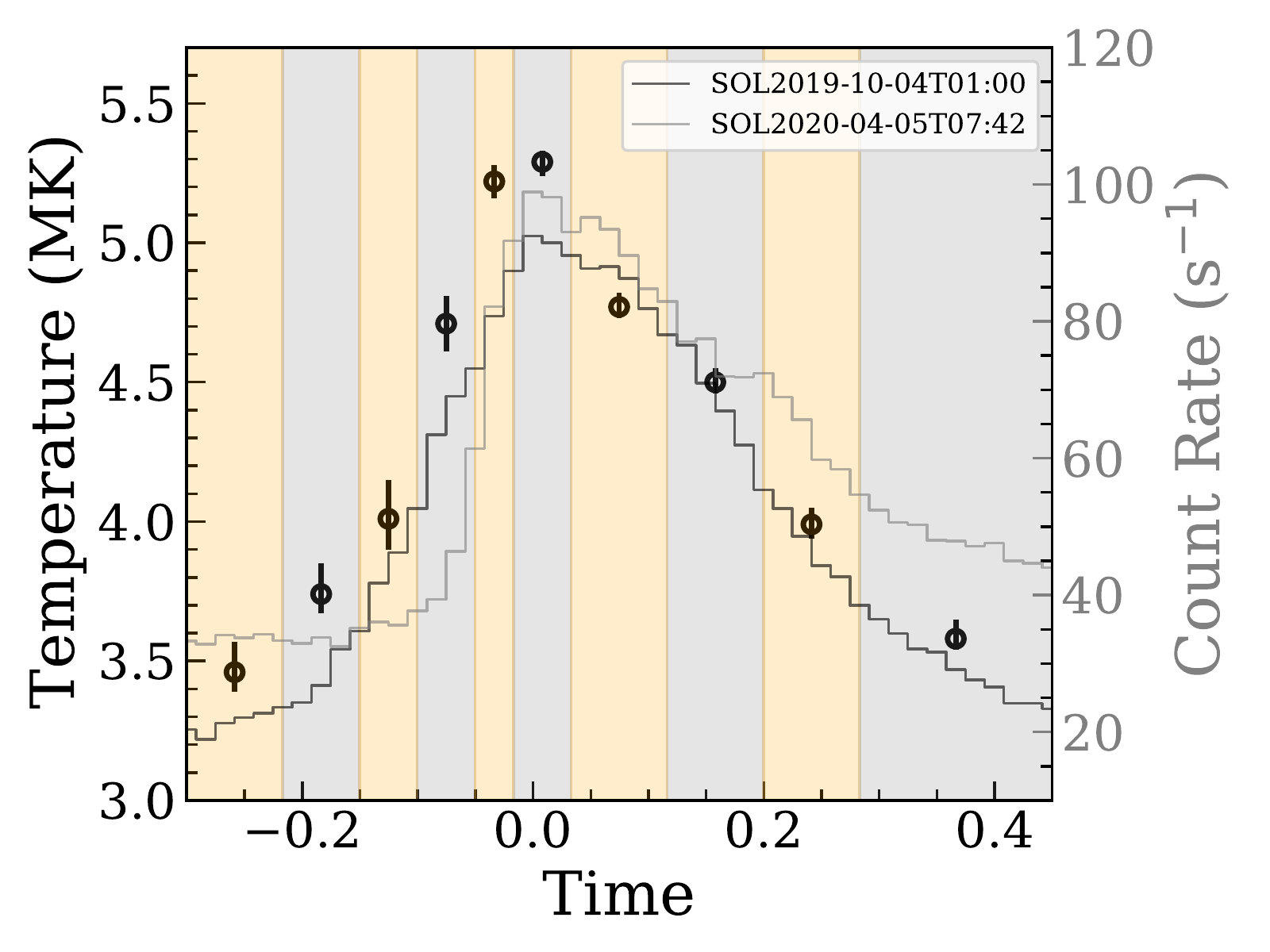} }}
    \hfill
    \subcaptionbox{}{{\includegraphics[width=0.25\textwidth]{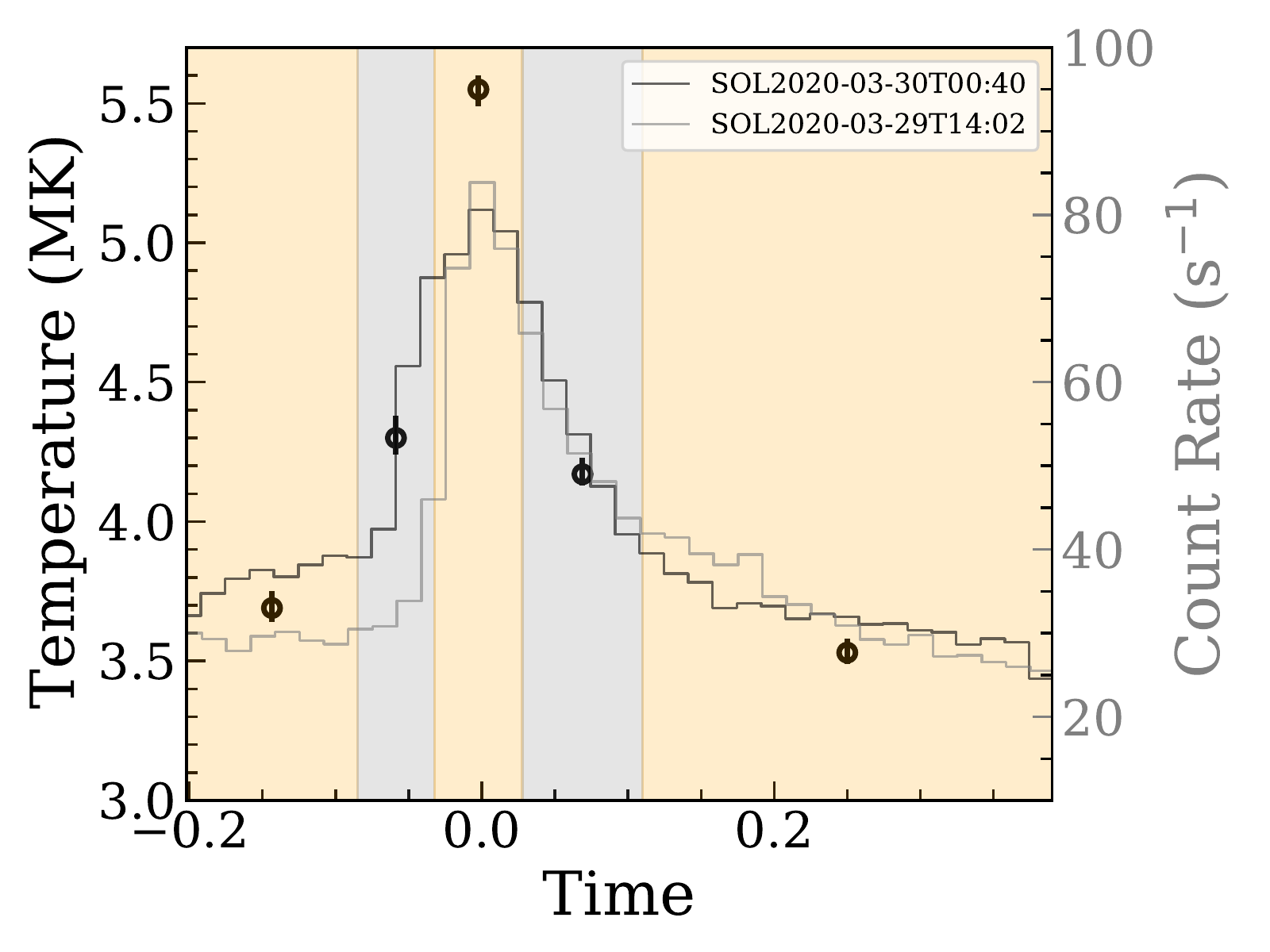} }}
    \hfill
    \subcaptionbox{}{{\includegraphics[width=0.25\textwidth]{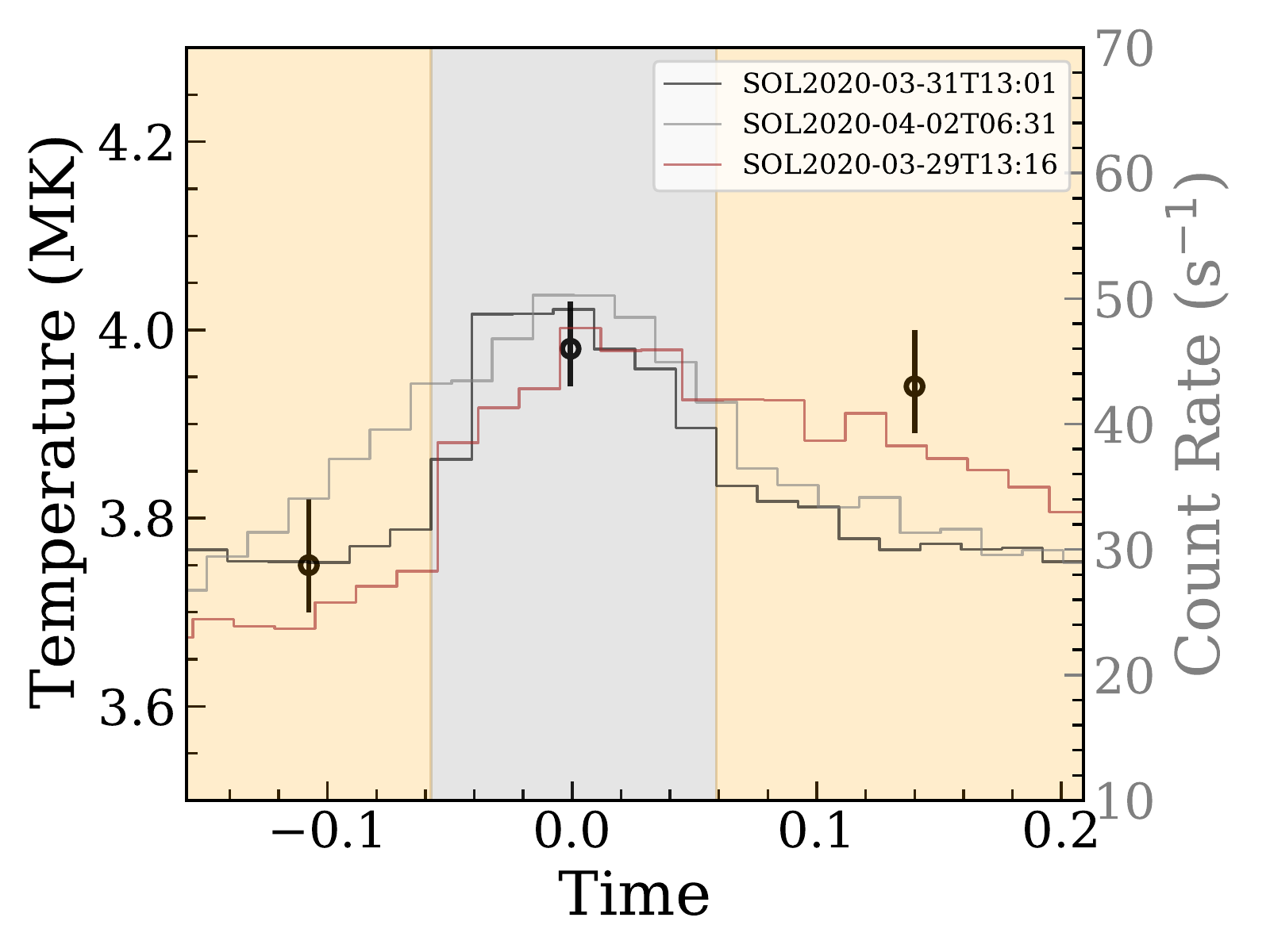} }} 
    
    \subcaptionbox{}{{\includegraphics[width=0.25\textwidth]{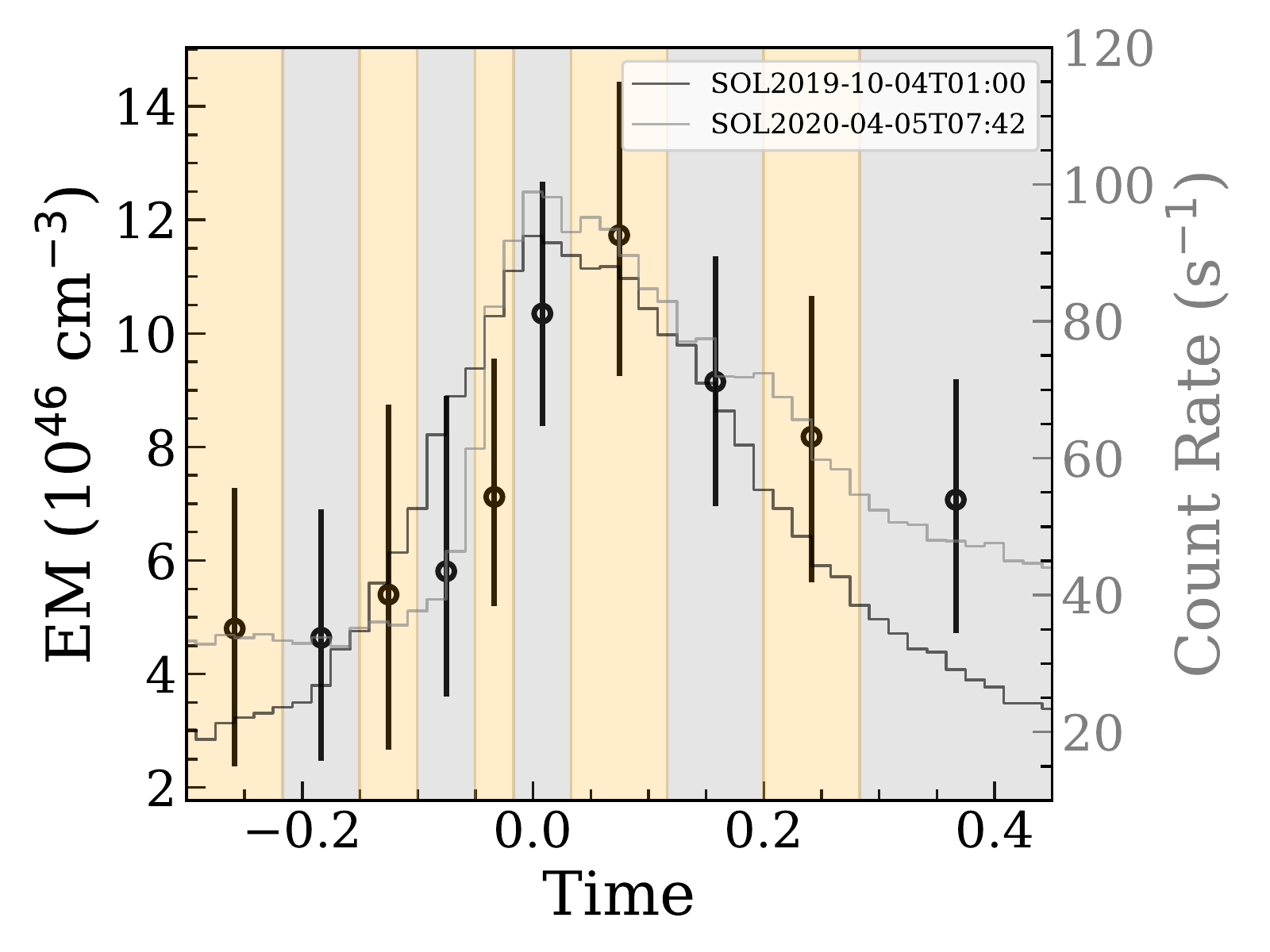} }}
    \hfill
    \subcaptionbox{}{{\includegraphics[width=0.25\textwidth]{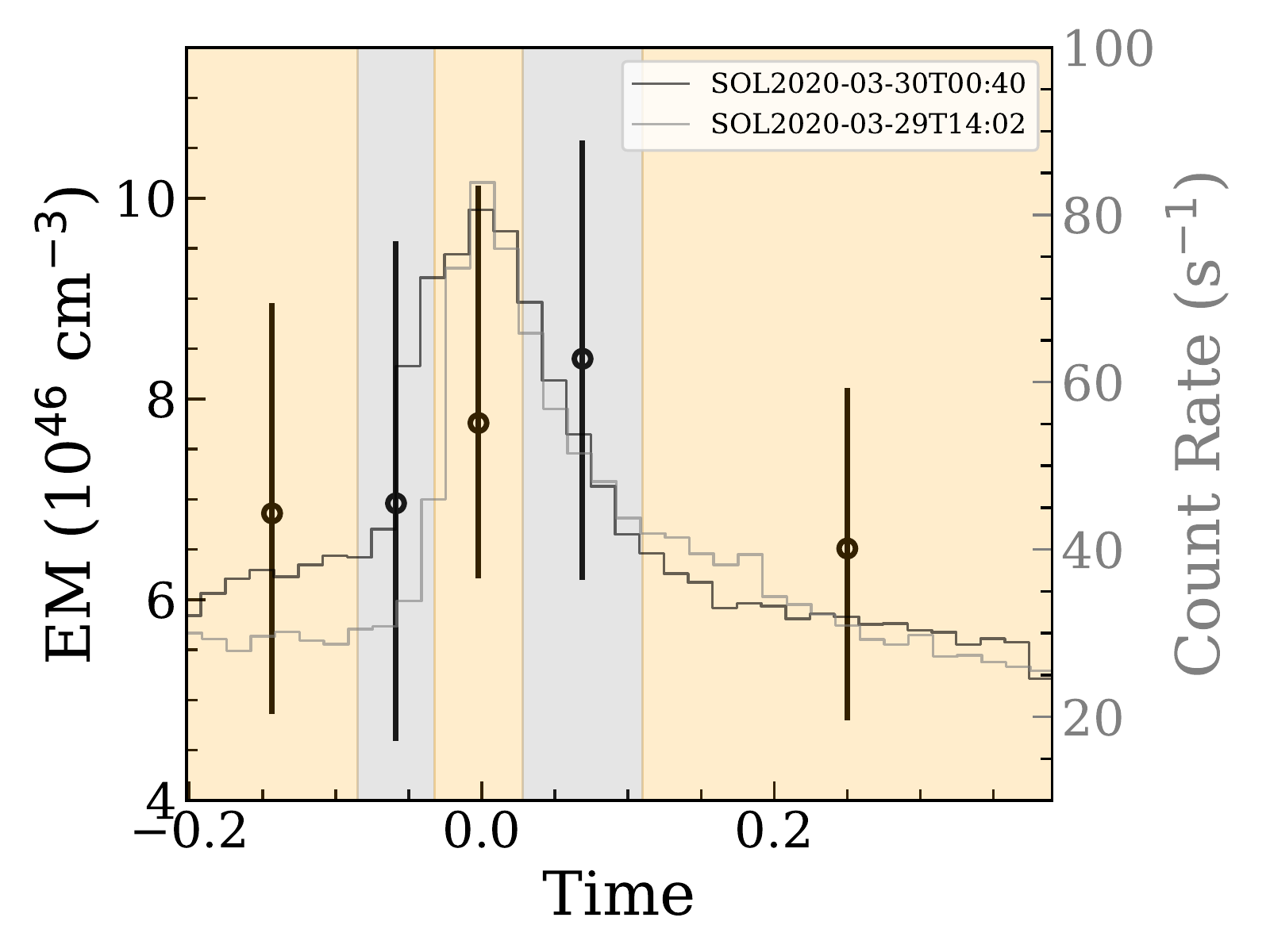} }}
    \hfill
    \subcaptionbox{}{{\includegraphics[width=0.25\textwidth]{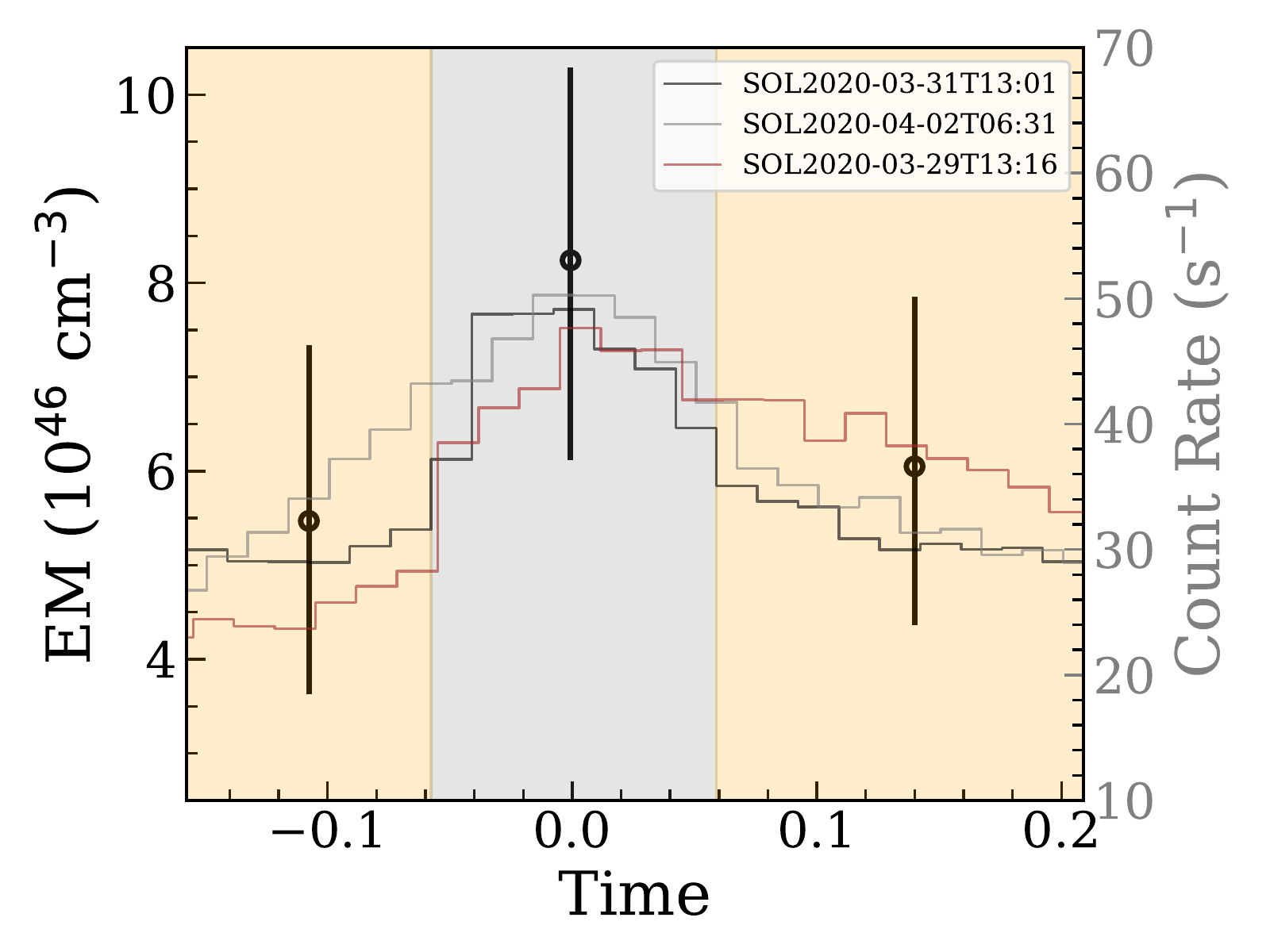} }}
    
    \subcaptionbox{}{{\includegraphics[width=0.25\textwidth]{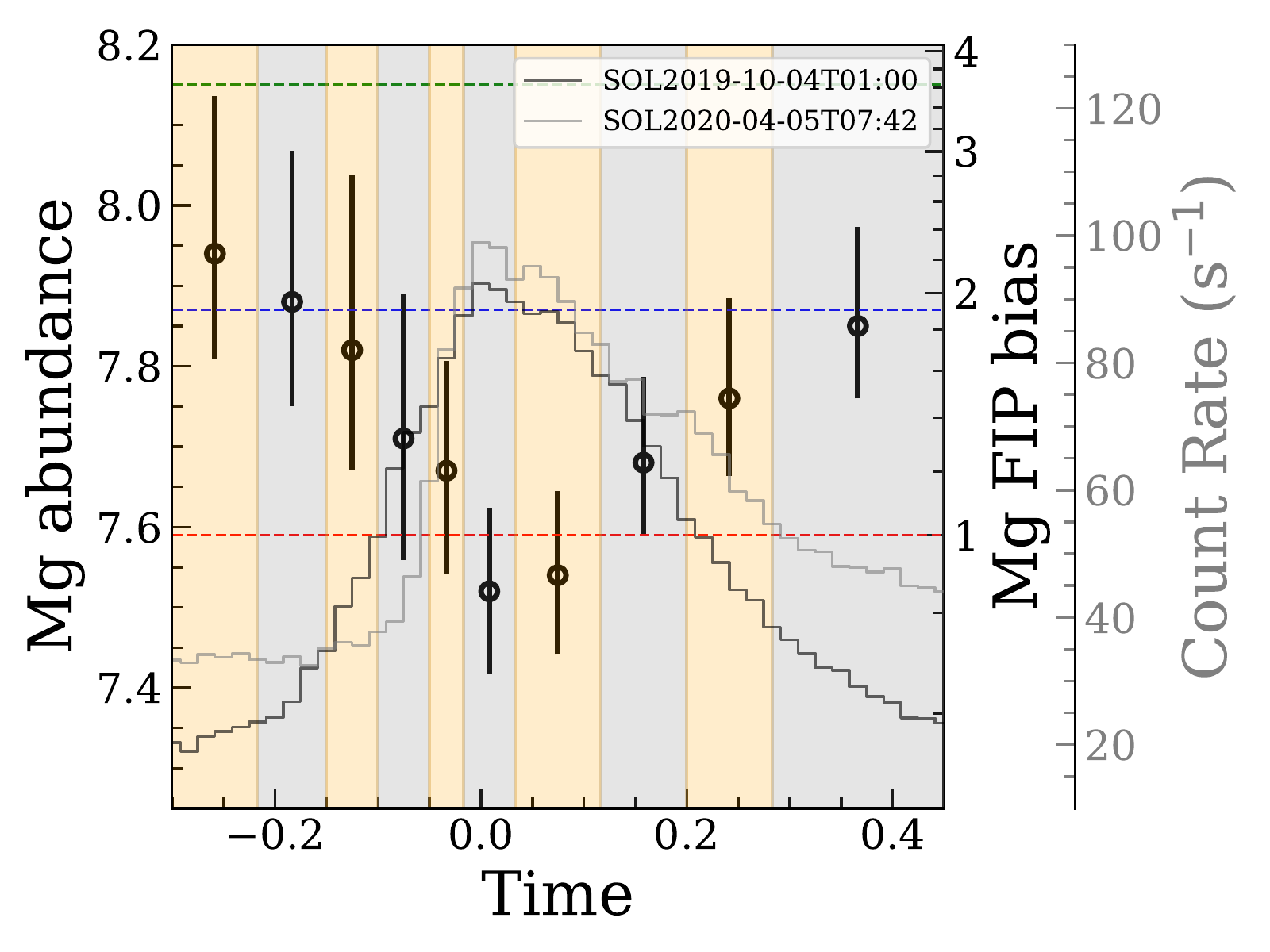} }}
    \hfill
    \subcaptionbox{}{{\includegraphics[width=0.25\textwidth]{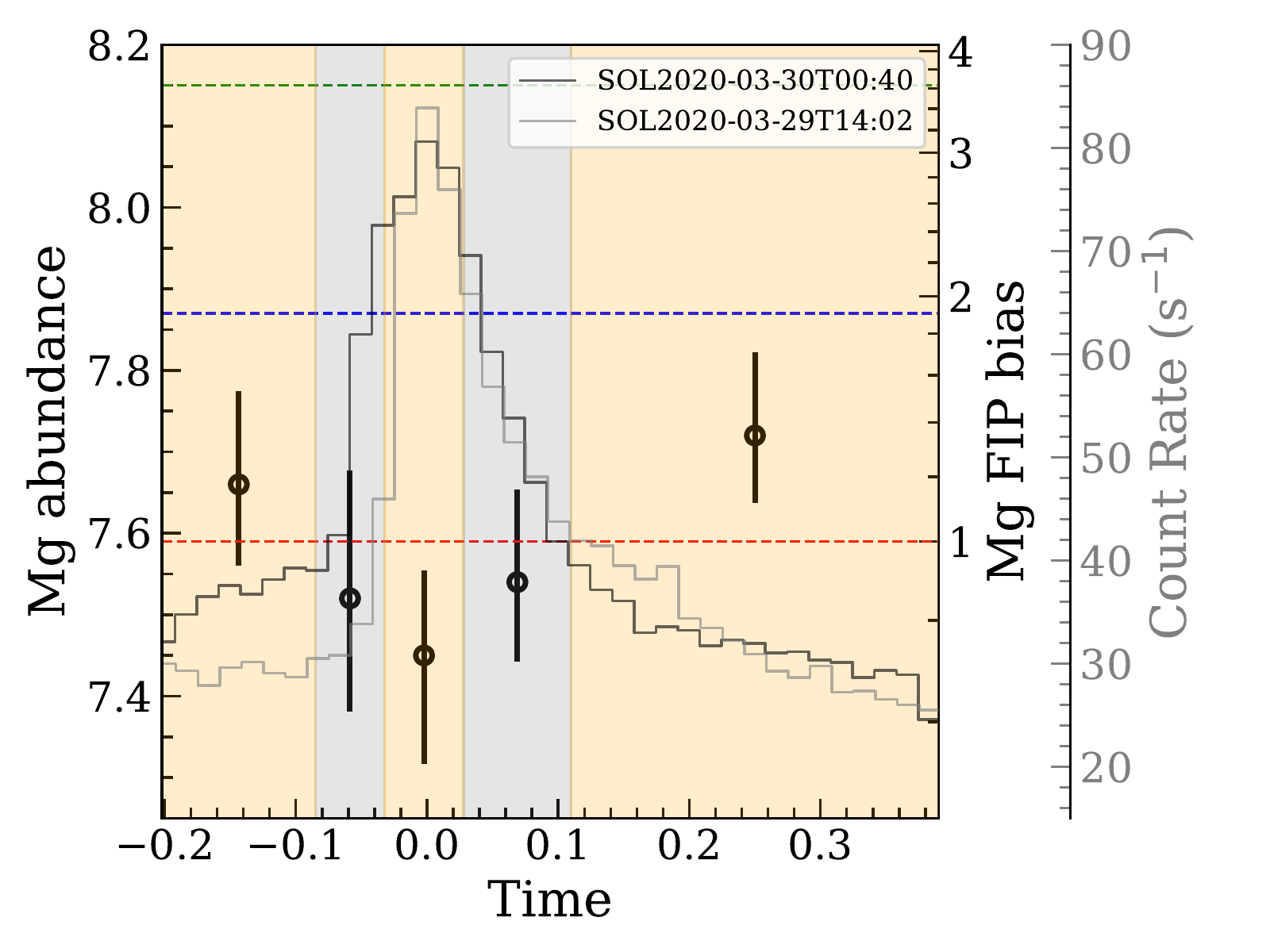} }}
    \hfill
    \subcaptionbox{}{{\includegraphics[width=0.25\textwidth]{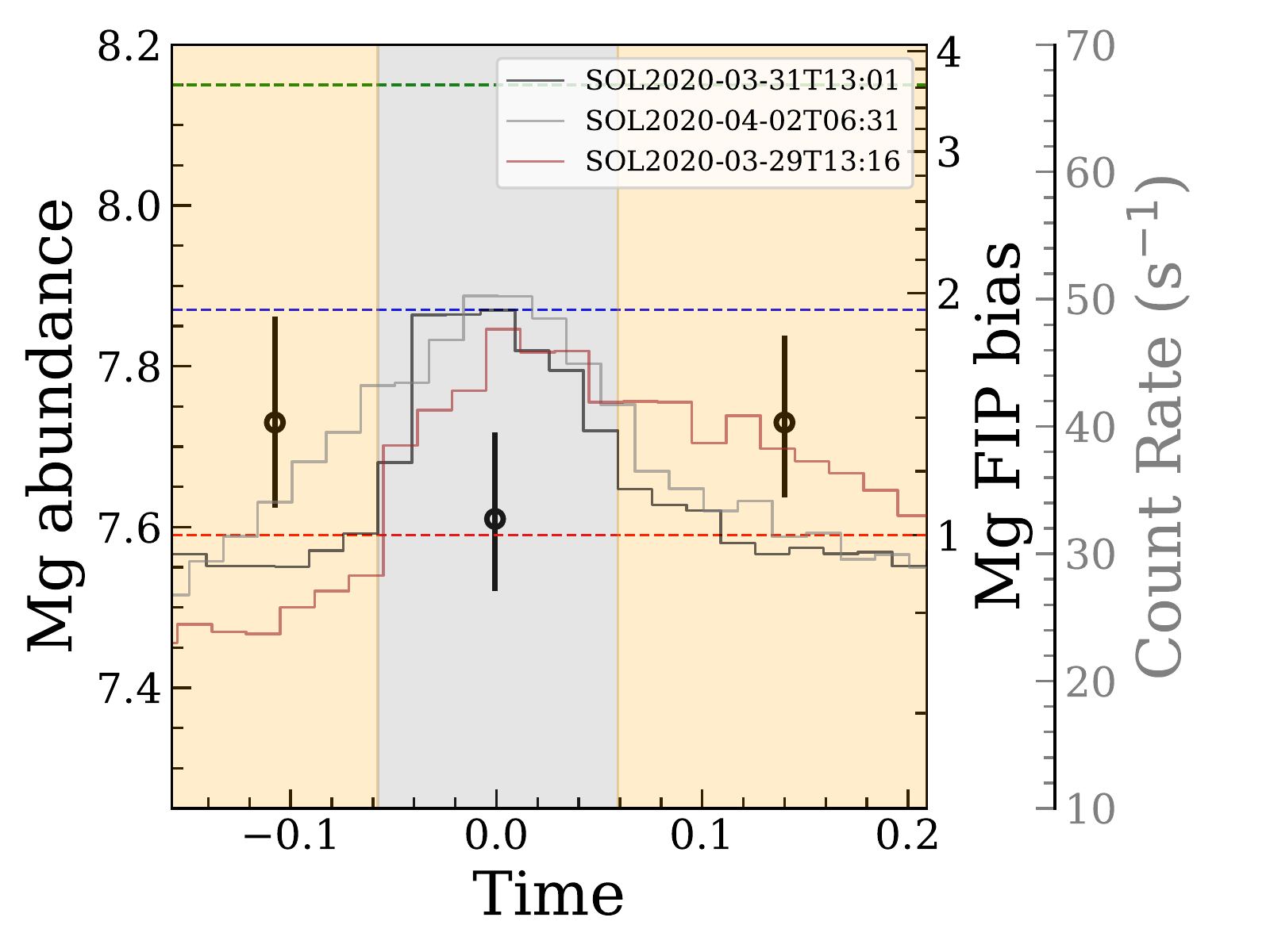} }}
    
    \subcaptionbox{}{{\includegraphics[width=0.25\textwidth]{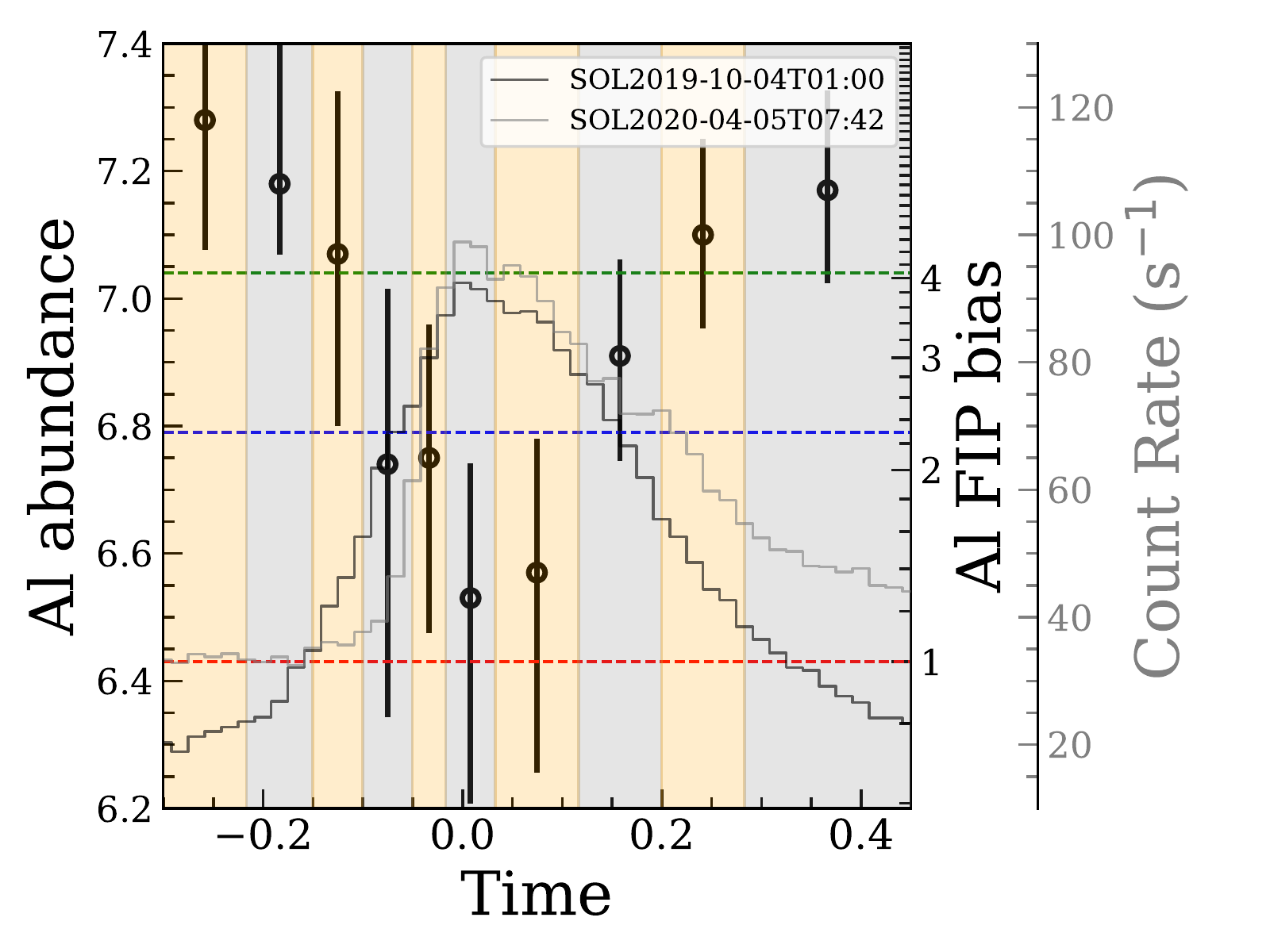} }}
    \hfill
    \subcaptionbox{}{{\includegraphics[width=0.25\textwidth]{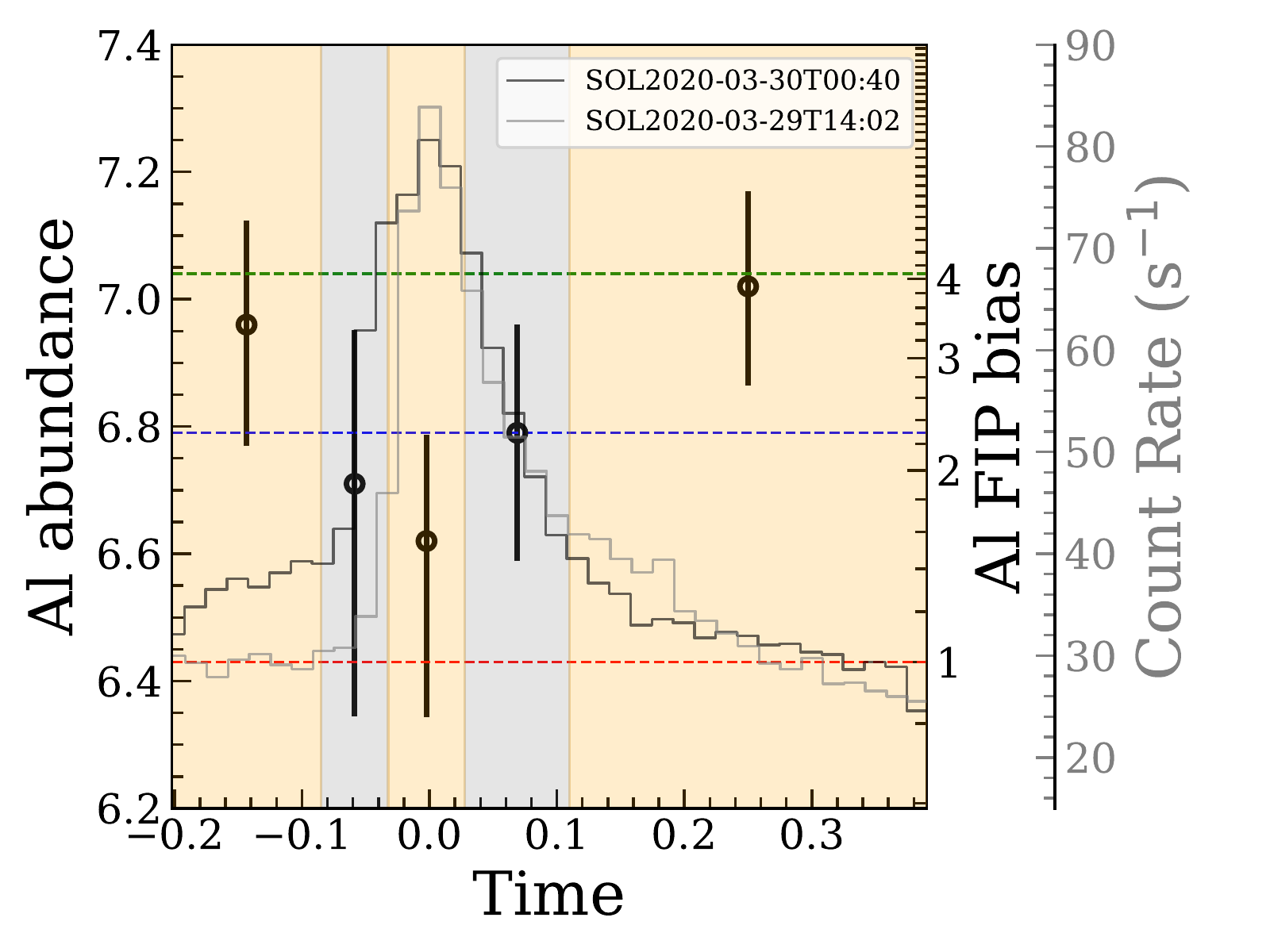} }}
    \hfill
    \subcaptionbox{}{{\includegraphics[width=0.25\textwidth]{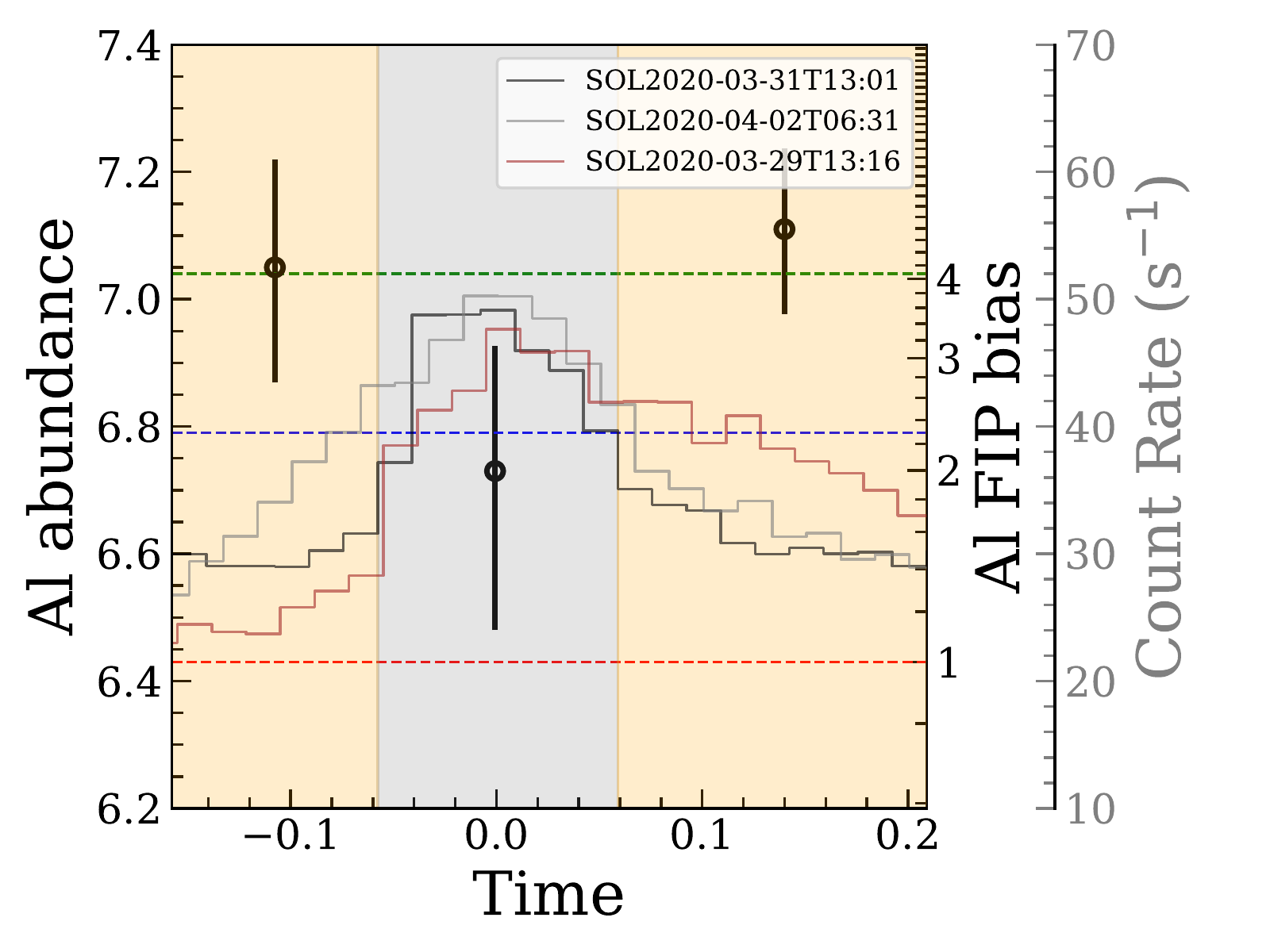} }}
    
    \subcaptionbox{}{{\includegraphics[width=0.25\textwidth]{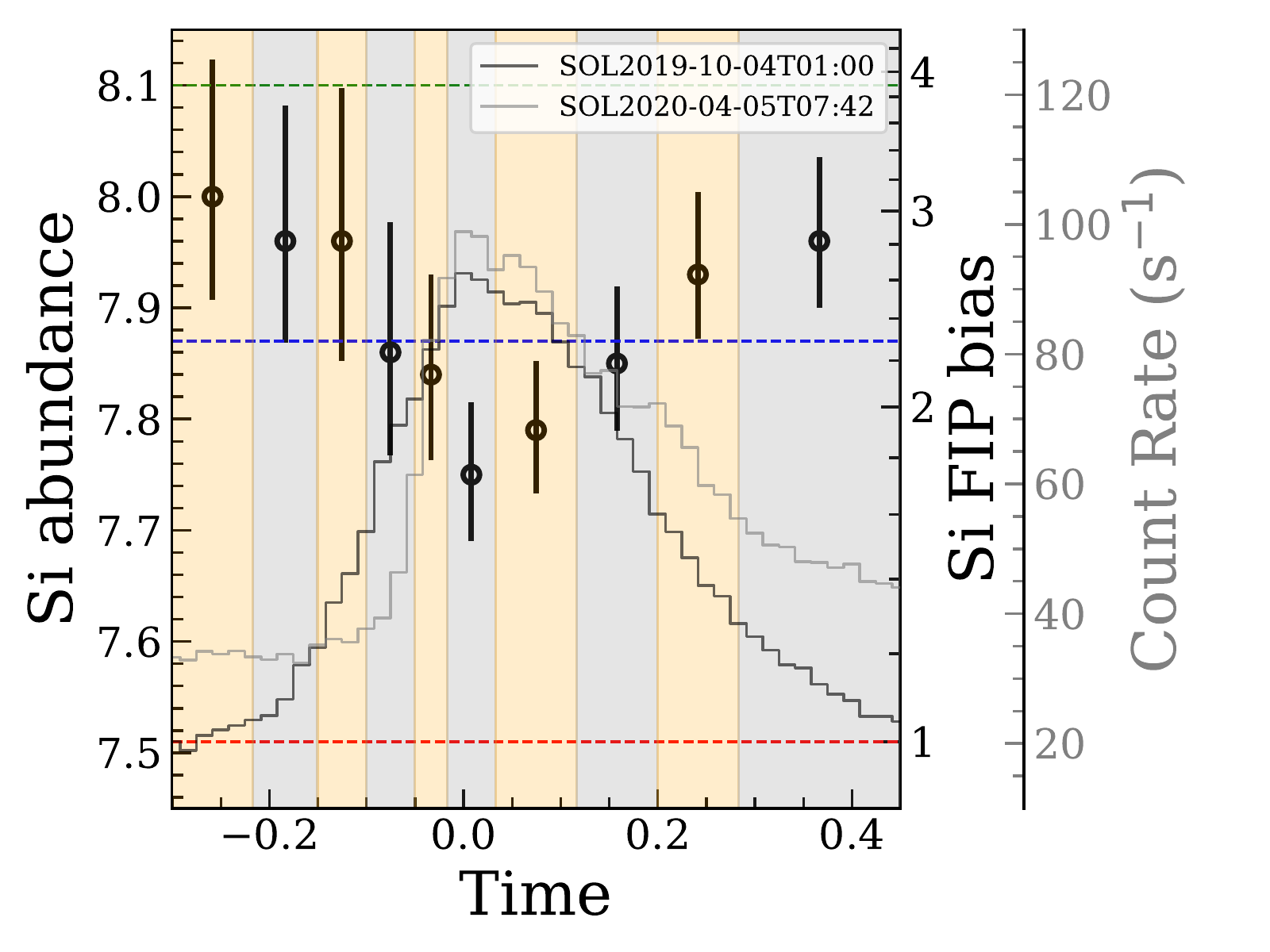} }}
    \hfill
    \subcaptionbox{}{{\includegraphics[width=0.25\textwidth]{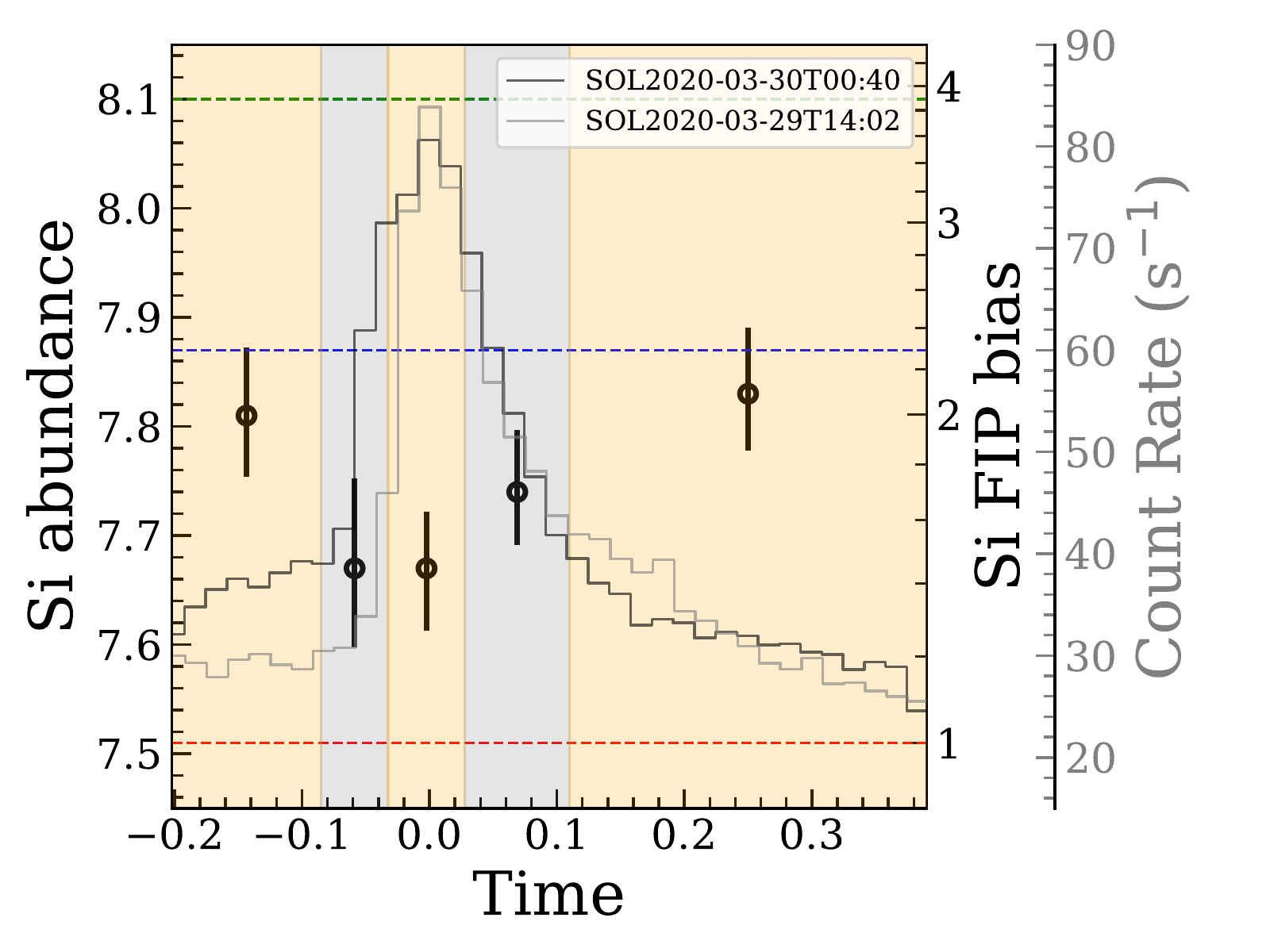} }}
    \hfill
    \subcaptionbox{}{{\includegraphics[width=0.25\textwidth]{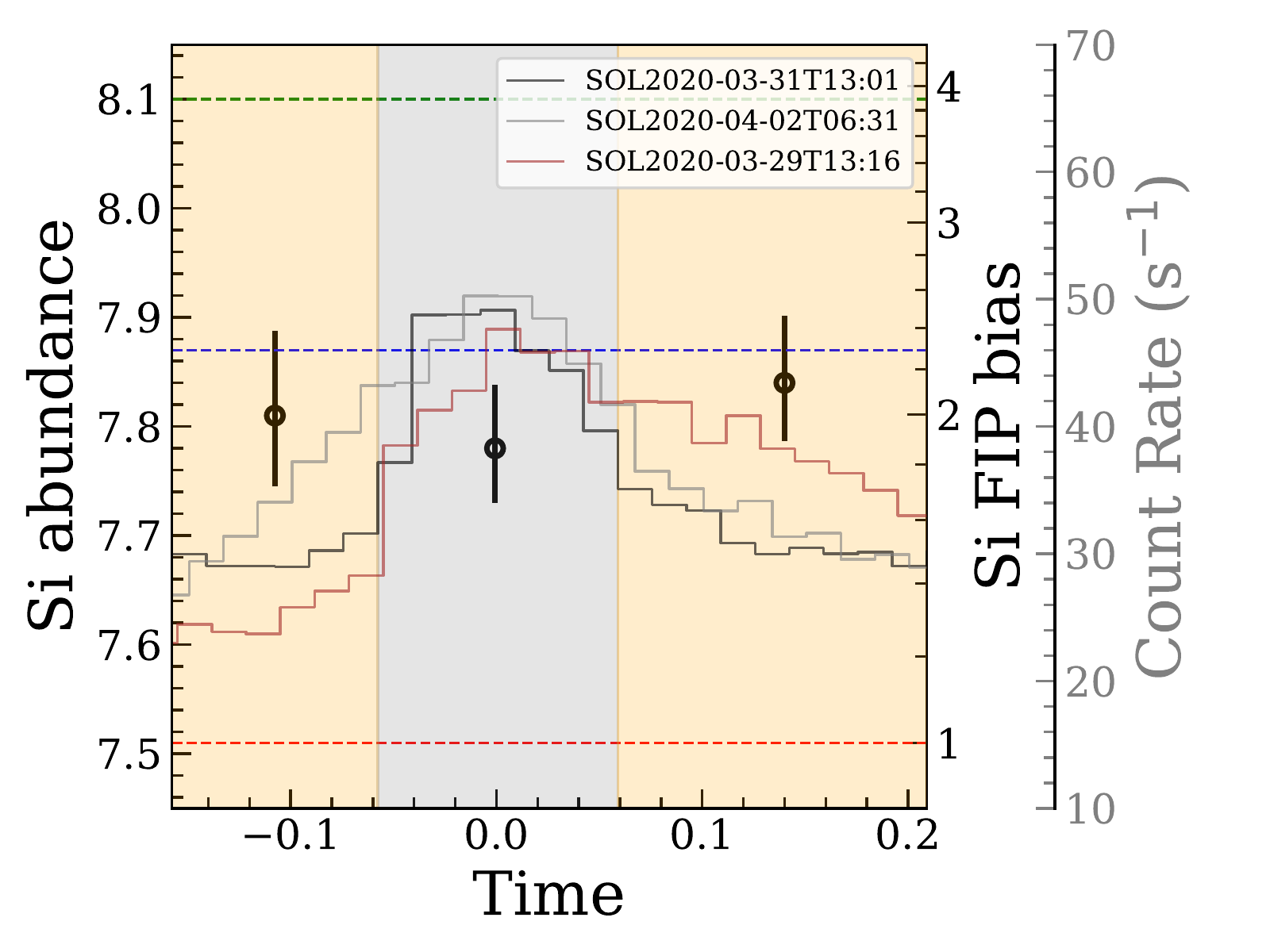} }}
    
    \subcaptionbox{}{{\includegraphics[width=0.25\textwidth]{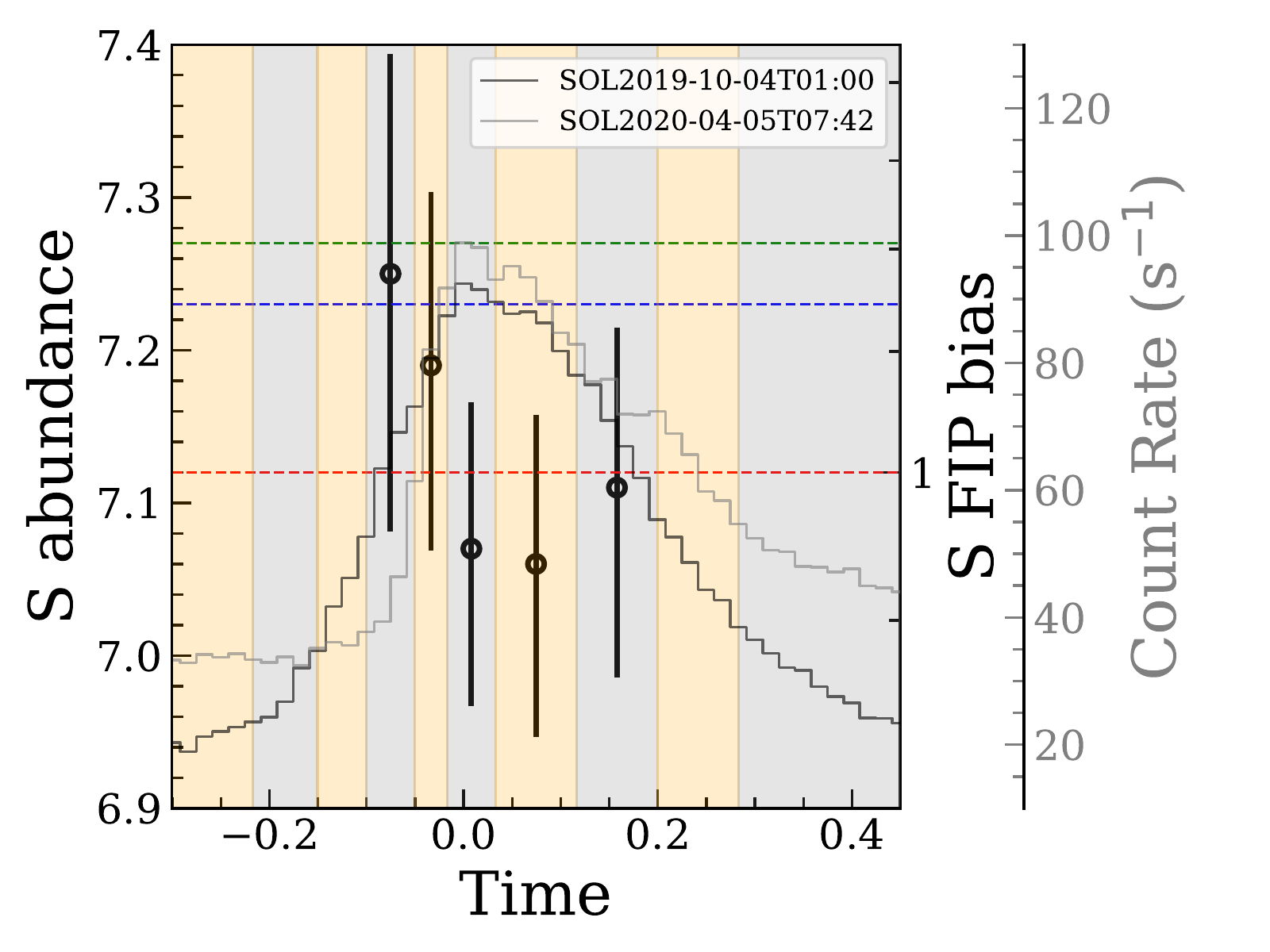} }}
    \hfill
    \caption{The above panels show the results of the time-resolved X-ray spectroscopy for three representative sets of flares (set 1, set 2 and set 3). Panels {\bf a-c} and {\bf d-f} show the variation of temperature and  emission measure. Panels {\bf g-i}, {\bf j-l}, {\bf m-o} and {\bf p} show  the variation of elemental abundances of Mg, Al, Si and S in logarithmic scale. In panels {\bf g-p}, the red horizontal dotted line represents the photospheric abundances from \cite{Asplund_2009} and the blue and green horizontal line represents the coronal abundances from \cite{Schmelz2012} and \cite{Feldman1992} respectively. The X-ray XSM light curves of all the flares of a set are over-plotted in grey (only in Panels {\bf a,d,g,j} and {\bf m}) in the background.}
    \label{fig8}
\end{figure}

Spectra is generated for each of these time bins, for all the flares of a set, in a similar way as discussed in Section~\ref{stat}. The spectra of corresponding bins of all the flares in a set are added using the `xsmaddspec' module of XSMDAS. The spectral analysis for each added spectra is carried out (Section~\ref{stat}) using the 1T `chisoth' model by keeping the temperature, emission measure and abundances of Mg, Al, and Si as free variables. 
The abundance of S was kept free only for selected time bins of set 1 of flares (two A3-class flares) as it was not possible to constrain its value in other sets due to poor statistics. The obtained best-fit parameters with the one sigma errors are plotted in Figure~\ref{fig8} for the three representative sets of flares. The evolution of temperature (Figure~\ref{fig9}), emission measure (Figure~\ref{fig10}) and abundances (Magnesium : Figure~\ref{fig11}, Aluminium: Figure~\ref{fig12}, Silicon: Figure~\ref{fig13}, Sulfur: Figure~\ref{fig14}) for all the other sets of flares is shown in the Appendix.

\section{Results and discussion} 
      \label{res} 

In this work, we have studied 65 A-class flares that were observed by XSM during the minimum of solar cycle 24. Analyzing the broadband spectra observed by XSM during these flares, we have derived the average plasma parameters (i.e., temperature, emission measure, and the abundances of low FIP elements Mg, Al, Si, and S). The results are shown in Figure~\ref{fig4}, and the same is tabulated in the Appendix (Table~\ref{tab2}). The different panels in Figure~\ref{fig4} represent the variation of temperature (a), emission measure (b), FIP bias of Mg (c), Al (d), Si (e), and S (f) with the flare number. 
The points with error bars in each panel represent the plasma parameters associated with each flare. The flares are grouped according to their sub-class, and the associated parameters of each group are shown by the same colour; e.g., the A1 class flares are shown by red error bars, whereas the A2 class flares are shown by green error bars, and so on. The horizontal line associated with each group represents the average value for all the flares in that group. The results show that the temperature and emission measure of the flaring plasma varies and scales with the flare sub-class, whereas the FIP bias of Mg, Al, Si, and S does not show a similar variation.
The flares-groups with lower activity (e.g., A1 class) are found to have intermediate FIP bias (specifically for Si and Al), whereas the higher class of flare-groups (e.g., A3-A6 class) have a lower FIP bias.
For comparison purposes, we also have shown the average plasma parameters for the higher B-class flares, which are observed by XSM and studied in detail by \cite{Mondal_2021}.

Earlier studies during the peak of larger solar flares have found that the FIP bias is close to unity (e.g., \citealp{Warren2014, Narendranath2014, Dennis_2015, Sylwester2015}),  which indicates that un-fractionated plasma materials from the lower solar atmosphere (e.g., beneath the chromosphere) evaporates through the flaring loops during the flaring process~\citep{Benz_2017}. In the present analysis, a near unity FIP bias for the integrated duration of the larger flares indicates the same phenomenon of evaporation of the un-fractionated plasma. 
During these flares, the emission from the hot ($>$ 3 MK) evaporated plasma is dominated over the surrounding quiescent AR emission of temperature $\sim$3 MK (Figure~\ref{fig4}a and Figure~\ref{fig4}b).
If the evaporation of the plasma from the depth of the lower atmosphere is a common phenomenon for the large as well as for the small flares, the flaring plasma should have a FIP bias of one.
But the smaller flares show intermediate FIP bias (specifically for Si and Al, as shown in Figure~\ref{fig4}).
Several factors could account for these findings.
First, and most likely, the enormous emission from the surrounding AR plasma (having higher FIP bias) is mixed with the smaller emission of the evaporated plasma (having unit FIP bias) from these tiny flares. 
The second possibility is that the plasma is being fractionated (i.e., ions are separating from neutrals) from a layer that is much lower than the layer from which the plasma is evaporating. 
To understand this further, we have carried out the time-resolved spectroscopic study of these small flares, as discussed later.

We also note that the FIP bias deviation between the smaller and larger class of flares is not the same for all the elements.
Figure~\ref{fig5} shows the variation of average FIP bias of the elements for each flare group with the FIP values.
The deviation is more for Al, which has the lowest FIP value among the four elements considered in this study, while the FIP bias deviation is the smallest for S, which has the highest FIP value. S is considered as an intermediate FIP element \citep{Feldman1992}. Thus, its FIP bias is intrinsically small for the coronal plasma \citep{Sylwester2012}, which explains our results of similar FIP bias of S for all the flare groups.

In addition to the time-integrated study of the A-class flares, it is interesting to see the evolution of the plasma parameters during these flares using the time-resolved spectroscopic analysis. However, poor statistics of the observed spectra from these small flares restrict the performance of the time-resolved spectroscopic analysis. Thus, in order to carry out the time-resolved spectral analysis, we have chosen multiple A-class flares having similar rise times, peak counts, and decay times (Figure~\ref{fig6}). After co-aligning all of these similar flares with respect to their peak time, we have added their spectra to increase the counting statistics. We perform the time-resolved spectroscopy of the added spectra and estimate the temporal evolution of the plasma parameters as stated in Section~\ref{time}. Here, we assume that the emission mechanism of the flaring plasmas with respect to the peak time is similar for all of these flares, which is a valid assumption for all the flares having similar X-ray light curves.
Results for three representative sets of flares are shown in Figure~\ref{fig8}. The panels {\bf a-c} show the evolution of temperature, {\bf d-f} show the evolution of emission measure, panels {\bf g-i}, {\bf j-l}, {\bf m-o} show the evolution of the abundances of Mg, Al, Si respectively. Panel {\bf p} shows the evolution of the S abundance during the peak of set 1 when the signal at the energy of the S line complex is good enough to constrain its abundance in the spectral fitting. In the background, the light curves of the flares are shown by solid grey lines. 
We find that both the temperature and emission measure follow the X-ray light curve of the flares with a maximum near the peak.
During the impulsive phase, the abundances of the elements reduce from the coronal values (as shown by dashed green and blue lines taken from \cite{Feldman1992} and \cite{Schmelz2012} respectively) towards the photospheric values (red dashed line: \cite{Asplund_2009}) and reach their minima during the flare peak. The elemental abundances quickly return to their preflare coronal values during the decay phase. From our observations, we find that the recovery rate of the coronal FIP bias is of the order of minutes. Typically, it varies between $\sim$ 4 to $\sim$ 10 minutes. The evolution of parameters is similar to that seen for the larger B-class of flares as reported by \cite{Mondal_2021}. 

The evolution of temperature and emission measures can be explained using the standard flare model (CSHKP: \citealp{Carmichael1964, Sturrock1966, Hirayama1974, Kopp1976}). Once the flaring mechanism is initiated by the process of reconnection, heat energy from the reconnection site flows down to the chromosphere and begins rapidly evaporating the chromospheric plasma into the loop with high velocity \citep{Antonucci1985}. The hot and dense chromospheric plasma fills the flaring loop, causing an increase in temperature and emission measures. The temperature and emission measure reaches its peak when the loop is mostly filled with high-density heated evaporated material \citep{Ryan2012, Klimchuk2017}. The rapid depletion of abundance can be explained by the evaporative up-flow, which transports unfractionated chromospheric material into the corona. However, the rapid recovery (in a time frame of minutes) of the coronal FIP bias during the flare decay stages is challenging to explain.


The Ponderomotive force model~\citep{Laming_2004, Laming2009, Laming2012, Laming2015} is the widely accepted  explanation of the FIP effect in closed-loop coronal structures (e.g., active regions).
In closed loops, this model suggests that downward propagating coronal  Alfv\'en waves, which could be generated by coronal nanoflares~\citep{Dahlburg_2016}, can reflect from the chromospheric heights and generate Ponderomotive force.
This force is directed upwards and acts only on the ions (not the neutrals) present in the chromospheric layers, causing a fractionation of the ions and the neutrals within the local plasma. The temperature and the density of the chromospheric plasma are such that only the low FIP elements are ionized and are present at the fractionation region, and thus the Ponderomotive force acts only on the ions of low FIP elements. The local fractionation  of the plasma at the chromospheric height can produce the FIP bias in a short time scale of a few minutes ($\sim$10 min)~\citep{Baker2015}. Once the fractionation occurs, the chromospheric fractionated plasma reaches the corona by the process of transportation or diffusion. 
\cite{Widing_2001} observationally measured that the FIP bias within the newly emerging AR loops is developed in a time scale of days, which is close to the diffusion time scale. 
In the present study, during the decay phase of the A-class flares, we find the coronal FIP bias is developed on a time scale of a few minutes  ($\sim$4 min to 10 min). This is possible by quickly transporting the fractionated chromospheric plasma to the corona or/and by increasing the fractionation rate of the chromospheric plasma. 
For these, \cite{Mondal_2021} suggested two possible scenarios. One is based on the evaporation process during the flare, which can transport the chromospheric fractionated plasma very quickly to the corona in the decay phase of the flares. The other scenario is based on the evaporation process as well as the flare-driven Alfv\'en waves.
In the decay phase, the reflection of the high amplitude flare-driven  Alfv\'en waves fractionates the chromospheric plasma at a much faster rate, and then this fractionated plasma is evaporated to the corona, causing a rapid development of the coronal FIP bias. 

As the FIP build-up in the decay phase of the few flares is happening in a few minutes time scale, we prefer the second scenario of \cite{Mondal_2021}, which relies on the evaporation velocity as well as the flare-driven Alfv\'en waves. 
In this case to explain both the depletion and recovery  of the FIP bias during the impulsive and decay phases, the heat transported from the flaring site at the corona should be faster than the flare-driven Alfv\'en waves. This could be possible by transporting the heat by the suprathermal electrons, which is known to have a faster speed than the Alfv\'en speed~\citep{Benz_2017}. Just after the initiation of the flare by the onset of the reconnection, heat is quickly transported to the chromosphere and starts rapid evaporation causing a depletion of the abundance until flare-peak. In the decay phase, high amplitude flare-driven Alfv\'en waves reach the chromosphere and fractionate the plasma at a faster rate. This fractionated plasma is carried by the evaporation processes to develop the coronal FIP bias at a faster rate.
It is to be noted that the flare-driven Alfv\'en waves are yet to be observed in the solar corona with our present generation instrumentation and techniques. However, our results indicate their existence even during the small A-class flares. To increase our understanding of this topic, a detailed simulation/modeling effort is needed further.

\section{Summary}
      \label{sum} 
 
We report the X-ray spectroscopic measurements of the plasma temperature, emission measure, and elemental (Mg, Al, Si, and S) abundances for a large number of A-class flares observed by XSM/Chandrayaan-2 during the minimum of solar cycle 24. Time-integrated spectral analysis during these flares shows that the average temperature and emission measure scales with the flare sub-class, whereas the FIP bias of the elements does not follow the same behavior. The higher A-class flares show a FIP bias of near unity, indicating the evaporation of the un-fractionated plasma materials from the lower solar atmosphere to the corona, while the lower A-class flares show an intermediate FIP bias, either due to the mixing-up of background solar emissions with the small flaring emissions, or/and the evaporated plasma materials are already fractionated.
We have also derived the time evolution of the parameters using the time-resolved spectroscopic analysis.
Due to the small emissions from these flares, it is challenging to carry out time-resolved spectroscopy using the sun-as-a-star observation of XSM. Thus, we have added the observed spectrum of multiple flares with similar properties (similar flare class, rise and decay times). 
We find that during the impulsive phase, the temperature and emission measure increases, while the elemental abundances drop towards photospheric values and reach a minimum near the peak.
In the decay phase, the abundances are seen to quickly return to their coronal values on a time scale of minutes, similar to the higher B-class flares, as reported earlier.
These observations suggest the existence and role of flare-driven Alfv\'en waves in increasing the fractionation rate and transporting the fractionated plasma to the corona through the evaporation processes, even in A-class flares.
Further simulations and the spectroscopic measurements of smaller flares with future instrumentation will help to increase our understanding of the topic.

\begin{acks}
XSM  was designed and developed by the Physical Research Laboratory (PRL), Ahmedabad, with support from the Space Application Centre(SAC), Ahmedabad, the U. R. Rao Satellite Centre (URSC), Bengaluru, and the Laboratory for Electro-Optics Systems (LEOS), Bengaluru. We thank various facilities and the technical teams of all the above centers and the Chandrayaan-2 project,  mission operations,  and ground segment teams for their support. The Chandrayaan-2 mission is funded and managed by the Indian Space Research Organisation (ISRO). We are very grateful to Mithun N.P.S and Dr. Santosh Vadawale for the helpful scientific discussions and important suggestions for this work. I acknowledge the support and resources provided during my visit to Physical Research Laboratory (PRL), Ahmedabad, thanks to Dr. Anil Bhardwaj. 

\end{acks}




\appendix

\begin{landscape}

\begin{longtable}{ c c c c c c c c c } \\ 

\caption{Plasma parameters obtained from the 1T spectral analysis for all the A-class flares considered in this study. Flare IDs corresponding to peak times of the flares is given in the format SOLyyyy-mm-ddThh:mm. Approximate GOES classes are mentioned according to the peak flux of the flare in the 1-8 $\textup{\AA}$ XSM flux. Errors correspond to the 1$\sigma$ limits associated with the parameters.}\\ \\
\hline
{\bf Sl no} & {\bf Flare ID} & \multicolumn{1}{p{3cm}}{\centering \bf Flare Class \\ (GOES approximation)} & \multicolumn{1}{p{3cm}}{\centering \bf Temperature \\ (MK)} &  \multicolumn{1}{p{3.5cm}}{\centering \bf Emission Measure \\ $\bf (\times 10^{46} cm^{-3}$)} & {\bf Mg FIP bias} & {\bf Al FIP bias} & {\bf Si FIP bias} & {\bf S FIP bias} \\ [0.5ex] \hline \hline \\
\endfirsthead
\hline
{\bf Sl no} & {\bf Flare ID} & \multicolumn{1}{p{3cm}}{\centering \bf Flare Class \\ (GOES approximation)} & \multicolumn{1}{p{3cm}}{\centering \bf Temperature \\ (MK)} &  \multicolumn{1}{p{3.5cm}}{\centering \bf Emission Measure \\ $\bf (\times 10^{46} cm^{-3}$)} & {\bf Mg FIP bias} & {\bf Al FIP bias} & {\bf Si FIP bias} & {\bf S FIP bias} \\ [0.5ex] \hline \hline \\
\endhead 

1 & SOL2019-09-30T12:21 & A3.3 & $6.98^{+0.67}_{-0.62}$ & $2.83^{+0.97}_{-0.79}$ & $1.38^{+0.28}_{-0.20}$ & $3.77^{+1.23}_{-0.95}$ & $2.09^{+0.26}_{-0.21}$ & $0.76^{+1.01}_{-0.53}$  \\ [0.3cm]
2 & SOL2019-09-30T12:42 & A2.3 & $5.95^{+0.48}_{-0.38}$ & $6.88^{1.81+}_{-1.84}$ & $1.20^{+0.29}_{-0.20}$ & $2.85^{+1.14}_{-0.87}$ & $1.71^{+0.24}_{-0.19}$ & $0.77^{+1.00}_{-0.55}$  \\ [0.3cm]  
3 & SOL2019-09-30T19:47 & A4.1 & $5.93^{+0.22}_{-0.21}$ & $10.05^{+1.50}_{-1.35}$ & $1.38^{+0.30}_{-0.20}$ & $2.83^{+1.27}_{-0.94}$ & $2.34^{+0.31}_{-0.26}$ & $1.18^{+1.56}_{-0.80}$  \\  [0.3cm]
4 & SOL2019-09-30T20:49 & A3.9 & $6.04^{+0.23}_{-0.26}$ & $11.52^{+2.11}_{-1.59}$ & $1.36^{+0.25}_{-0.16}$ & $3.22^{+1.07}_{-0.77}$ & $2.36^{+0.26}_{-0.20}$ & $0.95^{+1.15}_{-0.74}$  \\   [0.3cm]
5 & SOL2019-09-30T21:40 & A3.0 & $5.35^{+0.20}_{-0.19}$ & $7.57^{+1.26}_{-1.11}$ & $1.10^{+0.24}_{-0.19}$ & $2.49^{+0.89}_{-0.75}$ & $1.52^{+0.21}_{-0.17}$ & $0.63^{+0.80}_{-0.48}$  \\    [0.3cm]
6 & SOL2019-10-01T03:26 & A6.4 & $5.64^{+0.12}_{-0.13}$ & $20.69^{+2.15}_{-1.89}$ & $1.29^{+0.19}_{-0.13}$ & $3.06^{+0.87}_{-0.65}$ & $2.19^{+0.20}_{-0.15}$ & $0.85^{+0.98}_{-0.72}$  \\   [0.3cm]
7 & SOL2019-10-01T06:45 & A1.1 & $4.64^{+0.20}_{-0.18}$ & $4.18^{+0.88}_{-0.80}$ & $1.20^{+0.37}_{-0.23}$ & $2.89^{+1.28}_{-0.94}$ & $1.95^{+0.33}_{-0.24}$ & $0.95^{+1.23}_{-0.70}$  \\  [0.3cm]
8 & SOL2019-10-01T12:47 & A2.6 & $4.95^{+0.18}_{-0.18}$ & $7.32^{+1.39}_{-1.14}$ & $1.63^{+0.32}_{-0.25}$ & $4.36^{+1.28}_{-1.08}$ & $2.40^{+0.29}_{-0.25}$ & $1.06^{+1.34}_{-0.82}$  \\   [0.3cm]
9 & SOL2019-10-01T17:14 & A1.7 & $4.34^{+0.17}_{-0.14}$ & $9.11^{+1.57}_{-1.60}$ & $0.88^{+0.32}_{-0.21}$ & $1.21^{+1.06}_{-0.85}$ & $1.47^{+0.27}_{-0.21}$ & $0.79^{+1.09}_{-0.53}$  \\   [0.3cm]
10 & SOL2019-10-01T18:55 & A1.4 & $4.06^{+0.21}_{-0.17}$ & $6.67^{+1.52}_{-1.53}$ & $1.48^{+0.66}_{-0.39}$ & $4.61^{+2.35}_{-1.58}$ & $2.26^{+0.63}_{-0.41}$ & $1.31^{+1.79}_{-0.92}$ \\ [0.3cm]
11 & SOL2019-10-01T20:33 & A1.9 & $4.87^{+0.21}_{-0.19}$ & $5.65^{+1.17}_{-1.04}$ & $0.81^{+0.16}_{-0.11}$ & $1.66^{+0.71}_{-0.56}$ & $1.30^{+0.14}_{-0.12}$ & $0.75^{+0.90}_{-0.60}$ \\ [0.3cm]
12 & SOL2019-10-02T00:33 & A3.4 & $5.11^{+0.21}_{-0.21}$ & $11.50^{+2.37}_{-1.96}$ & $1.15^{+0.32}_{-0.22}$ & $2.60^{+1.12}_{-0.88}$ & $1.52^{+0.25}_{-0.19}$ & $0.57^{+0.77}_{-0.38}$ \\ [0.3cm]
13 & SOL2019-10-02T14:49 & A1.5 & $4.49^{+0.27}_{-0.22}$ & $6.09^{+1.69}_{-1.58}$ & $1.17^{+0.22}_{-0.16}$ & $2.85^{+0.95}_{-0.75}$ & $1.72^{+0.20}_{-0.16}$ & $0.74^{+0.92}_{-0.57}$ \\ [0.3cm]
14 & SOL2019-10-04T01:00 & A3.0 & $5.03^{+0.20}_{-0.17}$ & $8.57^{+1.54}_{-1.53}$ & $1.21^{+0.19}_{-0.13}$ & $3.16^{+0.91}_{-0.68}$ & $1.85^{+0.18}_{-0.14}$ & $0.72^{+0.87}_{-0.57}$ \\ [0.3cm]
15 & SOL2019-10-05T13:34 & A1.4 & $4.93^{+0.47}_{-0.42}$ & $4.07^{+1.82}_{-1.44}$ & $0.82^{+0.23}_{-0.17}$ & $2.42^{+1.00}_{-0.80}$ & $1.41^{+0.22}_{-0.17}$ & $0.75^{+0.96}_{-0.56}$ \\ [0.3cm]
16 & SOL2019-11-12T15:06 & A2.9 & $4.98^{+0.27}_{-0.25}$ & $7.76^{+2.10}_{-1.81}$ & $1.20^{+0.17}_{-0.11}$ & $3.00^{+0.84}_{-0.59}$ & $2.07^{+0.17}_{-0.13}$ & $0.90^{+1.01}_{-0.76}$ \\ [0.3cm]
17 & SOL2020-03-08T03:15 & A1.2 & $4.57^{+0.40}_{-0.35}$ & $3.41^{+1.55}_{-1.21}$ & $1.23^{+0.38}_{-0.23}$ & $2.25^{+1.34}_{-0.96}$ & $1.78^{+0.32}_{-0.24}$ & $1.09^{+1.46}_{-0.75}$ \\ [0.3cm]
18 & SOL2020-03-08T11:47 & A1.2 & $4.15^{+0.17}_{-0.12}$ & $7.54^{+1.18}_{-1.31}$ & $1.17^{+0.35}_{-0.23}$ & $2.67^{+1.30}_{-0.98}$ & $1.76^{+0.30}_{-0.23}$ & $0.83^{+1.14}_{-0.55}$ \\ [0.3cm] 
19 & SOL2020-03-08T12:42 & A1.2 & $4.39^{+0.26}_{-0.22}$ & $5.23^{+1.42}_{-1.30}$ & $1.28^{+0.35}_{-0.23}$ & $3.69^{+1.53}_{-1.14}$ & $2.20^{+0.36}_{-0.29}$ & $1.29^{+1.76}_{-0.87}$ \\ [0.3cm]
20 & SOL2020-03-08T13:19 & A1.2 & $3.90^{+0.16}_{-0.12}$ & $6.75^{+1.21}_{-1.30}$ & $0.96^{+0.41}_{-0.28}$ & $1.03^{+1.18}_{-0.93}$ & $1.47^{+0.37}_{-0.27}$ & $0.92^{+1.28}_{-0.62}$ \\ [0.3cm]
21 & SOL2020-03-08T15:11 & A1.2 & $3.63^{+0.12}_{-0.08}$ & $8.75^{+1.18}_{-1.47}$ & $1.17^{+0.22}_{-0.16}$ & $2.79^{+0.88}_{-0.70}$ & $1.77^{+0.20}_{-0.16}$ & $0.87^{+1.03}_{-0.72}$ \\ [0.3cm]
22 & SOL2020-03-11T09:36 & A1.1 & $3.69^{+0.11}_{-0.08}$ & $8.81^{+1.12}_{-1.39}$ & $1.08^{+0.22}_{-0.15}$ & $2.51^{+0.93}_{-0.71}$ & $1.79^{+0.21}_{-0.17}$ & $0.76^{+0.94}_{-0.59}$ \\ [0.3cm]
23 & SOL2020-03-11T11:50 & A1.3 & $4.96^{+0.40}_{-0.39}$ & $4.70^{+1.95}_{-1.43}$ & $1.18^{+0.28}_{-0.20}$ & $2.30^{+1.07}_{-0.83}$ & $1.79^{+0.26}_{-0.20}$ & $0.71^{+0.94}_{-0.49}$ \\ [0.3cm]
24 & SOL2020-03-11T23:35 & A1.1 & $4.30^{+0.32}_{-0.24}$ & $4.58^{+1.36}_{-1.34}$ & $0.97^{+0.50}_{-0.29}$ & $1.69^{+1.67}_{-1.16}$ & $2.05^{+0.59}_{-0.38}$ & $1.40^{+1.96}_{-0.96}$ \\ [0.3cm]
25 & SOL2020-03-29T13:16 & A1.2 & $4.58^{+0.26}_{-0.22}$ & $5.43^{+1.37}_{-1.24}$ & $1.21^{+0.50}_{-0.28}$ & $2.75^{+1.73}_{-1.20}$ & $1.93^{+0.42}_{-0.30}$ & $0.88^{+1.30}_{-0.53}$ \\ [0.3cm]
26 & SOL2020-03-29T14:02 & A2.7 & $5.49^{+0.31}_{-0.30}$ & $8.02^{+2.09}_{-1.73}$ & $0.94^{+0.18}_{-0.13}$ & $1.63^{+0.65}_{-0.54}$ & $1.62^{+0.17}_{-0.14}$ & $0.90^{+1.03}_{-0.78}$ \\ [0.3cm]
27 & SOL2020-03-29T18:43 & A1.7 & $4.29^{+0.17}_{-0.13}$ & $8.14^{+1.31}_{-1.45}$ & $1.21^{+0.36}_{-0.24}$ & $2.18^{+1.13}_{-0.87}$ & $2.09^{+0.36}_{-0.26}$ & $0.91^{+1.16}_{-0.69}$ \\ [0.3cm]
28 & SOL2020-03-30T00:44 & A2.6 & $5.02^{+0.27}_{-0.25}$ & $9.40^{+2.31}_{-1.97}$ & $1.28^{+0.24}_{-0.16}$ & $3.17^{+0.99}_{-0.76}$ & $1.90^{+0.22}_{-0.17}$ & $0.81^{+0.99}_{-0.64}$ \\ [0.3cm]
29 & SOL2020-03-30T06:01 & A1.8 & $3.94^{+0.15}_{-0.11}$ & $8.05^{+1.33}_{-1.49}$ & $1.27^{+0.42}_{-0.28}$ & $2.84^{+1.43}_{-1.08}$ & $2.11^{+0.42}_{-0.30}$ & $1.21^{+1.56}_{-0.90}$ \\ [0.3cm]
30 & SOL2020-03-30T10:40 & A1.3 & $4.21^{+0.17}_{-0.13}$ & $6.54^{+1.15}_{-1.20}$ & $1.07^{+0.64}_{-0.33}$ & $2.34^{+1.90}_{-1.21}$ & $1.77^{+0.57}_{-0.33}$ & $1.10^{+1.52}_{-0.75}$ \\ [0.3cm]
31 & SOL2020-03-30T16:56 & A1.9 & $4.90^{+0.16}_{-0.15}$ & $6.13^{+0.96}_{-0.88}$ & $0.61^{+0.24}_{-0.17}$ & $0.98^{+1.00}_{-0.82}$ & $1.32^{+0.25}_{-0.20}$ & $0.95^{+1.23}_{-0.70}$ \\ [0.3cm]
32 & SOL2020-03-30T18:28 & A2.2 & $5.05^{+0.28}_{-0.26}$ & $5.93^{+1.60}_{-1.33}$ & $1.13^{+0.23}_{-0.17}$ & $2.73^{+0.95}_{-0.74}$ & $2.00^{+0.24}_{-0.19}$ & $0.87^{+1.05}_{-0.70}$ \\ [0.3cm]
33 & SOL2020-03-30T21:28 & A1.0 & $3.64^{+0.17}_{-0.13}$ & $6.73^{+1.41}_{-1.48}$ & $1.00^{+0.19}_{-0.15}$ & $2.03^{+0.77}_{-0.63}$ & $1.67^{+0.19}_{-0.16}$ & $0.75^{+0.90}_{-0.61}$ \\ [0.3cm]
34 & SOL2020-03-31T08:09 & A1.2 & $3.52^{+0.09}_{-0.06}$ & $11.11^{+1.11}_{-1.61}$ & $0.98^{+0.19}_{-0.14}$ & $2.74^{+0.86}_{-0.67}$ & $1.78^{+0.20}_{-0.16}$ & $0.88^{+1.03}_{-0.72}$ \\ [0.3cm]
35 & SOL2020-03-31T09:28 & A1.5 & $3.92^{+0.15}_{-0.11}$ & $8.58^{+1.43}_{-1.60}$ & $1.06^{+0.22}_{-0.15}$ & $2.42^{+0.88}_{-0.67}$ & $1.75^{+0.21}_{-0.16}$ & $0.89^{+1.07}_{-0.72}$ \\ [0.3cm]
36 & SOL2020-03-31T13:01 & A1.2 & $4.03^{+0.30}_{-0.20}$ & $6.05^{+1.69}_{-1.80}$ & $1.28^{+0.23}_{-0.16}$ & $3.73^{+1.02}_{-0.78}$ & $2.07^{+0.23}_{-0.18}$ & $0.82^{+0.99}_{-0.66}$ \\ [0.3cm]
37 & SOL2020-04-01T05:15 & A1.3 & $3.95^{+0.17}_{-0.12}$ & $7.78^{+1.36}_{-1.53}$ & $1.44^{+0.37}_{-0.25}$ & $3.04^{+1.26}_{-0.95}$ & $2.41^{+0.39}_{-0.29}$ & $1.01^{+1.26}_{-0.78}$ \\ [0.3cm]
38 & SOL2020-04-01T07:47 & A1.0 & $3.99^{+0.22}_{-0.17}$ & $5.30^{+1.26}_{-1.28}$ & $1.01^{+0.27}_{-0.21}$ & $1.80^{+0.93}_{-0.79}$ & $1.75^{+0.29}_{-0.23}$ & $1.16^{+1.42}_{-0.92}$ \\ [0.3cm]
39 & SOL2020-04-01T20:11 & A1.7 & $4.18^{+0.18}_{-0.13}$ & $8.56^{+1.43}_{-1.63}$ & $1.13^{+0.24}_{-0.19}$ & $1.97^{+0.82}_{-0.70}$ & $1.64^{+0.23}_{-0.19}$ & $0.76^{+0.94}_{-0.60}$ \\ [0.3cm]
40 & SOL2020-04-02T02:04 & A3.1 & $4.95^{+0.16}_{-0.15}$ & $9.27^{+1.50}_{-1.39}$ & $0.97^{+0.16}_{-0.14}$ & $2.09^{+0.64}_{-0.56}$ & $1.41^{+0.15}_{-0.13}$ & $0.73^{+0.84}_{-0.62}$ \\ [0.3cm]
41 & SOL2020-04-02T06:31 & A1.2 & $4.08^{+0.30}_{-0.22}$ & $6.54^{+1.93}_{-1.90}$ & $1.08^{+0.44}_{-0.30}$ & $2.25^{+1.33}_{-1.06}$ & $1.45^{+0.34}_{-0.24}$ & $0.99^{+1.28}_{-0.74}$ \\ [0.3cm]
42 & SOL2020-04-02T07:32 & A4.1 & $5.29^{+0.16}_{-0.18}$ & $12.65^{+2.06}_{-1.66}$ & $0.93^{+0.38}_{-0.24}$ & $1.51^{+1.22}_{-0.91}$ & $1.68^{+0.36}_{-0.25}$ & $1.27^{+1.63}_{-0.95}$ \\ [0.3cm]
43 & SOL2020-04-02T16:23 & A1.6 & $4.17^{+0.17}_{-0.14}$ & $7.9^{+1.48}_{-1.54}$ & $1.00^{+0.15}_{-0.11}$ & $2.32^{+0.66}_{-0.52}$ & $1.88^{+0.17}_{-0.13}$ & $1.13^{+1.25}_{-1.02}$ \\ [0.3cm]
44 & SOL2020-04-04T03:09 & A6.2 & $5.60^{+0.29}_{-0.30}$ & $18.05^{+4.60}_{-3.50}$ & $0.97^{+0.45}_{-0.26}$ & $1.95^{+1.52}_{-1.09}$ & $1.51^{+0.38}_{-0.26}$ & $0.89^{+1.28}_{-0.57}$ \\ [0.3cm]
45 & SOL2020-04-04T03:38 & A2.0 & $4.83^{+0.39}_{-0.38}$ & $7.22^{+3.22}_{-2.28}$ & $0.91^{+0.27}_{-0.20}$ & $1.88^{+0.95}_{-0.78}$ & $1.47^{+0.26}_{-0.20}$ & $0.86^{+1.09}_{-0.66}$ \\ [0.3cm]
46 & SOL2020-04-04T06:03 & A2.1 & $4.49^{+0.42}_{-0.33}$ & $8.88^{+3.63}_{-3.10}$ & $0.33^{+0.21}_{-0.13}$ & \textendash & $0.82^{+0.19}_{-0.13}$ & $0.81^{+1.03}_{-0.62}$ \\ [0.3cm]
47 & SOL2020-04-04T06:23 & A4.5 & $5.49^{+0.34}_{-0.33}$ & $11.56^{+3.45}_{-2.74}$ & $1.02^{+0.19}_{-0.16}$ & $2.23^{+0.70}_{-0.65}$ & $1.54^{+0.18}_{-0.16}$ & $0.68^{+0.81}_{-0.56}$ \\ [0.3cm]
48 & SOL2020-04-04T07:02 & A3.6 & $5.25^{+0.22}_{-0.22}$ & $9.9^{+2.10}_{-1.77}$ & $0.64^{+0.20}_{-0.15}$ & $0.86^{+0.72}_{-0.62}$ & $1.25^{+0.19}_{-0.16}$ & $0.88^{+1.08}_{-0.70}$ \\ [0.3cm]
49 & SOL2020-04-05T02:14 & A1.1 & $3.67^{+0.17}_{-0.15}$ & $6.73^{+1.67}_{-1.46}$ & $0.62^{+0.21}_{-0.16}$ & $1.07^{+0.82}_{-0.68}$ & $1.20^{+0.21}_{-0.17}$ & $0.99^{+1.23}_{-0.78}$ \\ [0.3cm]
50 & SOL2020-04-05T07:42 & A3.0 & $5.06^{+0.19}_{-0.18}$ & $9.01^{+1.68}_{-1.48}$ & $0.97^{+0.29}_{-0.21}$ & $2.20^{+1.06}_{-0.85}$ & $1.75^{+0.31}_{-0.23}$ & $0.95^{+1.20}_{-0.73}$ \\ [0.3cm]
51 & SOL2020-04-05T11:03 & A1.8 & $4.32^{+0.22}_{-0.18}$ & $6.67^{+1.62}_{-1.60}$ & $0.62^{+0.13}_{-0.11}$ & $1.48^{+0.55}_{-0.50}$ & $1.01^{+0.11}_{-0.10}$ & $0.81^{+0.92}_{-0.70}$ \\ [0.3cm]
52 & SOL2020-04-06T03:57 & A1.2 & $3.68^{+0.25}_{-0.18}$ & $8.08^{+2.20}_{-2.27}$ & $0.99^{+0.22}_{-0.16}$ & $1.92^{+0.78}_{-0.63}$ & $1.99^{+0.26}_{-0.20}$ & $1.02^{+1.29}_{-0.87}$ \\ [0.3cm]
53 & SOL2020-04-06T04:17 & A1.4 & $4.08^{+0.40}_{-0.28}$ & $6.71^{+2.66}_{-2.48}$ & $1.00^{+0.20}_{-0.16}$ & $1.72^{+0.70}_{-0.61}$ & $1.74^{+0.21}_{-0.18}$ & $0.68^{+0.82}_{-0.56}$ \\ [0.3cm]
54 & SOL2020-04-06T13:23 & A1.0 & $3.44^{+0.19}_{-0.12}$ & $8.80^{+1.81}_{-2.14}$ & $1.04^{+0.18}_{-0.15}$ & $2.25^{+0.68}_{-0.59}$ & $1.79^{+0.19}_{-0.16}$ & $1.06^{+1.19}_{-0.93}$ \\ [0.3cm] 
55 & SOL2020-04-06T14:56 & A1.0 & $3.64^{+0.14}_{-0.09}$ & $8.58^{+1.28}_{-1.66}$ & $0.33^{+0.25}_{-0.17}$ & \textendash & $0.82^{+0.24}_{-0.17}$ & $0.76^{+1.04}_{-0.54}$ \\ [0.3cm]
56 & SOL2020-04-06T18:36 & A1.5 & $4.42^{+0.30}_{-0.25}$ & $5.82^{+1.89}_{-1.68}$ & $0.74^{+0.17}_{-0.15}$ & $0.97^{+0.62}_{-0.56}$ & $1.47^{+0.19}_{-0.16}$ & $0.87^{+1.03}_{-0.72}$ \\ [0.3cm]
57 & SOL2020-04-06T21:47 & A6.9 & $6.03^{+0.23}_{-0.24}$ & $14.26^{+2.70}_{-2.22}$ & $0.92^{+0.21}_{-0.17}$ & $2.12^{+0.78}_{-0.67}$ & $1.40^{+0.20}_{-0.16}$ & $0.91^{+1.08}_{-0.76}$ \\ [0.3cm]
58 & SOL2020-04-07T08:29 & A3.7 & $5.81^{+0.27}_{-0.28}$ & $8.26^{+2.01}_{-1.62}$ & $0.87^{+0.25}_{-0.20}$ & $1.70^{+0.81}_{-0.68}$ & $1.60^{+0.26}_{-0.21}$ & $1.10^{+1.30}_{-0.93}$ \\ [0.3cm]
59 & SOL2020-04-07T11:36 & A1.1 & $4.40^{+0.34}_{-0.29}$ & $5.79^{+2.00}_{-1.70}$ & $0.37^{+0.10}_{-0.10}$ & $0.61^{+0.43}_{-0.44}$ & $0.81^{+0.09}_{-0.09}$ & $0.67^{+0.77}_{-0.58}$ \\ [0.3cm]
60 & SOL2020-04-07T12:58 & A1.5 & $4.76^{+0.51}_{-0.41}$ & $5.01^{+2.47}_{-2.01}$ & $0.36^{+0.09}_{-0.08}$ & $0.60^{+0.44}_{-0.39}$ & $0.84^{+0.09}_{-0.08}$ & $0.68^{+0.78}_{-0.59}$ \\ [0.3cm]
61 & SOL2020-04-09T07:14 & A2.0 & $4.59^{+0.38}_{-0.31}$ & $7.79^{+2.80}_{-2.45}$ & $0.79^{+0.13}_{-0.12}$ & $1.42^{+0.51}_{-0.50}$ & $1.26^{+0.13}_{-0.12}$ & $0.57^{+0.67}_{-0.49}$ \\ [0.3cm]
62 & SOL2020-04-10T00:22 & A1.8 & $4.99^{+0.24}_{-0.24}$ & $5.42^{+1.39}_{-1.13}$ & $0.75^{+0.27}_{-0.20}$ & $1.36^{+0.92}_{-0.74}$ & $1.30^{+0.26}_{-0.20}$ & $0.77^{+0.99}_{-0.57}$ \\ [0.3cm]
63 & SOL2020-04-10T02:55 & A2.0 & $4.32^{+0.12}_{-0.09}$ & $8.33^{+0.97}_{-1.14}$ & $0.55^{+0.18}_{-0.15}$ & $0.68^{+0.63}_{-0.58}$ & $1.21^{+0.19}_{-0.16}$ & $0.67^{+0.83}_{-0.53}$ \\ [0.3cm]
64 & SOL2020-04-10T22:40 & A1.4 & $4.68^{+0.17}_{-0.14}$ & $5.36^{+0.85}_{-0.85}$ & $0.66^{+0.09}_{-0.08}$ & $1.42^{+0.37}_{-0.36}$ & $1.05^{+0.08}_{-0.07}$ & $0.56^{+0.61}_{-0.51}$ \\ [0.3cm]
65 & SOL2020-04-23T20:07 & A1.5 & $5.05^{+0.39}_{-0.35}$ & $5.31^{+1.66}_{-1.38}$ & $0.75^{+0.17}_{-0.15}$ & $1.26^{+0.59}_{-0.53}$ & $1.44^{+0.18}_{-0.16}$ & $0.83^{+0.96}_{-0.71}$  \\ [1ex]
\hline
\label{tab2}

\end{longtable}
\end{landscape}
 
\clearpage 
\begin{figure}[ht] 
    
    \subcaptionbox{}{{\includegraphics[width=0.43\textwidth]{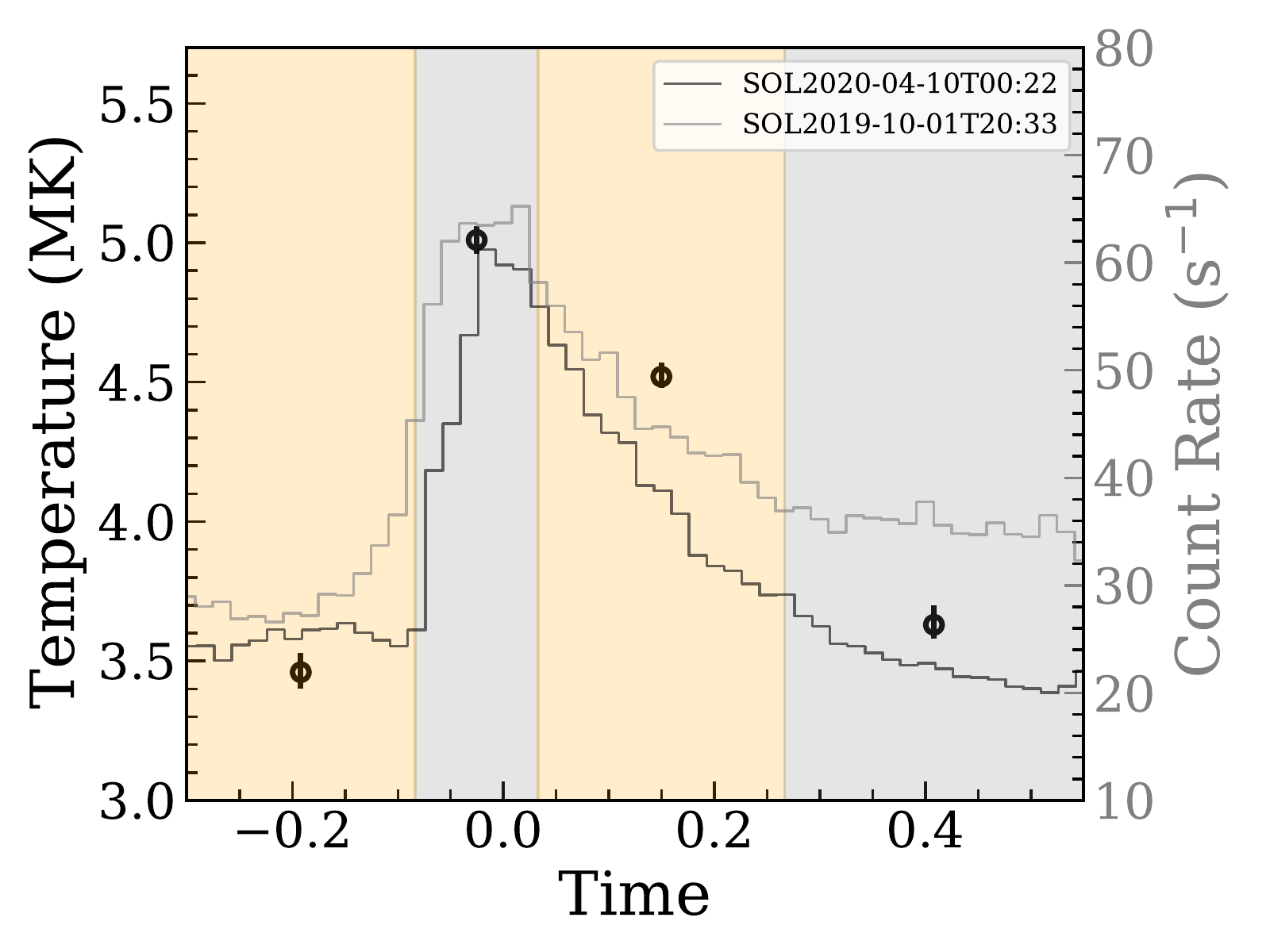} }}
    \hfill
    \subcaptionbox{}{{\includegraphics[width=0.43\textwidth]{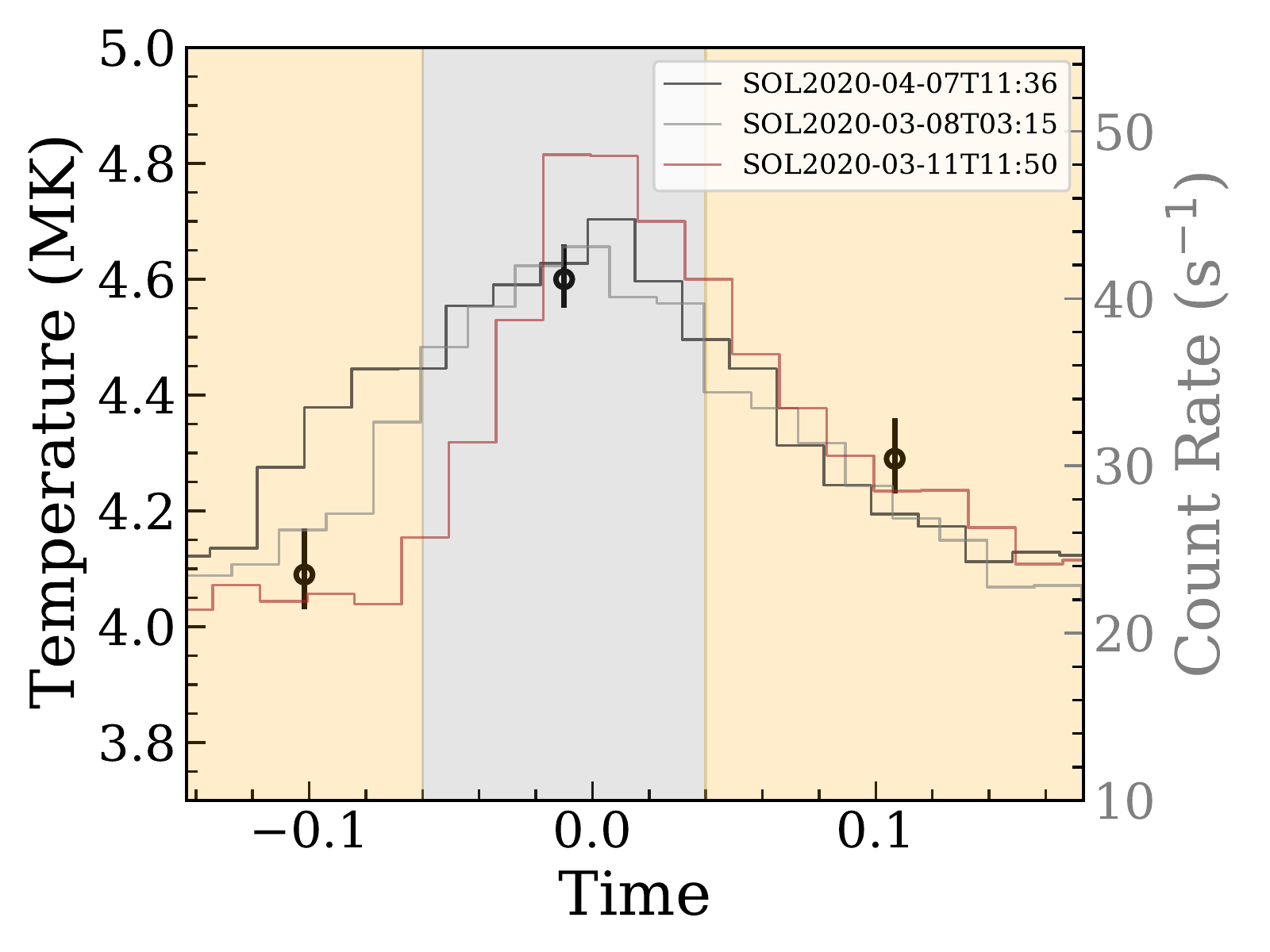} }}
    
    \subcaptionbox{}{{\includegraphics[width=0.44\textwidth]{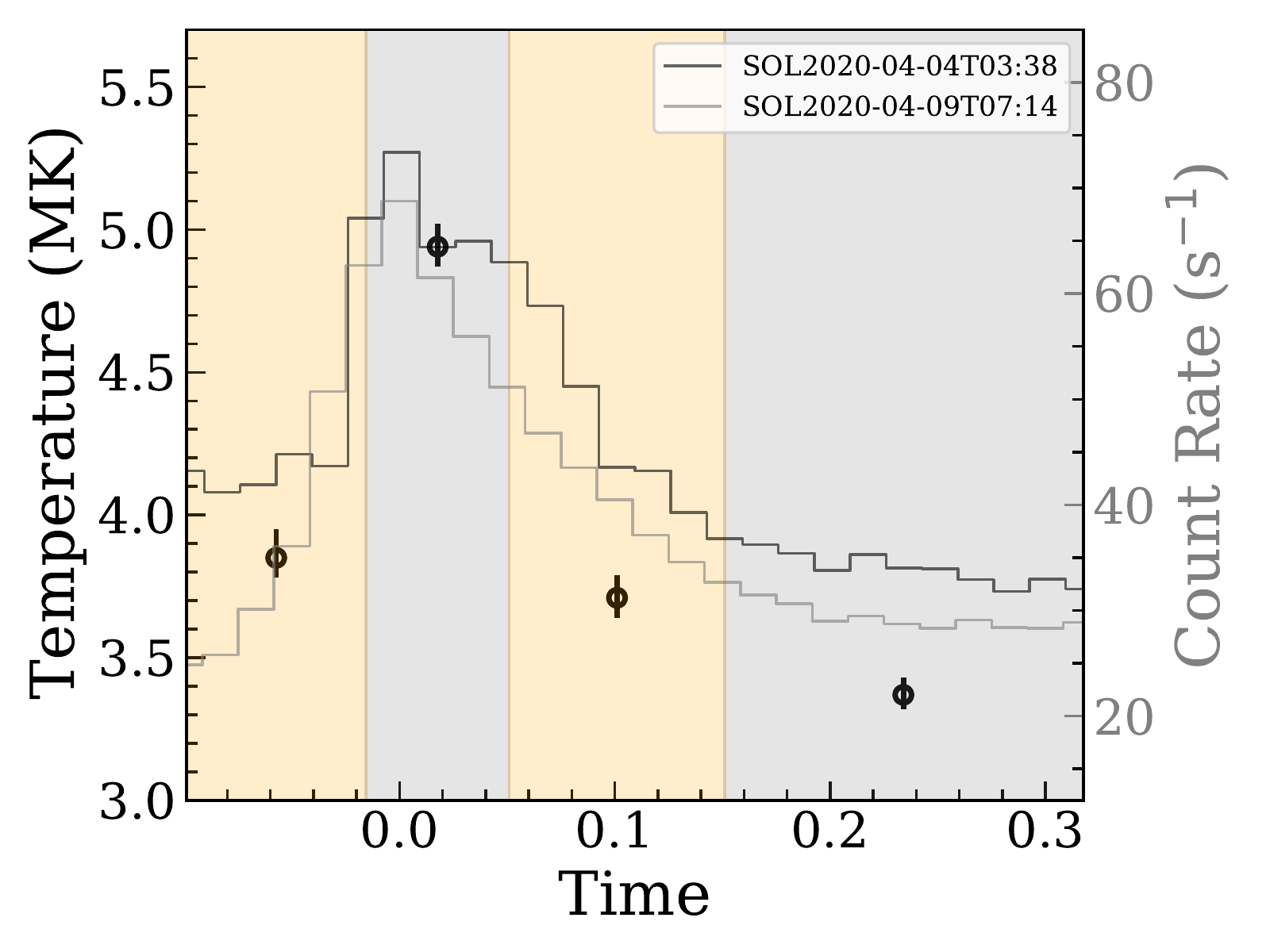} }} 
    \hfill
    \subcaptionbox{}{{\includegraphics[width=0.44\textwidth]{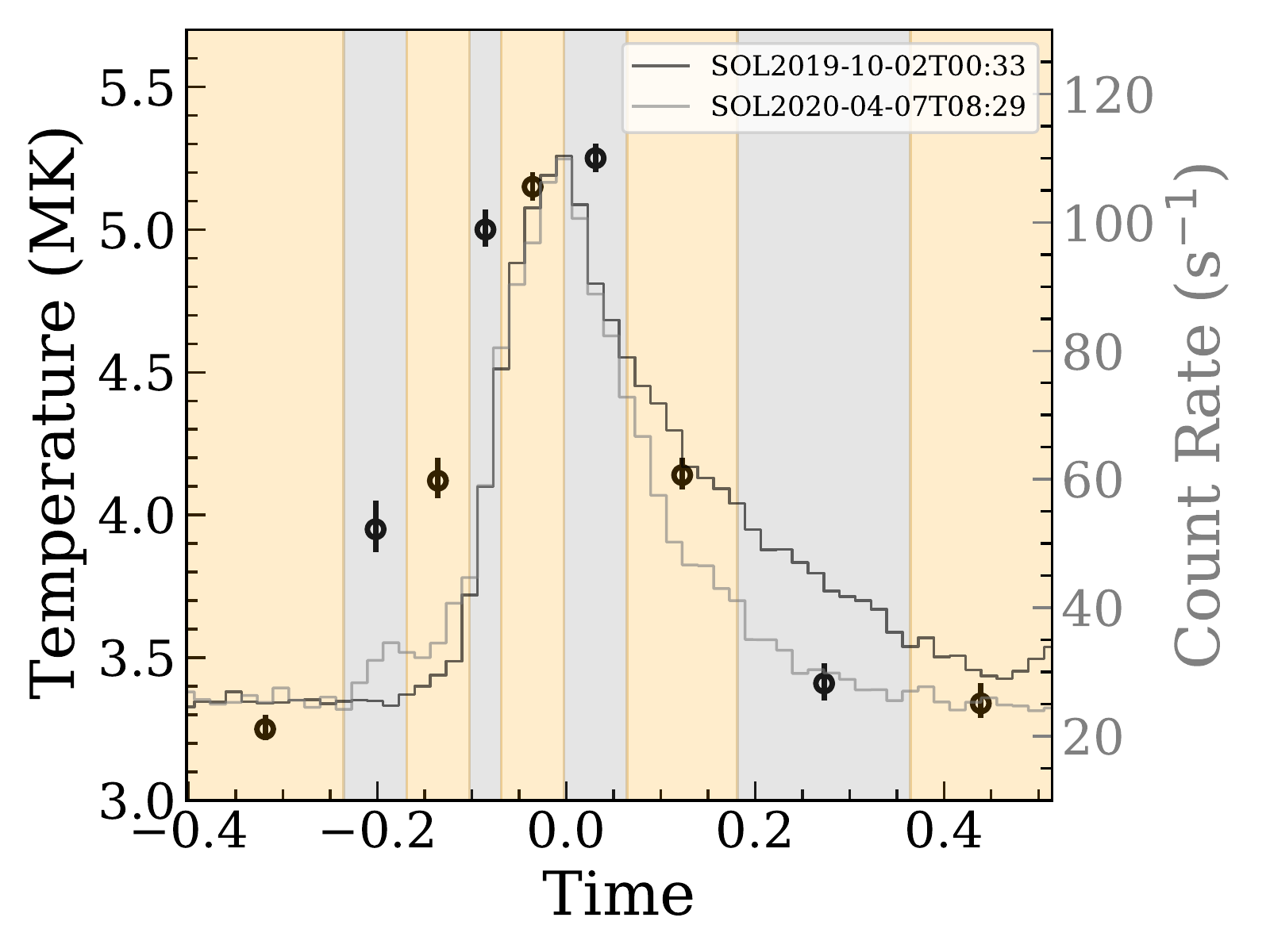} }}
    \caption{Evolution of temperature for the remaining sets of flares, similar to Figure~\ref{fig8} Panels {\bf a-c}}
    \label{fig9}
\end{figure}
 
\clearpage 
\begin{figure}[ht] 
    
    \subcaptionbox{}{{\includegraphics[width=0.43\textwidth]{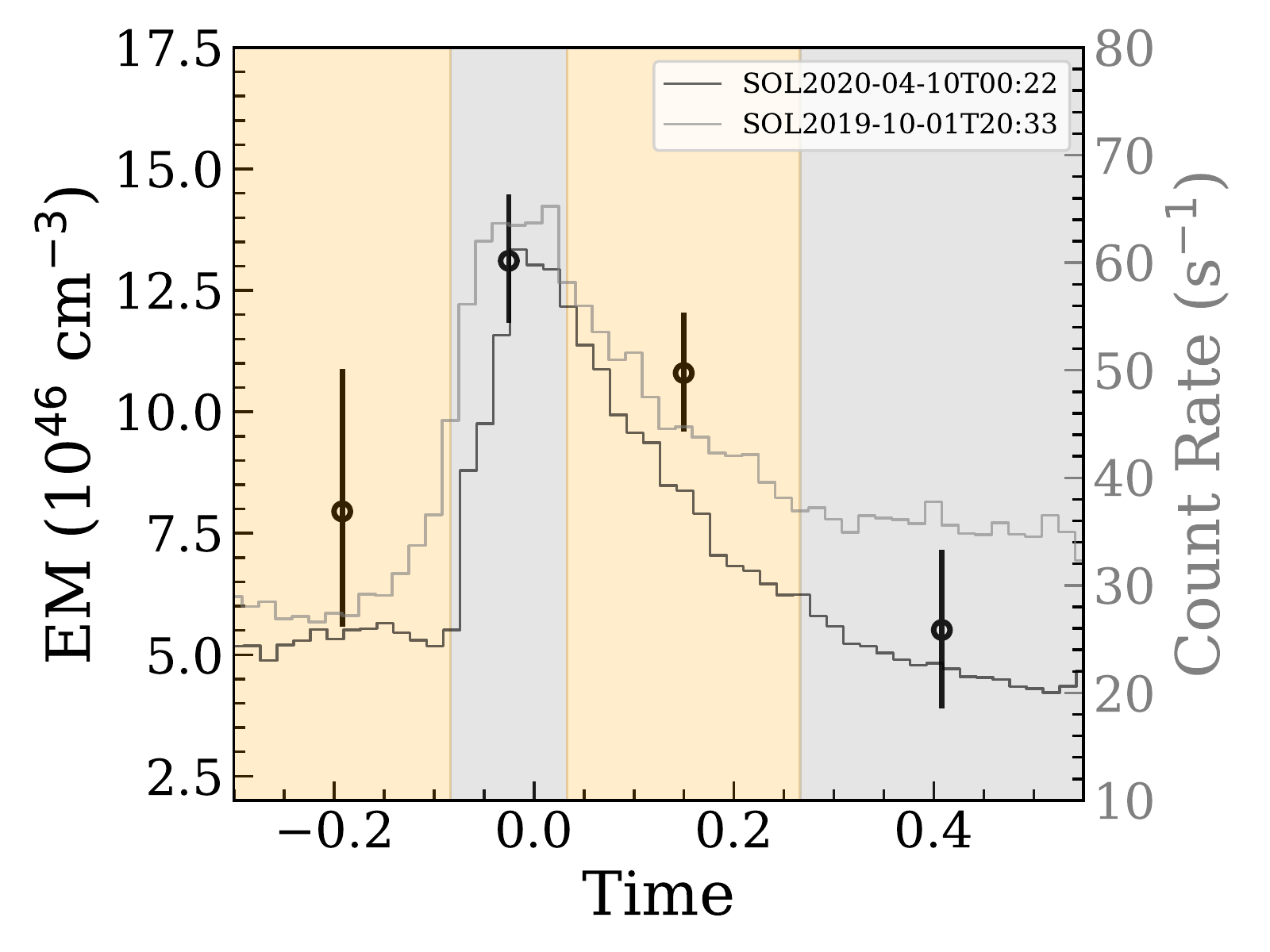} }}
    \hfill
    \subcaptionbox{}{{\includegraphics[width=0.43\textwidth]{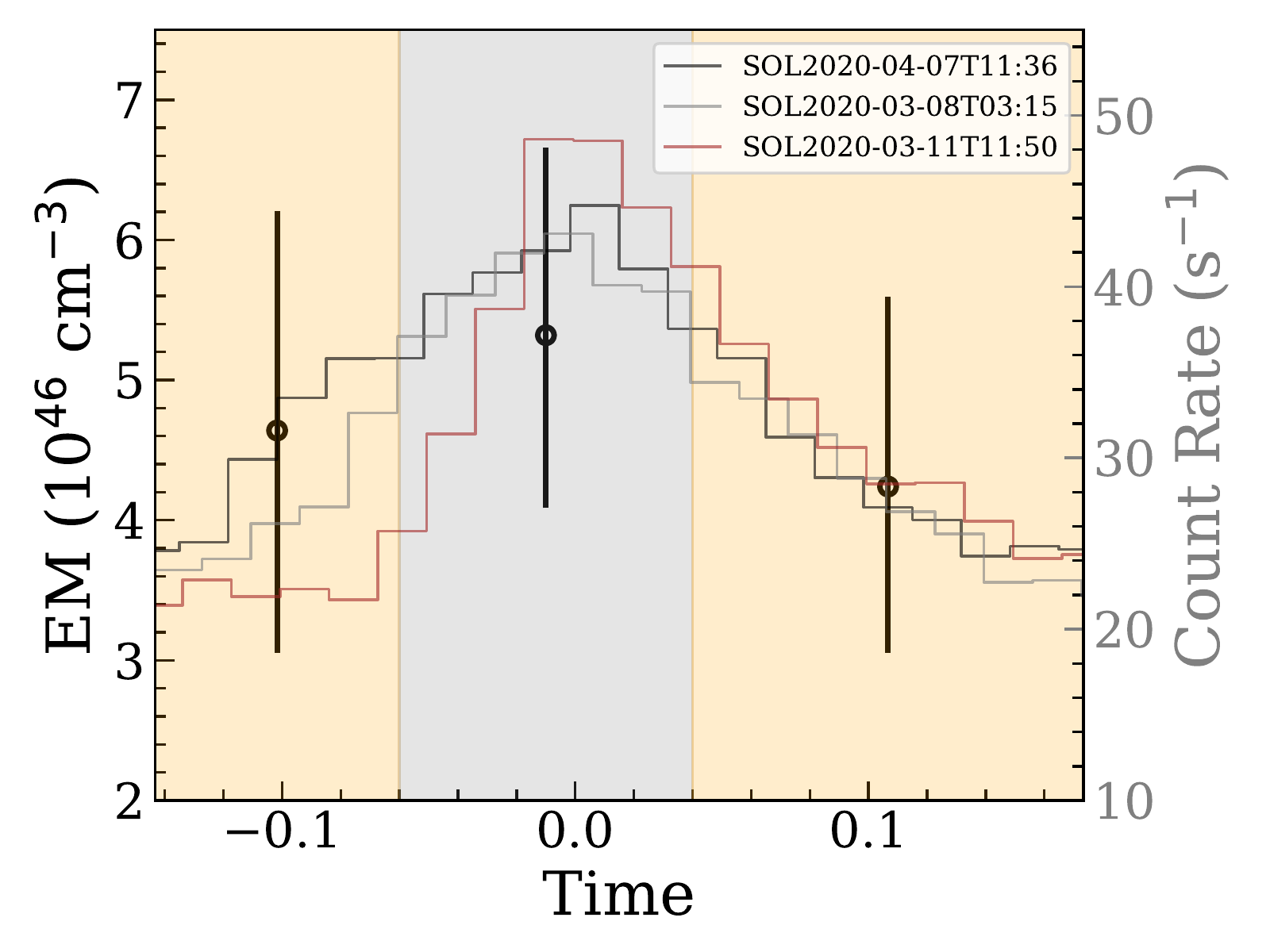} }}
    
    \subcaptionbox{}{{\includegraphics[width=0.44\textwidth]{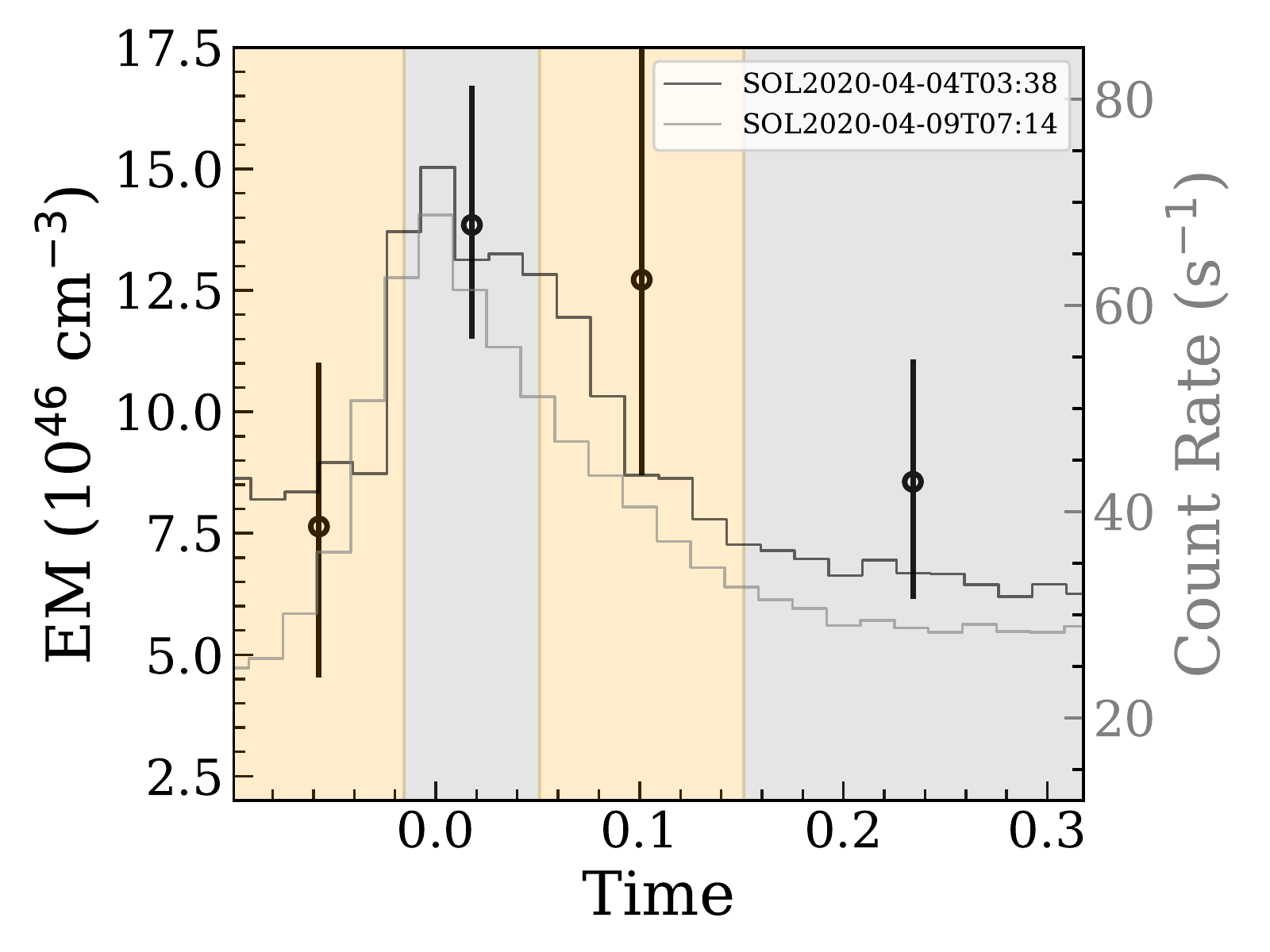} }} 
    \hfill
    \subcaptionbox{}{{\includegraphics[width=0.44\textwidth]{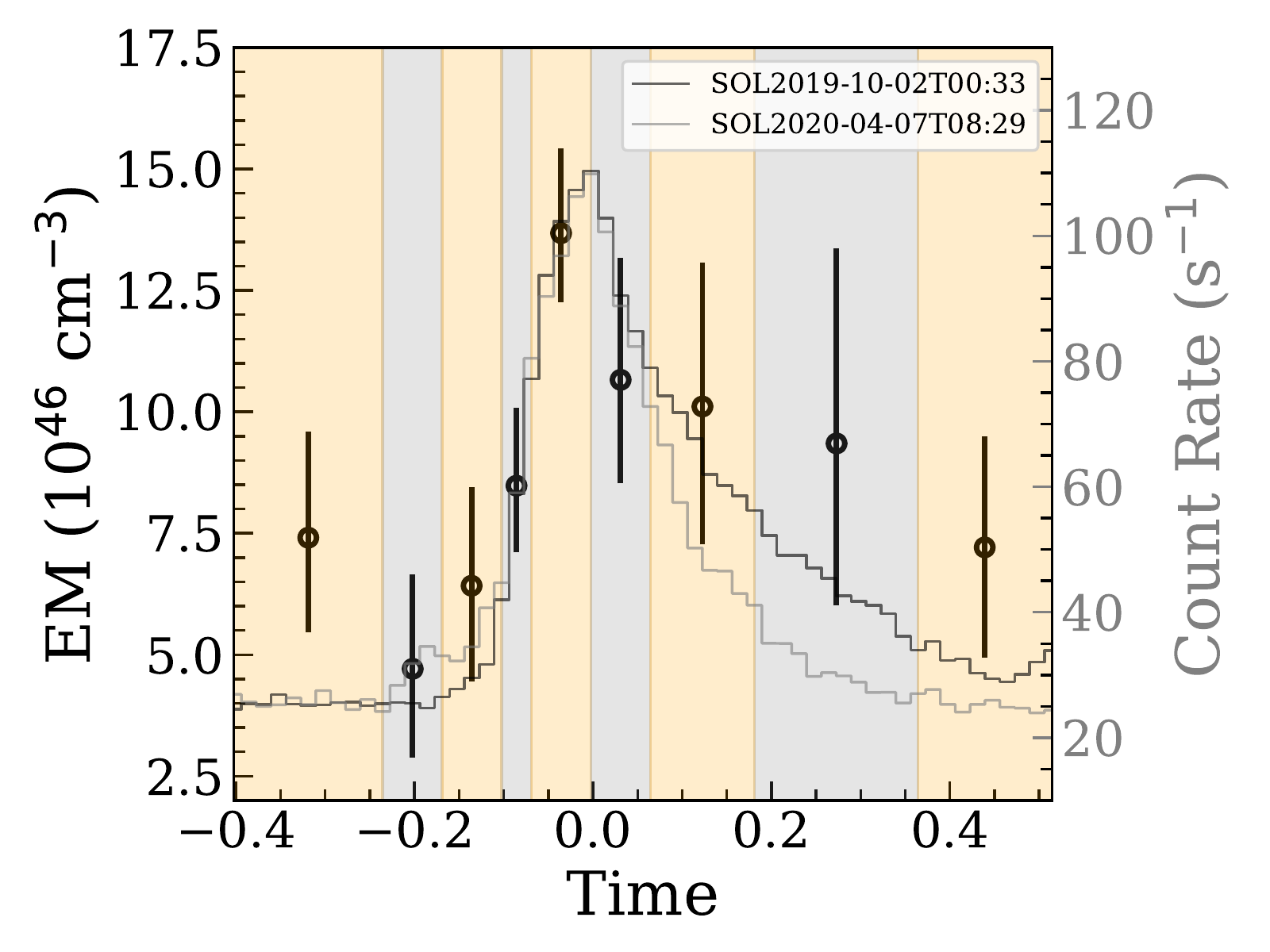} }}
    \caption{Evolution of emission measure for the remaining sets of flares, similar to Figure~\ref{fig7} Panels {\bf d-f}}
    \label{fig10}
\end{figure}

\clearpage
\begin{figure}[ht] 
    
    \subcaptionbox{}{{\includegraphics[width=0.43\textwidth]{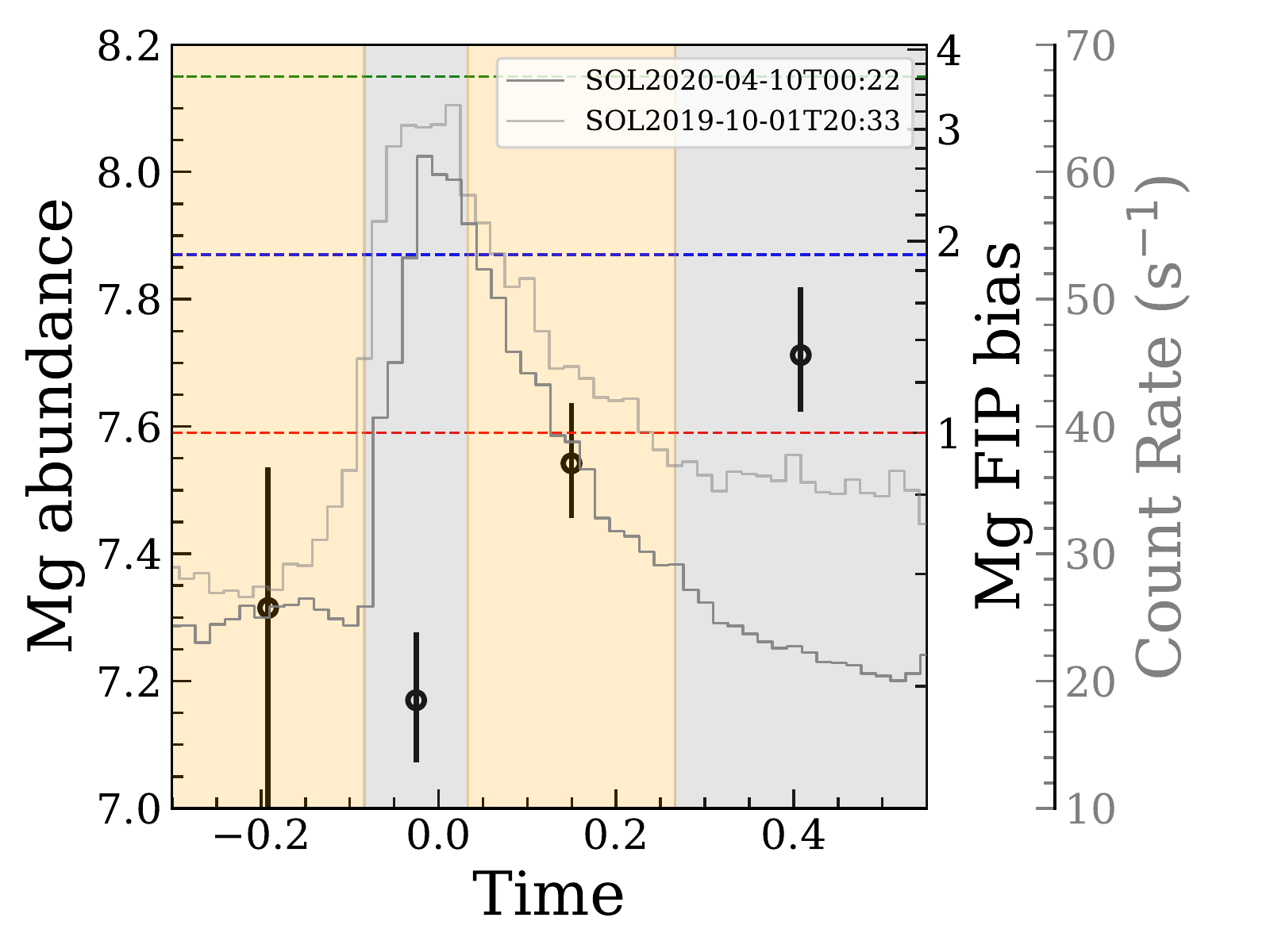} }}
    \hfill
    \subcaptionbox{}{{\includegraphics[width=0.43\textwidth]{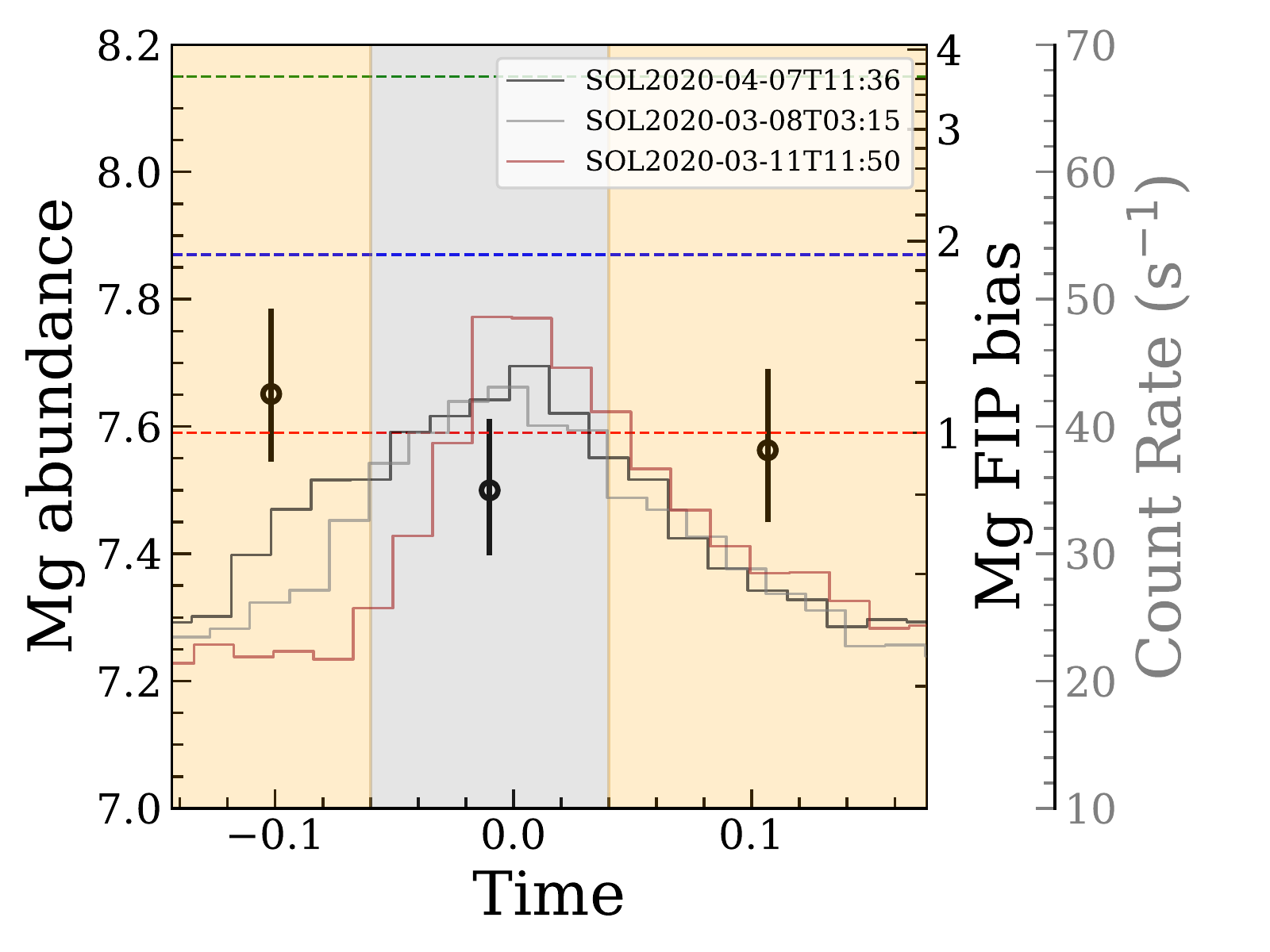} }}
    
    \subcaptionbox{}{{\includegraphics[width=0.44\textwidth]{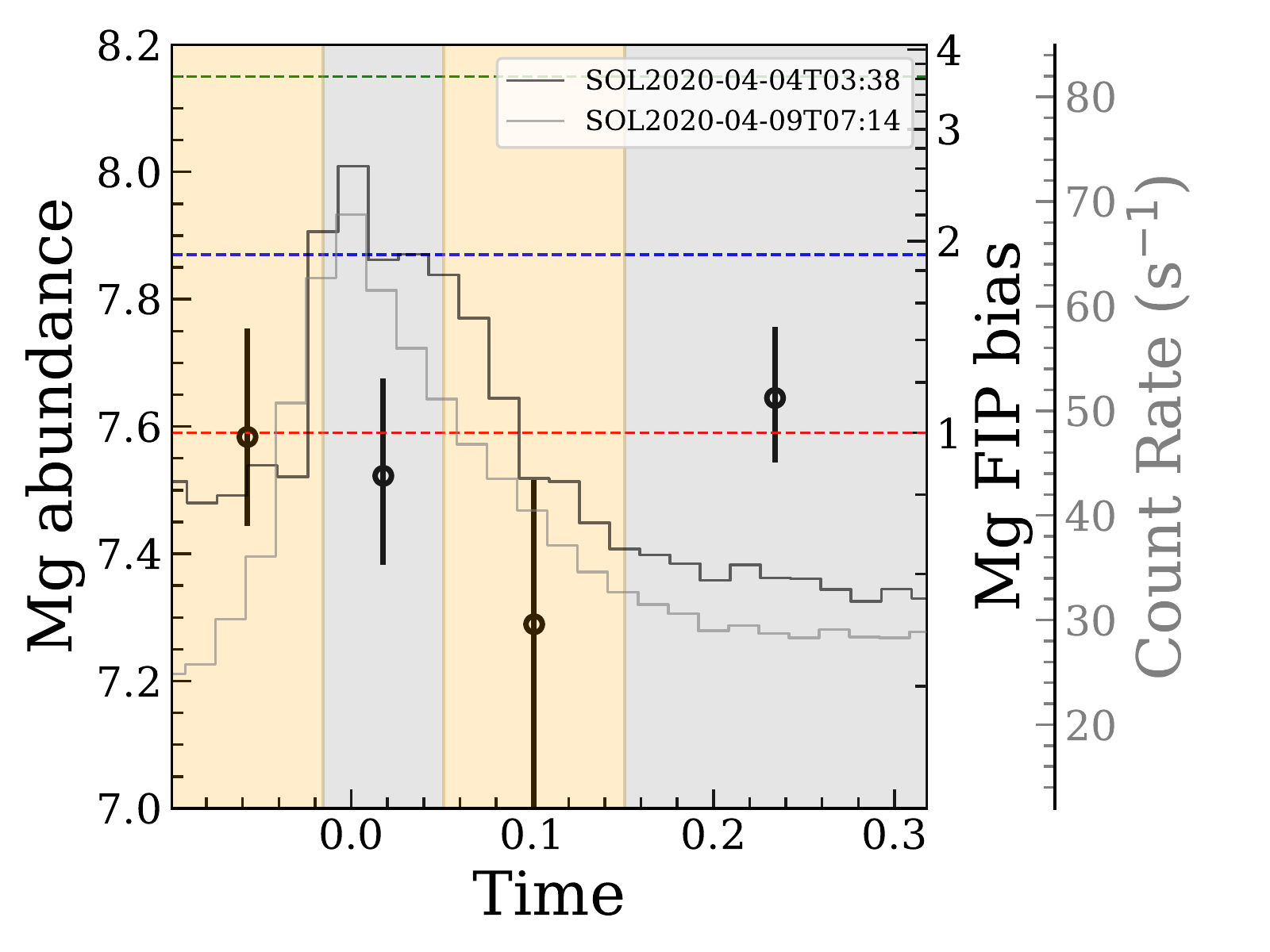} }} 
    \hfill
    \subcaptionbox{}{{\includegraphics[width=0.44\textwidth]{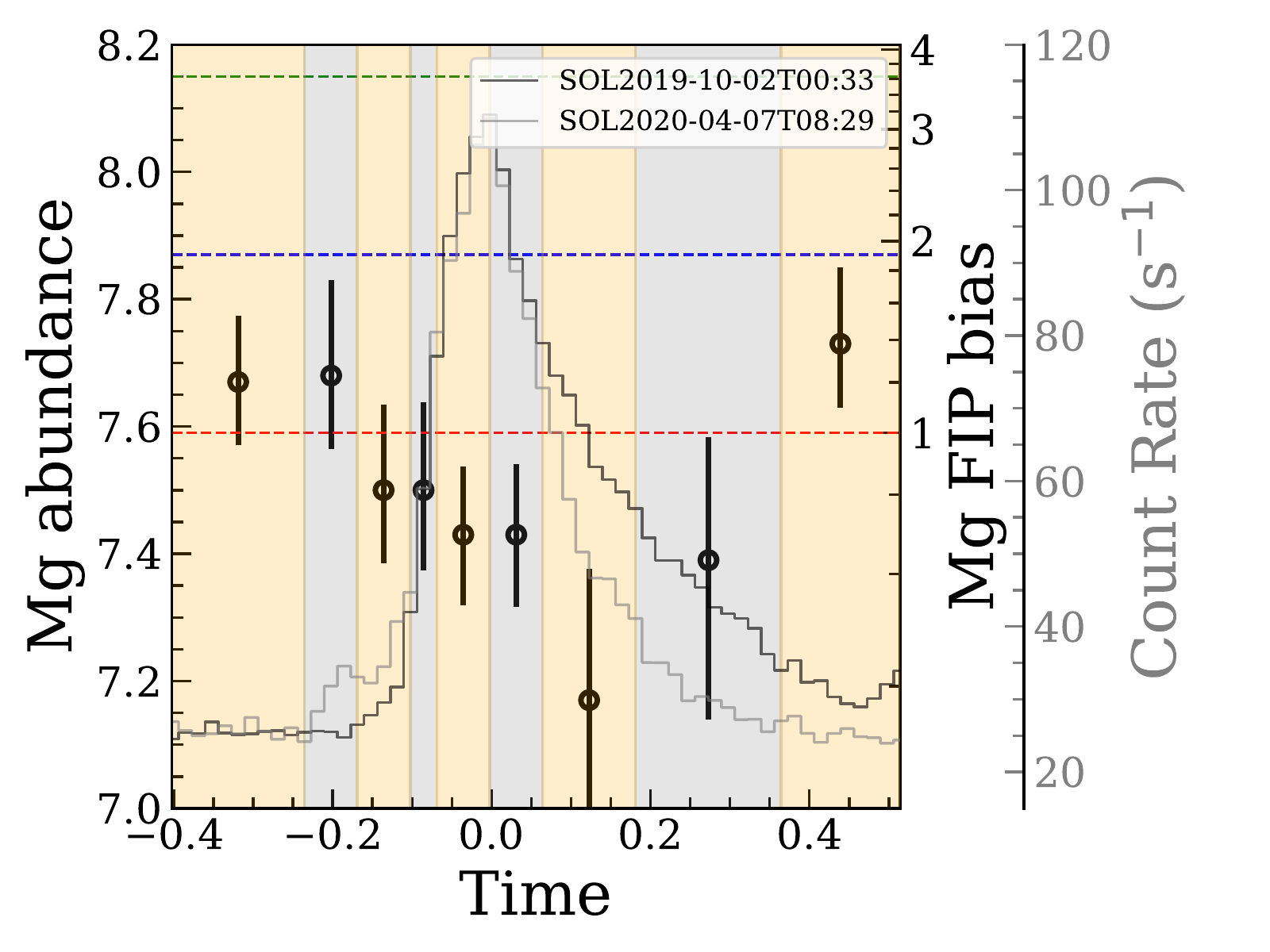} }}
    \caption{Evolution of absolute Magnesium abundance for the remaining sets of flares, similar to Figure~\ref{fig7} Panels {\bf g-i}}
    \label{fig11}
\end{figure}
 
\clearpage
\begin{figure}[ht] 
    
    \subcaptionbox{}{{\includegraphics[width=0.43\textwidth]{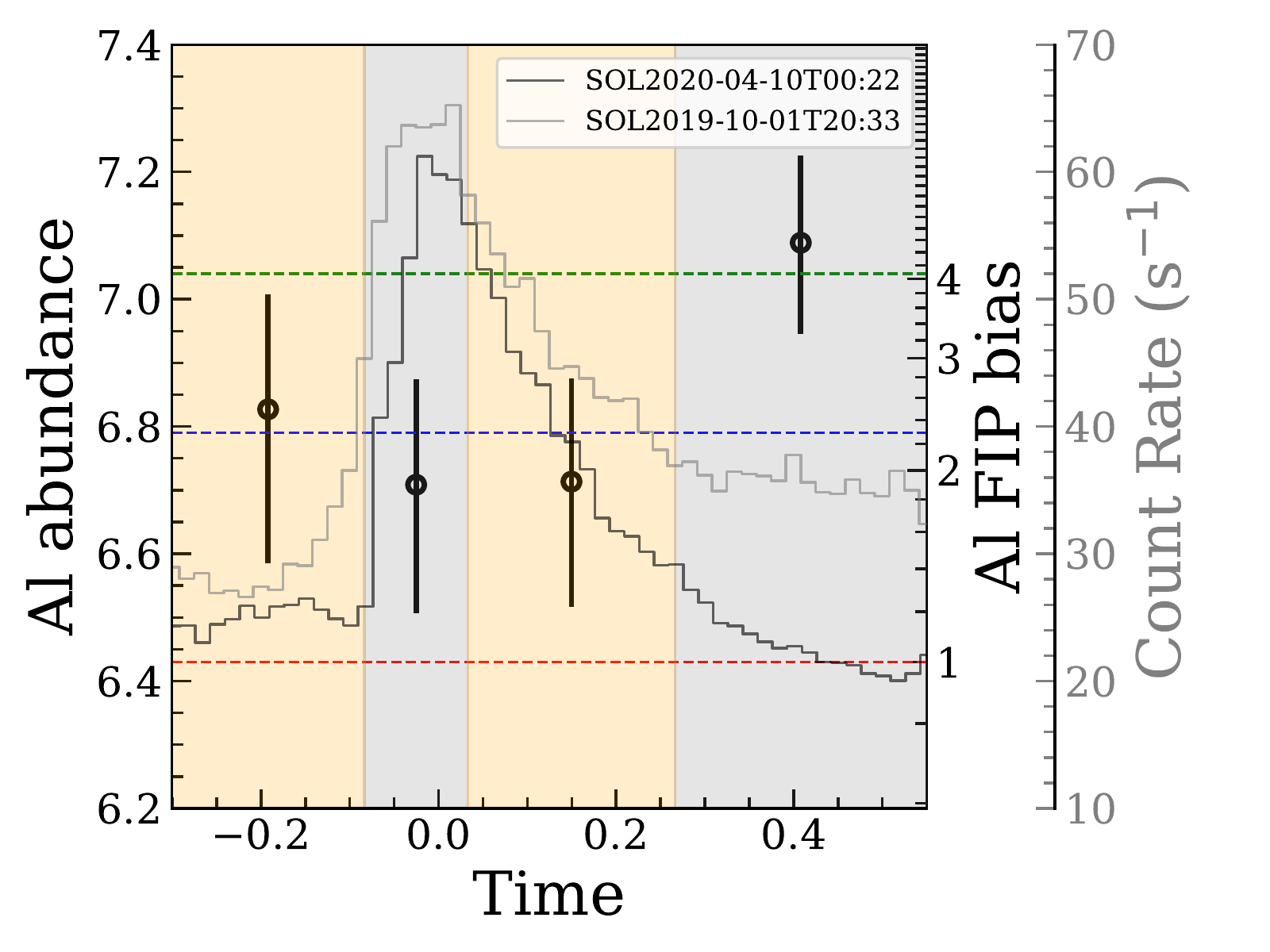} }}
    \hfill
    \subcaptionbox{}{{\includegraphics[width=0.43\textwidth]{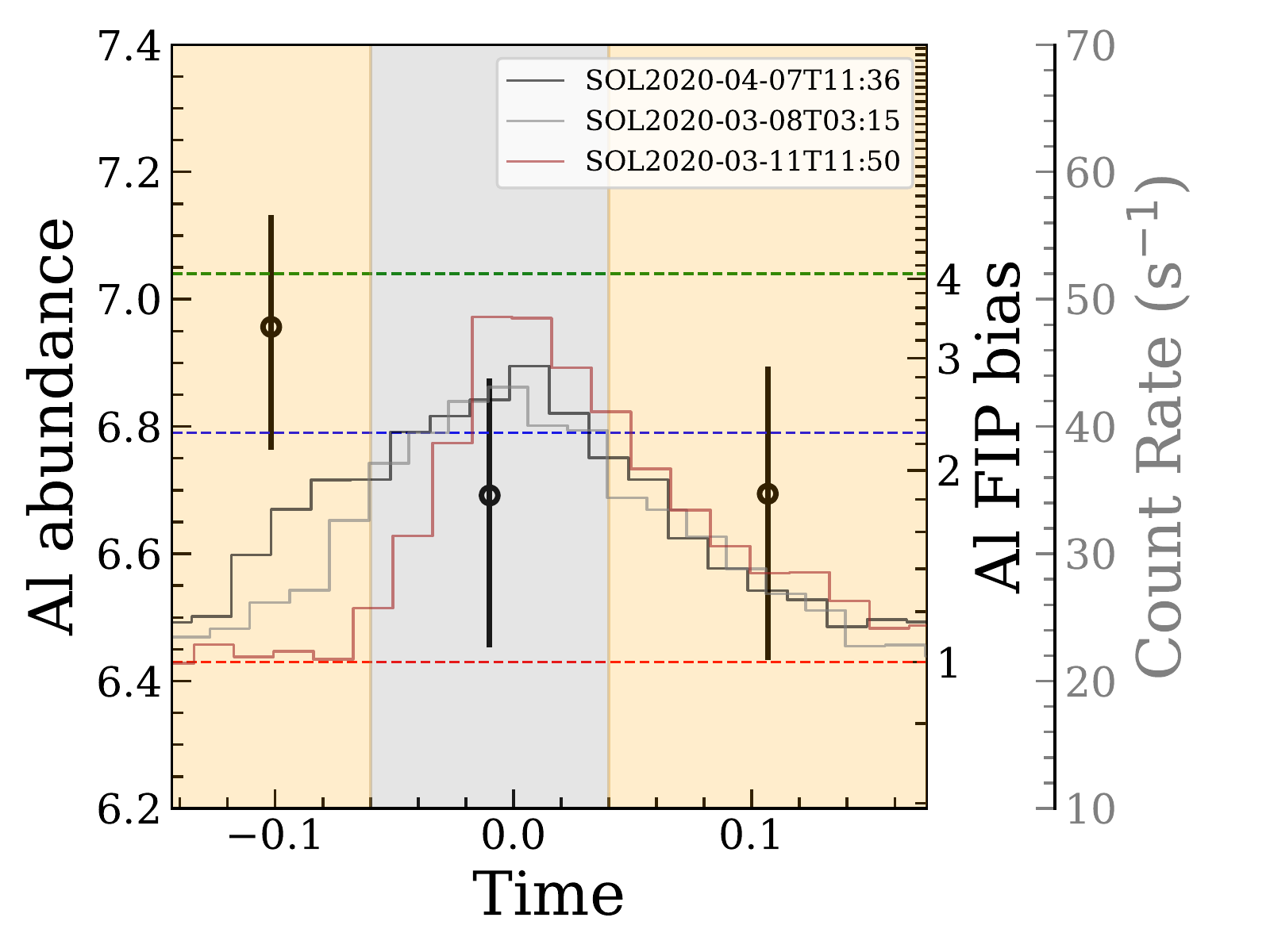} }}
    
    \subcaptionbox{}{{\includegraphics[width=0.44\textwidth]{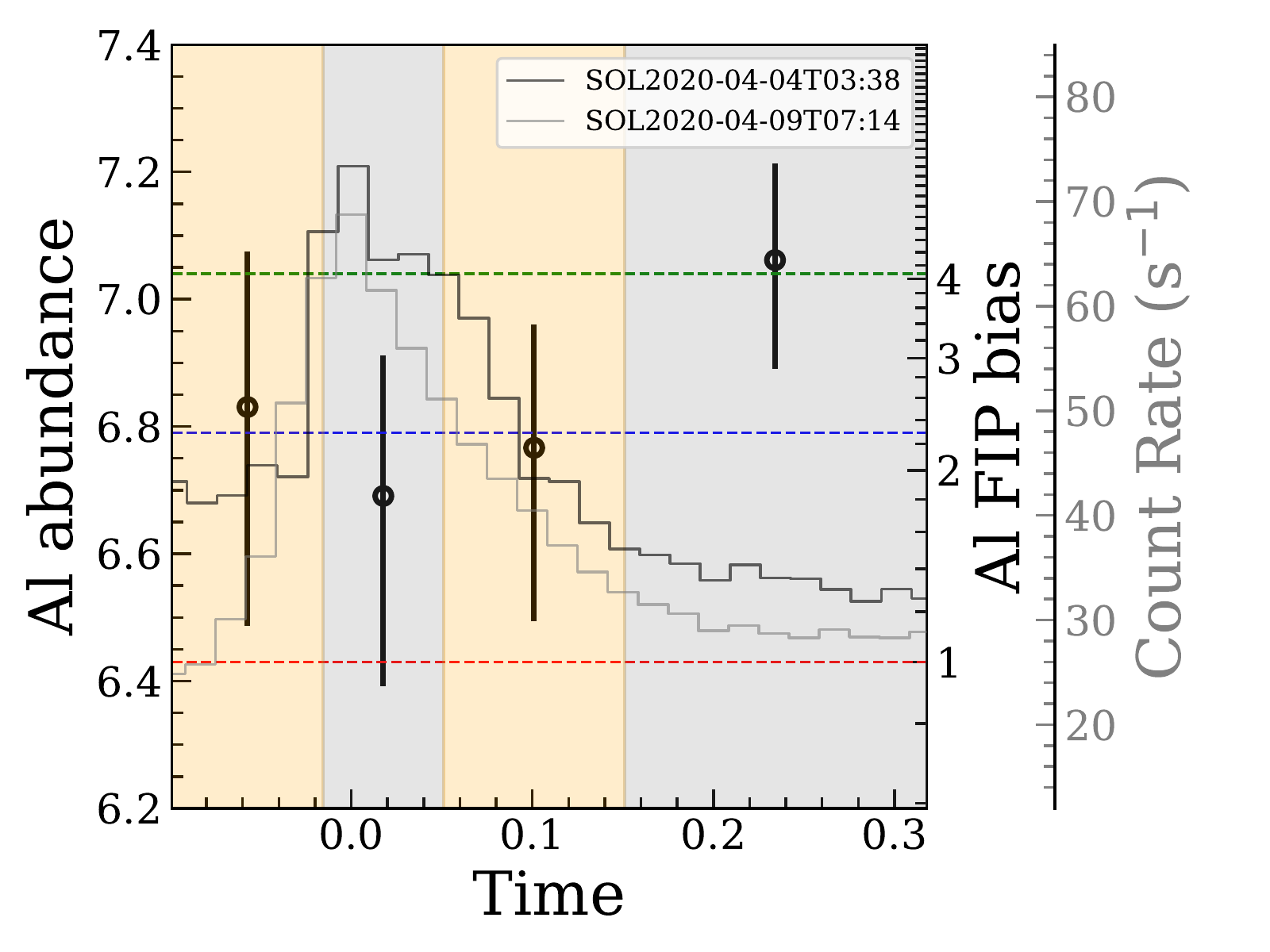} }} 
    \hfill
    \subcaptionbox{}{{\includegraphics[width=0.44\textwidth]{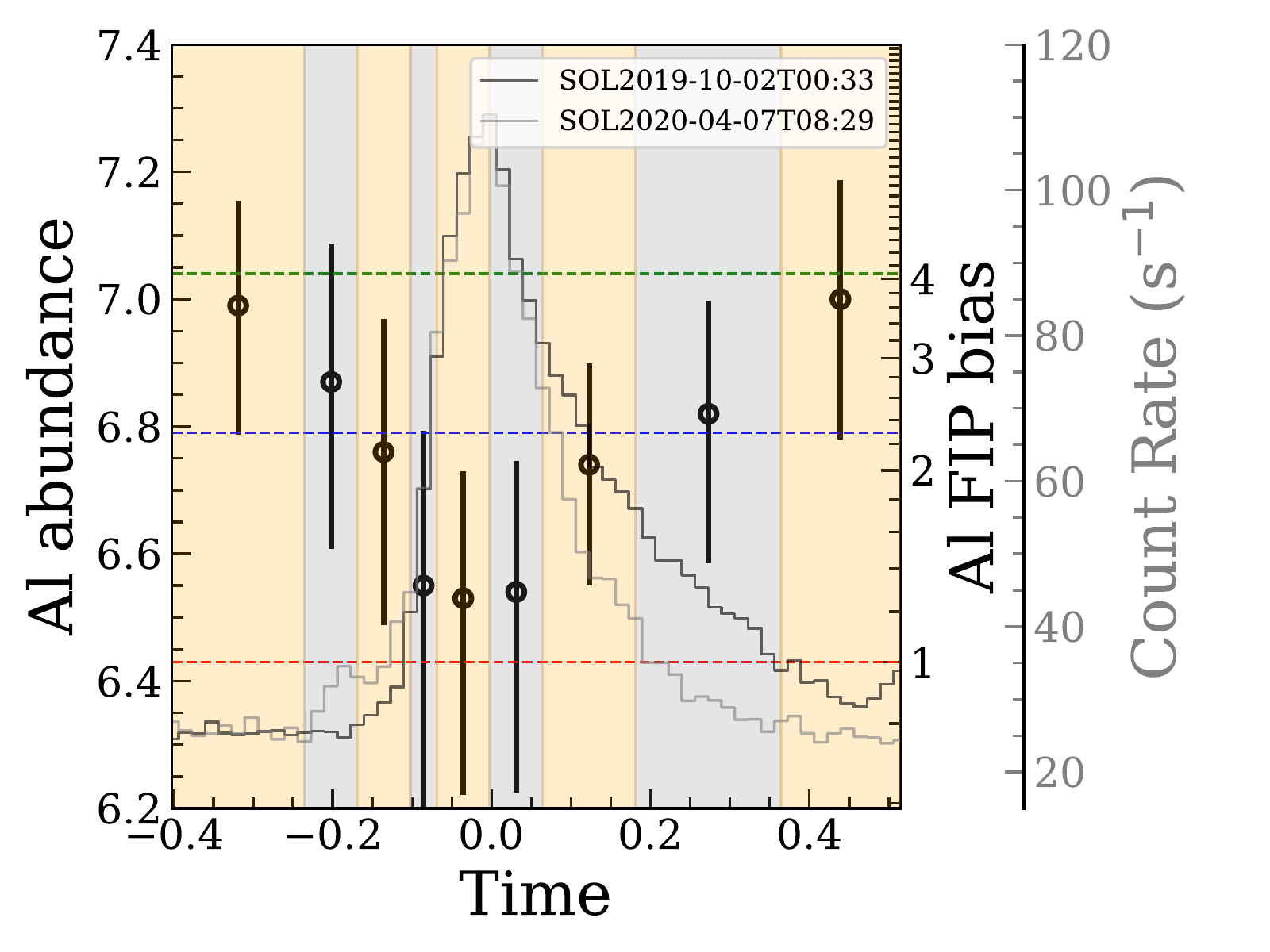} }}
    \caption{Evolution of absolute Aluminium abundance for the remaining sets of flares, similar to Figure~\ref{fig7} Panels {\bf j-l} }
    \label{fig12}
\end{figure}
 
\clearpage
\begin{figure}[ht] 
    
    \subcaptionbox{}{{\includegraphics[width=0.43\textwidth]{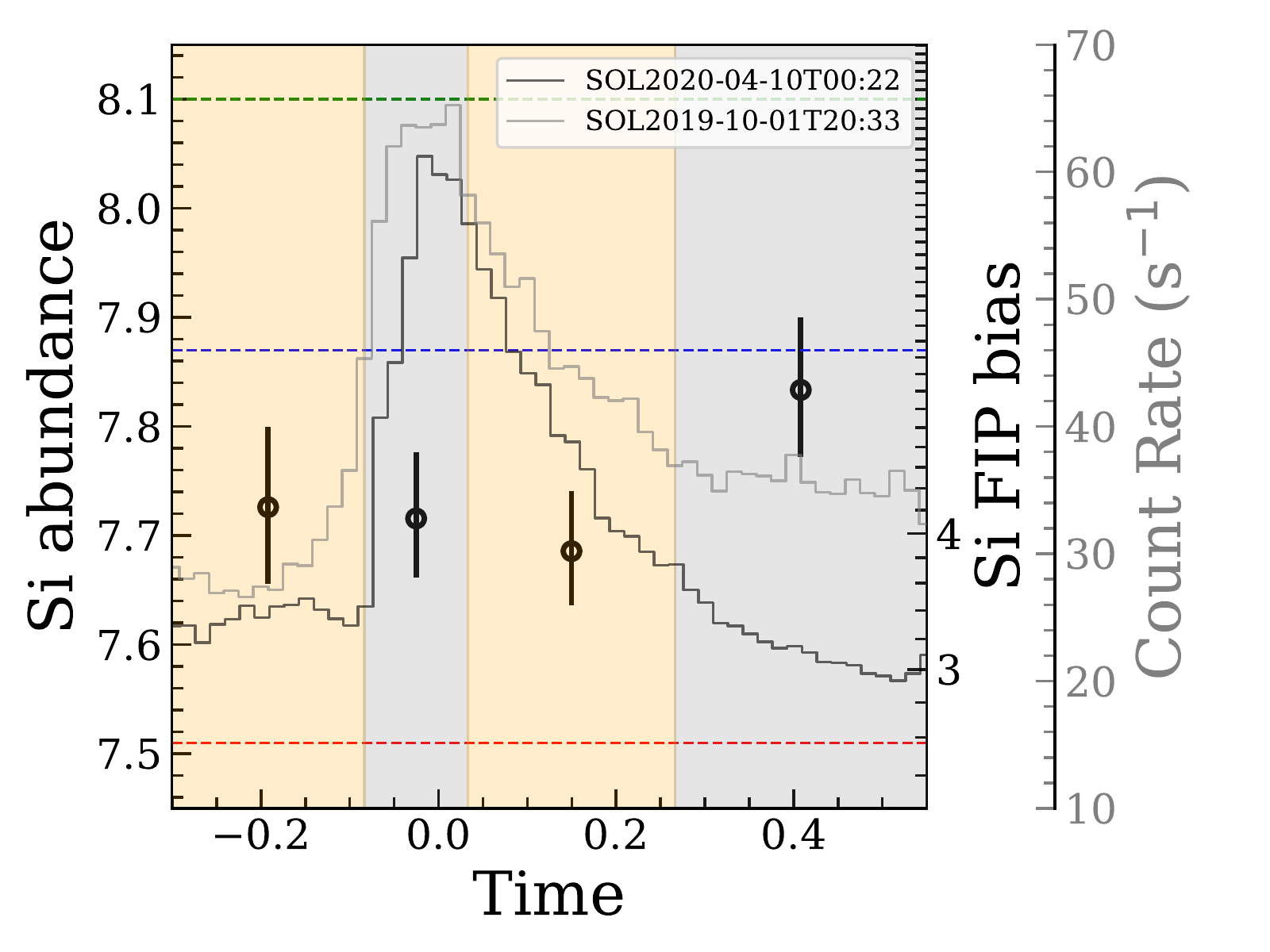} }}
    \hfill
    \subcaptionbox{}{{\includegraphics[width=0.43\textwidth]{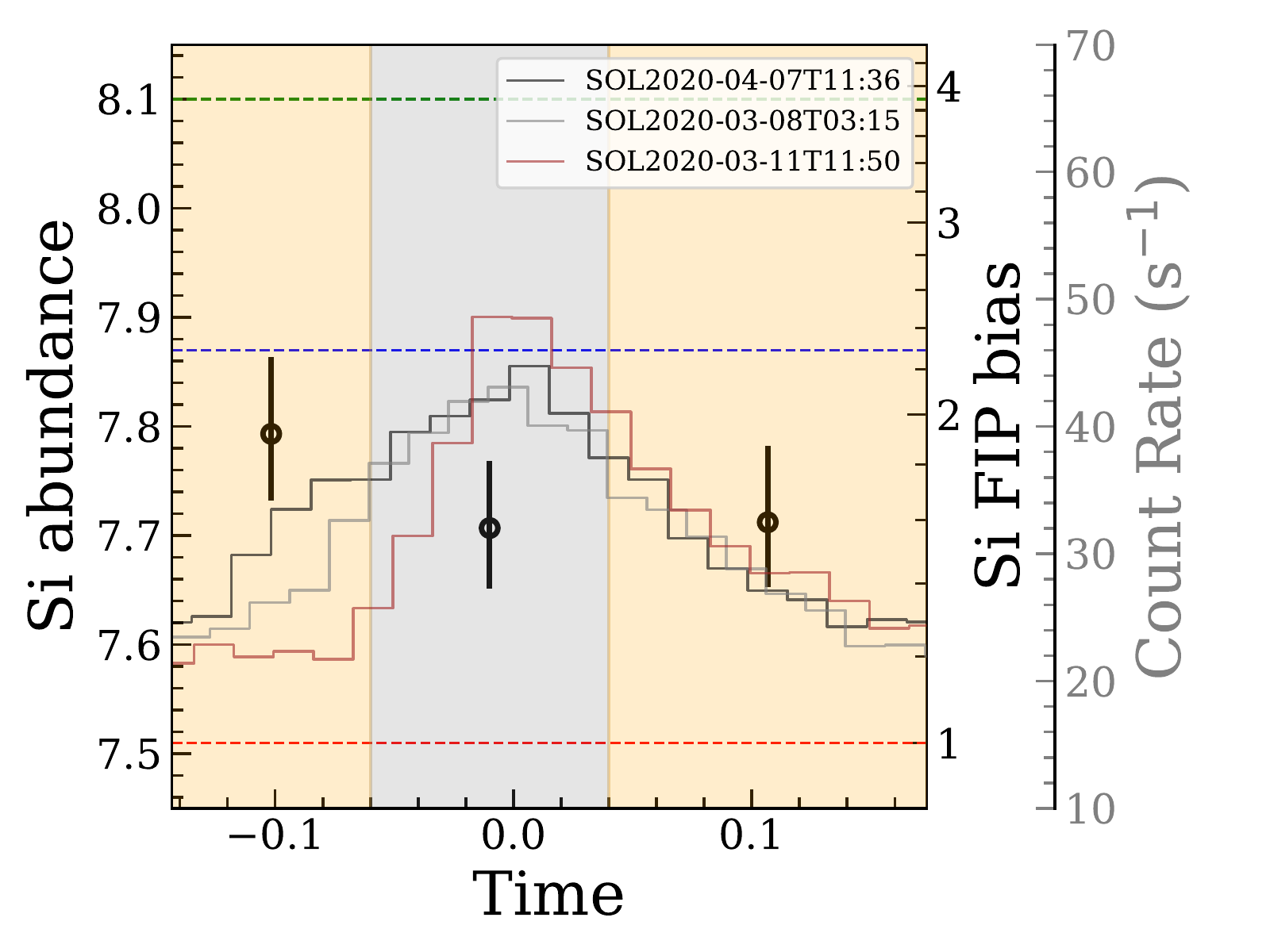} }}
    
    \subcaptionbox{}{{\includegraphics[width=0.44\textwidth]{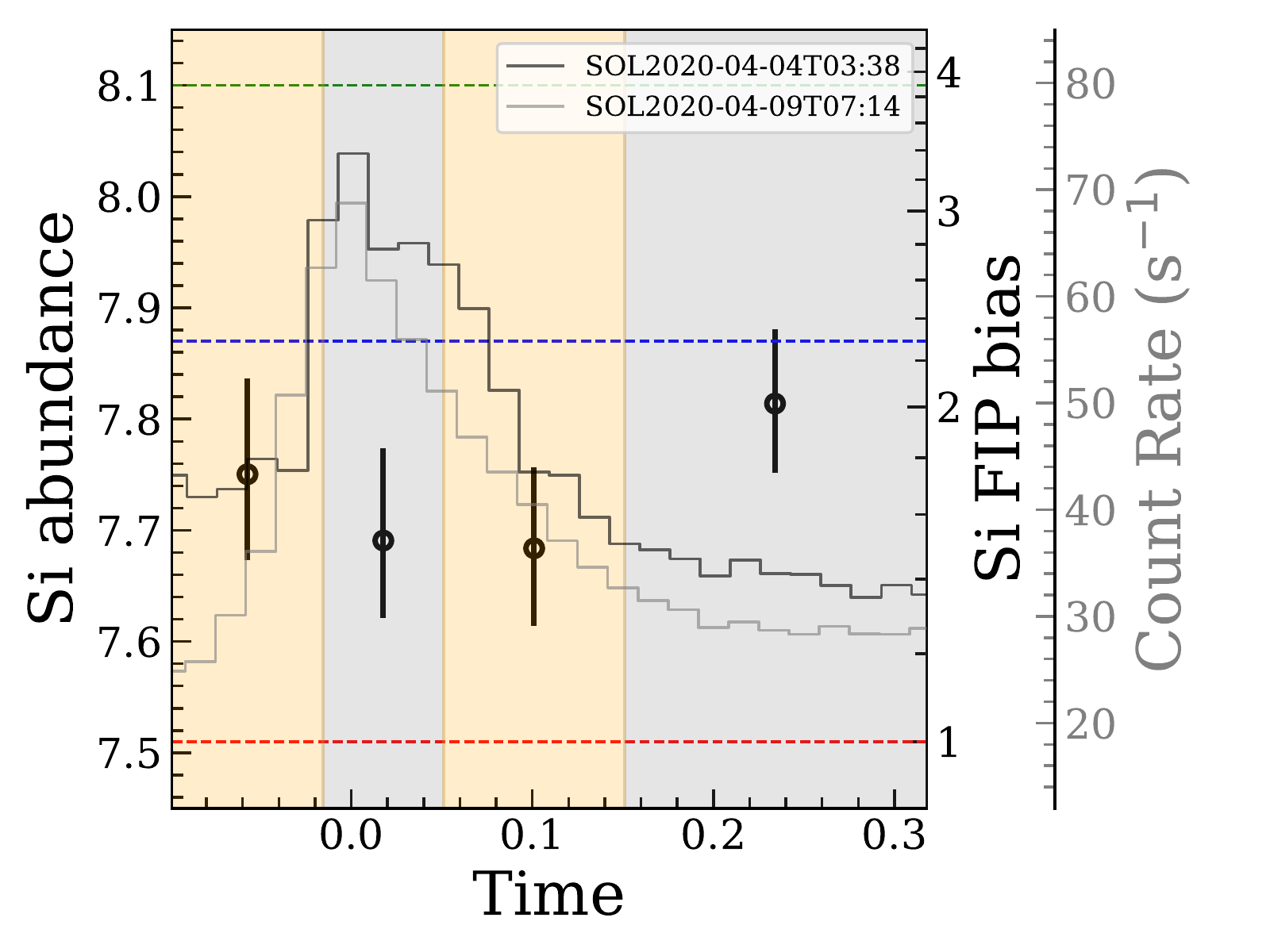} }} 
    \hfill
    \subcaptionbox{}{{\includegraphics[width=0.44\textwidth]{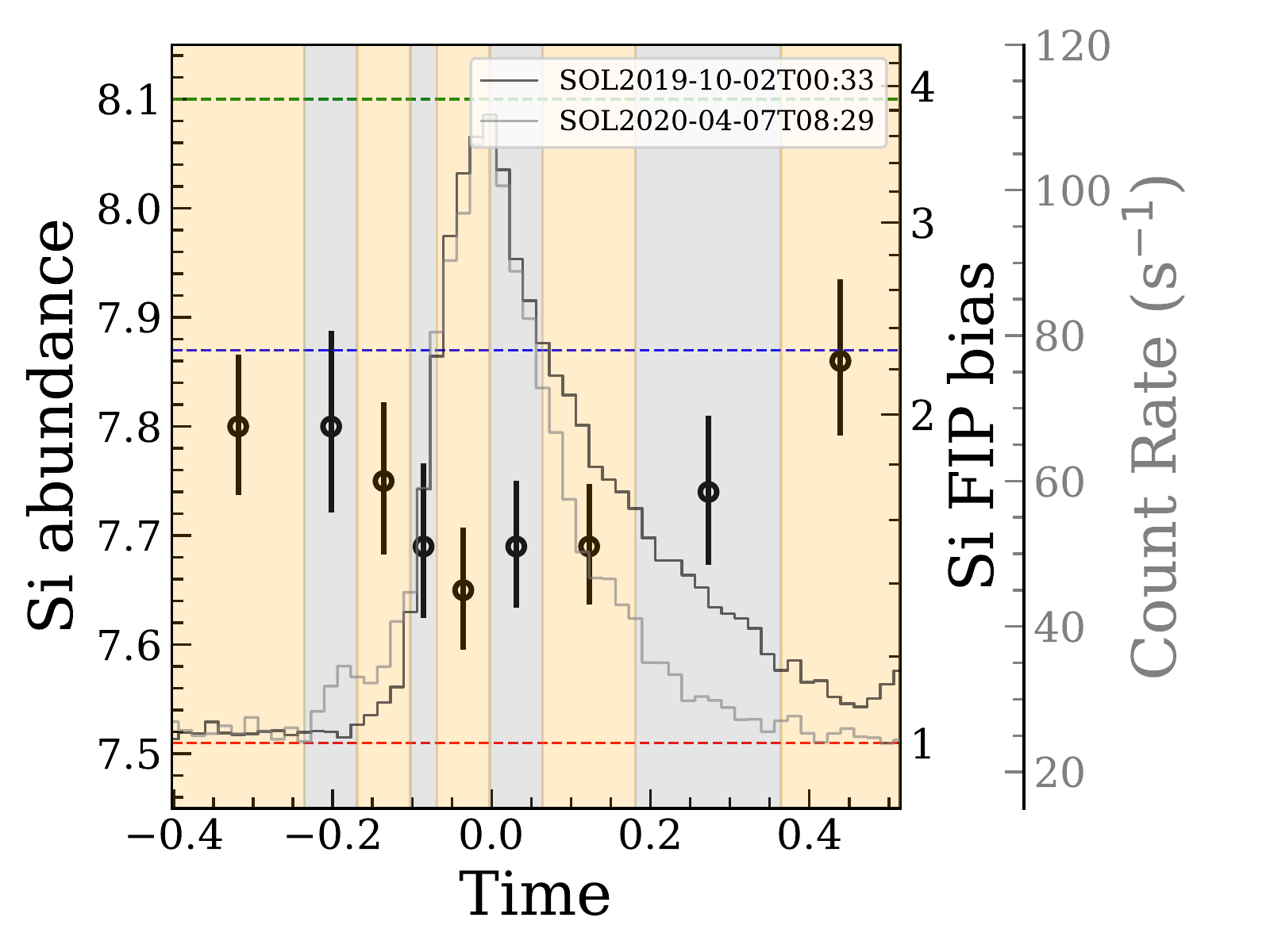} }}
    \caption{Evolution of absolute Silicon abundance for the remaining sets of flares, similar to Figure~\ref{fig7} Panels {\bf m-o} }
    \label{fig13}
\end{figure}
 
\begin{figure}[ht]    
\centerline{\includegraphics[width=0.44\textwidth,clip=]{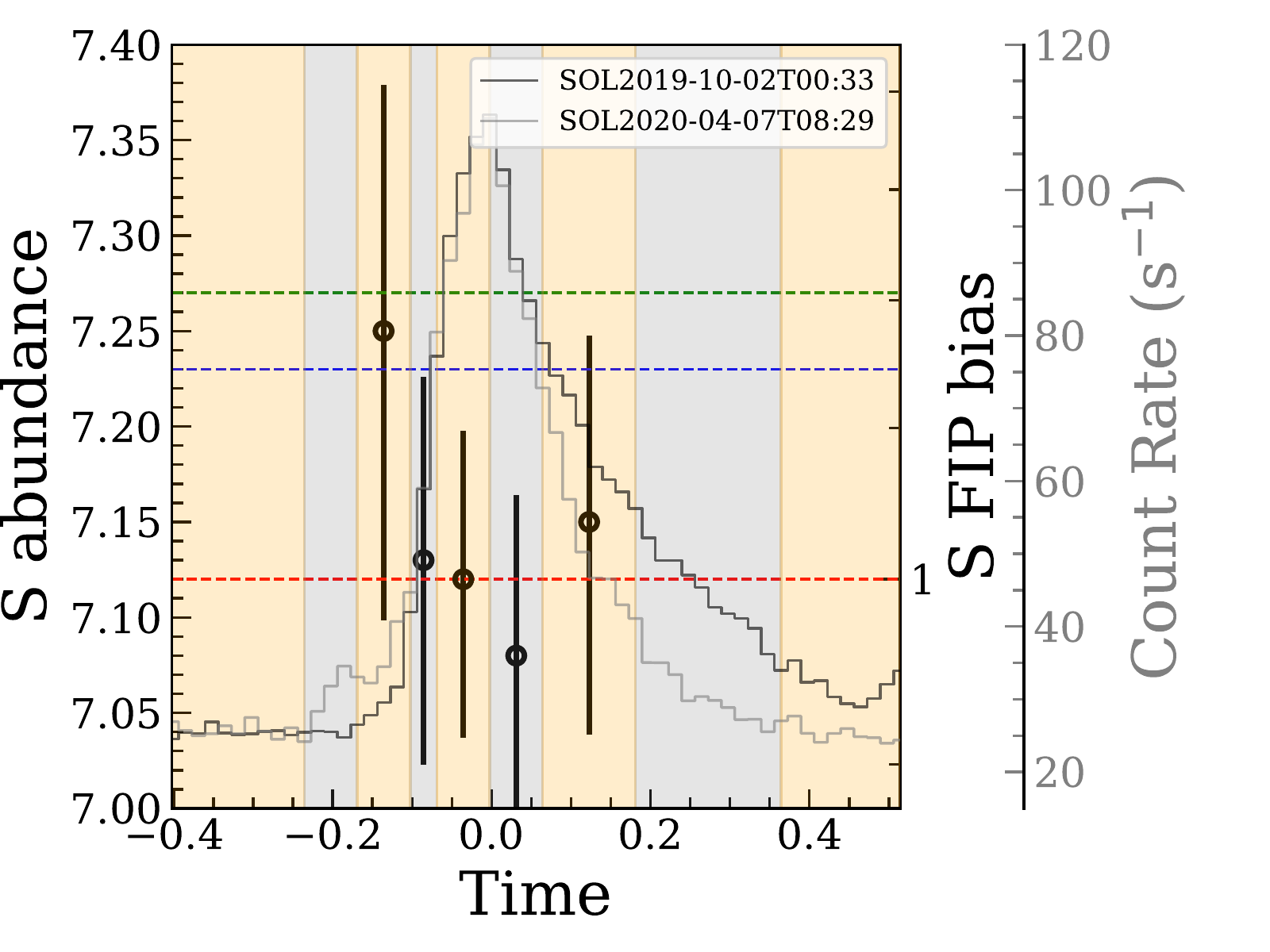}}
\caption{Evolution of absolute Sulfur abundance for the last set of flares, similar to Figure~\ref{fig7} Panel {\bf p}}
   \label{fig14}
\end{figure}

\newpage
\bibliographystyle{spr-mp-sola}
\bibliography{library}

\end{article} 

\end{document}